\documentclass[12pt]{article}
\pdfoutput=1

\newlength{\abstractwidth}
\usepackage{epsf}
\usepackage{color}
\usepackage{graphicx}
\usepackage{hyperref}
\hypersetup{colorlinks=true, citecolor=blue, linktoc=page}
\usepackage{dsfont}

\usepackage{tikz}
\usetikzlibrary{calc} 
\usetikzlibrary{patterns,snakes} 
\usetikzlibrary{decorations.pathreplacing} 
\usetikzlibrary{decorations.markings} 
\usetikzlibrary{decorations.pathmorphing} 
\usetikzlibrary{positioning}
\usetikzlibrary{arrows.meta}

\usepackage{amsmath, nccmath}
\usepackage{amsmath, amsthm}
\usepackage{amssymb}
\usepackage{amsfonts}
\usepackage{latexsym}
\usepackage{mathtools}
\usepackage[all,cmtip]{xy}

\usepackage{subfigure}

\renewcommand{\thefootnote}{\fnsymbol{footnote}}
\renewcommand{\thanks}[1]{\footnote{#1}}
\newcommand{\starttext}{
\setcounter{footnote}{0}
\renewcommand{\thefootnote}{\arabic{footnote}}}

\numberwithin{equation}{section}
\newcommand{\bea}{\begin{eqnarray}}
\newcommand{\eea}{\end{eqnarray}}
\newcommand{\be}{\begin{eqnarray}}
\newcommand{\ee}{\end{eqnarray}}
\def\beq{\begin{equation}}
\def\eeq{\end{equation}}

\newcommand{\bma}{\begin{matrix}}
\newcommand{\ema}{\end{matrix}}

\def\cA{{\cal A}}
\def\cB{{\cal B}}
\def\cC{{\cal C}}

\def\cG{{\cal G}}

\def\cM{{\cal M}}
\def\cN{{\cal N}}
\def\cO{{\cal O}}

\def\cQ{{\cal Q}}

\def\cU{{\cal U}}
\def\cV{{\cal V}}
\def\cW{{\cal W}}

\def\cY{{\cal Y}}

\def\mN{\mathfrak{N}}

\def\mo{\mathfrak{o}}

\def\ms{\mathfrak{s}}

\def\CC{{\mathbb C}}

\def\RR{{\mathbb R}}
\def\ZZ{{\mathbb Z}}

\def\Re{{\rm Re \,}}
\def\Im{{\rm Im \,}}
\def\tr{{\rm tr}}

\def\det{{\rm det \,}}

\def\half{{1\over 2}}

\def\p{\partial}

\def\a{\alpha}
\def\b{\beta}

\def\ep{\varepsilon}

\def\G{\Gamma}

\def\pbw{\p _{\bar w}}

\def\vol{{\rm vol}}
\def\tz{{\tilde z}}
\def\tw{{\tilde w}}
\def\sw{{w^\star}}

\DeclareMathOperator*{\Res}{Res}

\def\no{\nonumber}
\def\sm{\smallskip}

\definecolor{Cyan}{cmyk}{1.,0,0,0}
\definecolor{Magenta}{cmyk}{0,1.,0,0}
\definecolor{Yellow}{cmyk}{0,0,1.,0}
\definecolor{White}{cmyk}{0,0,0,0}
\definecolor{Orange}{cmyk}{0,0.61,0.87,0}
\definecolor{RedOrange}{cmyk}{0,0.77,0.87,0}
\definecolor{Red}{cmyk}{0,1.,1.,0}
\definecolor{Purple}{cmyk}{0.45,0.86,0,0}
\definecolor{Violet}{cmyk}{0.79,0.88,0,0}
\definecolor{Blue}{cmyk}{1,0.5,0,0}
\definecolor{ProcessBlue}{cmyk}{0.96,0,0,0}
\definecolor{GreenYellow}{cmyk}{0.6,0,1.,0}
\definecolor{Black}{cmyk}{0,0,0,1}
\usepackage{fix-cm}

\begin{document}
\starttext
\setcounter{footnote}{0}


\vskip 0.3in

\begin{center}

{\Large \bf Type IIB supergravity solutions for polarized IKKT}

\vskip 0.1in

\vskip 0.4in

{\large Eric D'Hoker${}^{(a)}$, Michael Gutperle${}^{(a)}$ and Christoph F.~Uhlemann${}^{(b)}$} 

\vskip 0.2in

{ \sl ${}^{(a)}$Mani L. Bhaumik Institute for Theoretical Physics}\\
{\sl  Department of Physics and Astronomy}\\
{\sl University of California, Los Angeles, CA 90095, USA}

 {\tt \small dhoker@physics.ucla.edu, gutperle@physics.ucla.edu}

\vskip 0.15in

{ \sl ${}^{(b)}$Theoretische Natuurkunde, Vrije Universiteit Brussel and}\\
{\sl  The International Solvay Institutes, Pleinlaan 2, B-1050 Brussels, Belgium}

{\tt \small christoph.uhlemann@vub.be}

\vskip 0.1in

\vskip 0.2in

\begin{abstract}
\vskip 0.1in

The general local solutions to complexified Type IIB supergravity which are invariant under the complex Lie superalgebra $F(4)$, and which were derived and classified in arXiv:2508.14959, are supplemented here by the construction of their flux potentials and field strengths. We then focus on the Type IIB$^\star$ real form of complex Type IIB supergravity, where we obtain solutions with spacetime metric of the warped product form $dS_{1,5}\times S^2\times\Sigma$. We systematically derive boundary and regularity conditions under which the $dS_{1,5}\times S^2\times\Sigma$ solutions can be analytically continued to regular $S^6\times S^2\times\Sigma$ solutions in Euclidean Type IIB supergravity. Using these results, we construct explicit global solutions. These solutions include explicit realizations of recent proposals for a holographic description of the polarized IKKT model, and we clarify the charges carried by these solutions. We also obtain novel solutions in the same regularity class with different brane charges.

\end{abstract}
\end{center}

\newpage

\setcounter{tocdepth}{2} 
\tableofcontents

\baselineskip=15pt
\setcounter{equation}{0}
\setcounter{footnote}{0}

\newpage

\section{Introduction}
\setcounter{equation}{0}
\label{sec:1}

Matrix models provide an intriguing setup in which spacetime along with string theory or M-theory can emerge as effective descriptions from the basic matrix degrees of freedom. The Banks-Fischler-Shenker-Susskind (BFSS) theory \cite{Banks:1996vh} conjecturally provides a non-perturbative formulation of M-theory based on the matrix quantum mechanics of multiple D0 branes in Type IIA superstring theory. The matrix model of  Ishibashi-Kawai-Kitazawa-Tsuchiya (IKKT) \cite{Ishibashi:1996xs}  was conceived as a discretized non-perturbative formulation of the Type IIB superstring. It is defined by reducing ten-dimensional $SU(N)$ super Yang-Mills theory to zero dimensions. This model also describes the non-abelian collective coordinates of $N$ D-instantons \cite{Gibbons:1995vg} and its partition function can be related to D-instanton-induced higher derivative  corrections to the Type IIB action  \cite{Green:1997tv,Krauth:1998xh,Moore:1998et}. 

\sm

In both putative non-perturbative formulations of string theory and M-theory, geometric descriptions are expected to emerge from the matrix degrees of freedom in the large~$N$ limit. The two models cannot be considered, however, to enjoy an equal footing. In the relation between the BFSS theory and M-theory, space is emergent while time is already present in the matrix quantum mechanics. For the IKKT model, on the other hand, any geometric description has to be fully emergent, since neither space nor time are present in the formulation of the matrix model. While the predictions of the BFSS matrix theory have been widely tested against predictions from holography (see \cite{Lin:2025iir} for a review from a modern point of view), much less is known about the IKKT model. Specifically, without time, the IKKT model lacks a natural notion of energy and D-instantons do not exhibit the decoupling limit familiar from well-established holographic descriptions \cite{Ooguri:1998pf}. 

\sm

Progress was made recently by considering a deformation of the IKKT model with Euclidean signature. The deformation was introduced in \cite{Bonelli:2002mb} and the resulting model is referred to as \textit{polarized IKKT model} \cite{Hartnoll:2024csr}. The deformation is similar in spirit  to the Berenstein-Maldacena-Nastase (BMN) \cite{Berenstein:2002jq} deformation of the BFSS matrix model. The polarized IKKT model is invariant under an $\ms \mo (7) \oplus \ms \mo (3)$ subalgebra of the $\ms \mo (10)$ symmetry of the undeformed model  as well as under 16 supersymmetries \cite{Hartnoll:2024csr,Hartnoll:2025ecj}. Holographic descriptions were proposed in \cite{Hartnoll:2024csr,Komatsu:2024bop}.\footnote{Other recent approaches to IKKT holography can be found in \cite{Ciceri:2025maa,Ciceri:2025wpb,Hahner:2026mfg}.} In particular, the symmetry properties have led to the suggestion that the polarized IKKT model possesses a holographic dual in Type IIB supergravity whose spacetime is $S^6 \times S^2$ warped over a Riemann surface \cite{Komatsu:2024bop,Komatsu:2024ydh}. 

\sm

More precisely, two avenues have been explored in the study of possible holographic duals to the polarized IKKT model. A first avenue, pursued in \cite{Hartnoll:2024csr},  is based on analyzing the dynamics of a probe D1 brane in the background of a solution dubbed \textit{the cavity}.  A second avenue, pursued in  \cite{Komatsu:2024bop,Komatsu:2024ydh}, is based on the construction of fully back-reacted $S^6\times S^2\times\Sigma$ solutions to the field equations by analytic continuation from the local form of the known $AdS_{1,5}\times S^2\times\Sigma$ solutions of \cite{DHoker:2016ujz,DHoker:2017mds}. The unsettled fate of supersymmetry in the second approach and the inherent limitations of the probe brane construction used in the first approach, combined with the presence of disagreements between the two avenues regarding the involved brane charges, motivates a more systematic investigation.

\sm

Additionally, there are practical motivations. The constructions of \cite{Komatsu:2024bop,Komatsu:2024ydh}  are based on an electrostatic reformulation of the  $AdS_{1,5} \times S^2\times\Sigma$ solutions of \cite{DHoker:2016ujz,DHoker:2017mds} in which the two holomorphic functions that parametrize all solutions in the original construction are represented by a single harmonic function. This is achieved by gauge fixing the conformal reparametrization symmetry of the solutions, through which one holomorphic function is adopted as a complex coordinate on $\Sigma$, as previously employed e.g.\ in \cite{Legramandi:2021uds}. A downside to this approach is that it obscures the natural action of $SL(2)$ symmetry under which the pair of holomorphic functions forms a doublet. The gauge fixing further complicates a systematic classification of admissible solutions, as it conflates singular behavior in the holomorphic functions with singularities resulting from the gauge fixing. A more systematic understanding and classification of fully back-reacted solutions to Type~IIB supergravity, whose symmetries match those of the polarized IKKT model,  without seeking recourse to gauge fixing conformal symmetry, is therefore  desirable.

\sm

In this paper we present such a systematic classification. To do so, we first review the symmetries of the polarized IKKT model. Its Lorentzian version has a bosonic symmetry, $\ms \mo (1,6) \, \oplus \, \ms \mo (3)$ or $\ms \mo (7) \, \oplus \, \ms \mo (1,2)$ depending on where the time direction resides, that fits with its 16 linearly realized supersymmetries into one of the real forms of the Lie superalgebra $F(4;\CC)$ (see \cite{DHoker:2025nid} and references therein). Its Euclidean version, which is obtained from the Lorentzian version by analytically continuing the time direction, has bosonic symmetry $\ms \mo (7) \oplus \ms \mo (3)$. The analytic continuation modifies the reality conditions on spinors and there exists no real form of $F(4;\CC)$, or of any other complex Lie superalgebra with 16 supersymmetries, with maximal bosonic subalgebra $\ms \mo (7) \oplus \ms \mo (3)$~\cite{DHoker:2025nid}. 

\sm
 
Nonetheless, there exist purely bosonic field configurations in the Euclidean IKKT model which are invariant under 16 ``complexified supersymmetries", even though these transformations do not preserve the integration contour of the Euclidean action.  These configurations in the Euclidean model may be regarded as analogous to half-BPS solutions. In conjunction with contour deformations they allow for efficient computations of the integral using supersymmetric localization. This situation is analogous, for example, to the features of Euclidean $\mathcal N=4$ SYM and its study using supersymmetric localization~\cite{Pestun:2007rz}.

\sm

A compelling holographic Type IIB dual to the polarized IKKT model must, at the very least, match all symmetries. We shall now explain how the symmetries may be matched perfectly. All local solutions to complexified Type~IIB$_\CC$ supergravity  on product spaces  $\cM_{6 \CC} \times \cM_{2\CC}$ warped over a Riemann surface $\Sigma$ were constructed explicitly  in \cite{DHoker:2025nid}. Real forms of the complex solutions exist in real forms of Type IIB supergravity, and all real form solutions in all Type IIB real forms were classified there. As expected, each half-BPS real form solution corresponds to a real form of $F(4;\CC)$.
This includes the solutions to standard Type IIB supergravity with Minkowski signature that are invariant under one of the real forms of  $F(4;\CC)$, namely the $AdS_{1,5} \times S^2$ solutions of \cite{DHoker:2016ujz} and the $AdS_{1,1} \times S^6$ solutions of \cite{Corbino:2017tfl}, in both cases warped over a Riemann surface $\Sigma$ with Euclidean signature. It also includes many more real form solutions to other real forms of complex Type IIB supergravity. Here we will use these results and start from a real form of the solutions in Type IIB$^\star$ supergravity, and subsequently obtain $S^6\times S^2$ solutions in complex Type~IIB by analytic continuation.

\sm

The real form solutions with one time-like direction to the real forms Type IIB$^\star$ and Type IIB$^\prime$ of Type IIB$_\CC$ supergravity are those for which the real form of $F(4;\CC)$ has the following maximal bosonic subalgebra and corresponding spacetime,
\begin{align}
	& \ms \mo(1,6) ~ \oplus ~ \ms \mo(3) & & dS_{1,5} ~ \times ~ S^2 
	\no \\
	& \ms \mo(1,2)  ~ \oplus ~ \ms \mo(7) & & dS_{1,1} ~ \times ~ S^6
\end{align}
in both cases warped over a Riemann surface $\Sigma$ with Euclidean signature. Either class of solutions may be analytically continued to a geometry of the form $S^6 \times S^2$ warped over a surface $\Sigma$ with Euclidean signature. The analytic continuation maps solutions of the Lorentzian signature supergravity Bianchi identities and bosonic field equations to solutions of the Euclidean signature supergravity Bianchi identities and bosonic field equations, so that we obtain genuine  $S^6 \times S^2$ solutions. The fate of the 16 supersymmetries of the Lorentzian solutions upon analytic continuation  is parallel to what happened for the polarized IKKT model: there exists no real form of $F(4;\CC)$ whose maximal bosonic subalgebra is $\ms \mo (7) \oplus \ms \mo (3)$, as is reflected in the fact that the 16 spinor solutions to the BPS equations do not satisfy the reality conditions for Euclidean signature. Nonetheless, these 16 solutions to the BPS equations  exist and may be viewed as generalized supersymmetries, in parallel to the situation for the Euclidean polarized IKKT model.

\sm

Whether or not the Lorentzian signature solutions with $dS_{1,5} \times S^2 $ or $dS_{1,1}  \times  S^6$ symmetric spaces could be regarded as candidate holographic duals for the Lorentzian polarized IKKT models with the corresponding signatures is an interesting question whose study lies beyond the present paper. Certainly, their full Lie superalgebras match in both cases.

\subsection*{Summary of results}

The $F(4,\CC)$-invariant solutions to complexified Type IIB supergravity were constructed locally  in \cite{DHoker:2025nid} by solving the complexified BPS equations for 16 complex Weyl spinors.
The general local solution is parametrized by two pairs  of  functions $\cA_{1,2}(w)$ and $\tilde \cA_{1,2}(\tilde w)$  which are locally holomorphic with respect to independent complex coordinates $w$ and $\tilde w$ of the complexified Riemann surface, respectively.

\sm

Here, we complete the earlier construction of these solutions by explicitly solving the Bianchi identities in terms of NSNS and RR gauge potentials. Having the gauge potentials in explicit form will facilitate the analysis of the various (brane) charges of concrete solutions, while solving  the Bianchi identities will fix a sign variable associated with a branch choice that was left undetermined in \cite{DHoker:2025nid}.
The real form solutions may be obtained by imposing suitable involution conditions on the two  complex coordinates $w, \tilde w$ as well as on the pairs  of locally holomorphic functions $\cA_{1,2}(w)$ and $ \tilde \cA_{1,2}(\tilde w)$.

\sm

In order to match the symmetries of the polarized IKKT model, we shall consider solutions to the  Type IIB$^\star$ real form of complex Type IIB supergravity whose spacetime is a product space $dS_{1,5} \times S^2$ warped over a Riemann surface $\Sigma$ with Euclidean signature. These solutions are invariant under 
$\ms \mo(1,6)  \oplus  \ms \mo(3)$ and are parametrized by a pair of locally holomorphic functions $(\cA_1, \cA_2)$ on $\Sigma$, while the pair $(\tilde \cA_1, \tilde \cA_2)$ of the complex solution is determined in terms of $(\cA_1, \cA_2)$ by the reality conditions. Using both holomorphic functions (as opposed to eliminating one in favor of a particular gauge choice for conformal invariance) will allow for an efficient analysis of all the admissible singularities,  asymptotic regions, and permissible brane sources in the solutions.

\sm

With these building blocks for the construction of $dS_{1,5}\times S^2\times\Sigma$ solutions in hand, we  proceed to building global regular solutions with spacetime  $S^6\times S^2$ warped over a Riemann surface $\Sigma$ with the topology of the upper half plane. We note that the regularity conditions imposed on the $S^6\times S^2$ solutions are qualitatively different from those that were imposed on its $AdS_6 \times S^2$ or $AdS_2 \times S^6$ counterparts. The $AdS_6$ and $AdS_2$ components never collapse to zero size, while  both the $S^6$ and the $S^2$ factors of the $S^6\times S^2$ solutions will be allowed to collapse to zero size on the boundary of $\Sigma$. 

\sm 

In all solutions we construct, the boundary of $\Sigma$ is partitioned into a finite number of intervals on which the boundary conditions on $(\cA_1, \cA_2)$ alternate between $S^6$ collapsing and $S^2$ collapsing, with the transition points corresponding to square root branch points in the functions $(\cA_1,\cA_2)$. These alternating boundary segments lead to 3-cycles, 7-cycles and 9-cycles. We further show that regularity of the supergravity solutions requires the presence of at least one singularity in the functions $(\cA_1, \cA_2)$ on the boundary of $\Sigma$ at which an asymptotic region emerges. So the geometries are genuinely non-compact. These singularities include higher-order poles on $\partial\Sigma$ at which the geometry decompactifies, and which either lead to 3-cycles with $(p,q)$ 5-brane charge or to 7-cycles with $(p,q)$ string charge, as well as branch point singularities on $\partial\Sigma$ which lead to an asymptotically flat region and D-instanton charge.

\sm

As the simplest case, we recover the \textit{cavity} background of \cite{Hartnoll:2024csr}, which corresponds to partitioning $\partial\Sigma$ into two intervals and having one asymptotic region where the solution asymptotically becomes S-dual to a deformation of the D-instanton solution. We refine the discussion of the regularity properties of this solution in Type IIB supergravity and of its behavior under S-duality. In addition, we provide new explicit solutions with $\partial\Sigma$ partitioned into two segments, which have asymptotic regions with either $(p,q)$ 5-brane or $(p,q)$ string charge. They have the same symmetries but different asymptotics compared to the solutions of both \cite{Hartnoll:2024csr} and \cite{Komatsu:2024bop}, and we leave their physical interpretation open.

\sm

We then focus on solutions with a larger number of alternating boundary segments and a single asymptotic region with D-instanton asymptotics. We will give a fully explicit form for generic solutions with two segments where $S^6$ collapses and two segments where $S^2$ collapses, with the holomorphic functions $(\cA_1, \cA_2)$ expressed in terms of elliptic integrals. In the language of \cite{Komatsu:2024bop} these are single-disc solutions, for which explicit expressions were previously given in a certain limiting case. For solutions with more alternating boundary segments, corresponding to multi-disc configurations in the language of \cite{Komatsu:2024bop}, we explicitly construct the differentials $\partial\cA_{1,2}$, in terms of which the functions $\cA_{1,2}$ can be obtained as hyper-elliptic integrals. These explicit solutions should facilitate further detailed studies of the proposed duality with the polarized IKKT model.

\sm

As a first application, we clarify the charges carried by the solutions that were proposed to be dual to the polarized IKKT model in \cite{Komatsu:2024bop}. Fixing the asymptotic behavior to approach the D-instanton solution partially fixes an $SL(2,\RR)$ frame, and we can discuss individual D1, F1 and D5, NS5 charges. We find that our solutions generally carry all of these charges, including  F1 charge. Comparing with the solutions of \cite{Komatsu:2024bop}, we confirm their general parameter count and the relations between the D1, NS5 and D-instanton charges, all of which were crucial in establishing a correspondence with the polarized IKKT model. However, contrary to the conclusion in \cite{Komatsu:2024bop} that the general solution carries zero  F1 charge, we find this charge to be generally non-vanishing. As a consistency check, we will revisit the arguments in \cite{Komatsu:2024bop} and show that, for the limiting case of the single disc solutions for which explicit expressions were given there, a direct evaluation of the charge formulas provided  in \cite{Komatsu:2024bop} also produces non-vanishing F1 charge.

\sm

These results give a refined perspective on the relation between the two proposed holographic descriptions for the polarized IKKT model in \cite{Hartnoll:2024csr} and in \cite{Komatsu:2024bop}. This relation was previously discussed in the latter reference, whose conclusions may be summarized as follows. The behavior of the backreacted solutions of \cite{Komatsu:2024bop} in the asymptotically flat region is S-dual to the asymptotics of the cavity background used in \cite{Hartnoll:2024csr}. The probe D1 brane used in \cite{Hartnoll:2024csr} would therefore correspond to F1 charge in the backreacted solutions. Since \cite{Komatsu:2024bop} concluded that the backreacted solutions do not have F1 charge, the proposals appeared incompatible. Our conclusion in this paper, however, is that the backreacted solutions do carry F1 charge. This eliminates the immediate incompatibility between the charges. Whether the proposals can ultimately be connected or not remains to be seen.

\subsection*{Organization}

In section \ref{sec:2}, we review the BPS and field equations of Type IIB$_\CC$ supergravity and its $F(4;\CC)$ invariant solutions of \cite{DHoker:2025nid} and extend those results by giving explicit formulas for all flux fields, the associated potentials, and the charges in terms of the functions $\cA_{1,2}(w)$ that parametrize the solutions. 
Furthermore, we establish the conditions for collapsing one of the complexified symmetric spaces. In section \ref{sec:3} we discuss $dS_{1,5}\times S^2\times\Sigma$ solutions of Type IIB$^\star$ supergravity and determine the reality, 
regularity and positivity conditions. In section~\ref{sec:4} we construct solutions with geometry $S^6 \times S^2$ warped over a Riemann surface with Euclidean signature in terms of two holomorphic functions on the upper half complex plane. We implement the regularity and positivity conditions and classify admissible singularities of the holomorphic functions. Regular solutions are characterized by the number of boundary segments where $S^2$ or $S^6$ collapse, as well as further parameters. In section \ref{sec:L1-sol} we present the simplest solution with two such segments. We make contact with the cavity solution of \cite{Hartnoll:2024csr} and present new solutions with different asymptotics.  In section~\ref{sec:sol}, solutions with four segments are constructed explicitly in terms of elliptic functions. We discuss the charges and present illustrative numerical examples.

\sm

Appendix \ref{sec:A} contains useful intermediate formulas, while the derivation of the flux potentials is relegated to appendix \ref{sec:potentials}. In appendix \ref{app:cB-polylogs} we express $\cB$ for the solutions of section \ref{sec:sol} in terms of elliptic polylogarithms. The relation of the solutions constructed here to those found in \cite{Komatsu:2024bop,Komatsu:2024ydh} in gauge-fixed electrostatic form  is presented in appendix \ref{appc}.

\subsection*{Acknowledgments}

We thank Sean Hartnoll, Shota Komatsu and João Penedones for correspondence.
The research of ED and MG was supported, in part, by NSF grant PHY-26-09924.

\clearpage

\section{Complexified  Type IIB supergravity}
\setcounter{equation}{0}
\label{sec:2}

 In this section we shall review the BPS, Bianchi and field equations for Type IIB$_\CC$ supergravity, expressed in the Einstein frame metric. We will further review its general  local solutions on a complexified spacetime of the form $\cM_{6  \CC} \times \cM_{2  \CC}$ warped over a complexified Riemann surface $\Sigma _\CC$ that are invariant under the complex Lie superalgebra $F(4,\CC)$. These solutions were obtained for the metric and dilaton/axion field as well as the field strengths in \cite{DHoker:2025nid} and will be supplemented here by providing also their flux potentials.

\subsection{Complexified Type IIB}

Type IIB$_\CC$ supergravity is obtained from standard Type IIB supergravity with Minkowski signature by expressing the latter in terms of real bosonic fields and Majorana-Weyl spinor fermion fields, and then promoting each of the bosonic fields to a complex field and each Majorana-Weyl spinor field to a complex Weyl spinor field \cite{DHoker:2025nid}. The dependence of the BPS equations, Bianchi identities and field equations on these complex fields is then holomorphic by construction, as complex conjugate fields never enter. 

\sm

The complexified BPS equations at vanishing fermion fields, in Einstein frame metric,  are given for the variation of the gravitino and dilatino fields, respectively,  as follows \cite{DHoker:2025nid},\footnote{To avoid a proliferation of symbols, we shall designate a complexified field by the same letter as the corresponding real field. Here and throughout, we shall follow the conventions and notations of \cite{DHoker:2025nid} so that $M,N,P$ are Einstein indices lowered or raised with the help of the metric $G_{MN}$ or its inverse; the $\Gamma$-matrices are given in appendix~A of \cite{DHoker:2025nid}; the dot product of two $n$-forms $S,T$ is defined by $S \cdot T = S^{M_1 \cdots M_n} T_{M_1 \cdots M_n}$ and the local supersymmetry transformation spinor $\ep$ is a doublet of complex Weyl spinors of chirality $\Gamma ^{11} \ep = \ep$ in a basis of real Pauli matrices given by $s^1 = \sigma ^1, s^2=i \sigma ^2$ and $s^3 = \sigma ^3$. }
\bea
\label{2.a.1}
0 &=& \nabla  _M  \ep - { e^{ \phi}  \over 4} F_{(1)M} s^2  \ep 
-{e^{- \phi/2} \over 96} \Big \{  \Gamma_M (\G \cdot H_{(3)} ) + 2 (\G \cdot H_{(3)} ) \G_M \Big \} s^3 \ep
\no \\ &&
- {1 \over 480}(\G \cdot  F_{(5)})  \Gamma_M s^2 \ep
-{e^{ \phi/2 } \over 96} \Big \{  \Gamma_M (\G \cdot \tilde F_{(3)}  ) 
+ 2 ( \G \cdot \tilde F_{(3)}  ) \G_M \Big \} s^1  \ep
\no \\
0 & = & 
(\G \cdot \p \phi) s^1\ep 
- e^{ \phi}  (\G \cdot F_{(1)}) s^3 \ep 
+ {e^{- \phi/2} \over 12 }  ( \G \cdot H_{(3)}) s^2 \ep 
+ {e^{\phi/2} \over 12} (\G \cdot \tilde F_{(3)}  )  \ep
\eea
Here, $\phi$ is the dilaton, $\nabla_M$ is the covariant derivative on spinors with respect to the metric $G_{MN}$ and the field strengths $F_{(1)}, \tilde F_{(3)}, F_{(5)}$ and $H_{(3)}$ obey the Bianchi identities,
\begin{align}
\label{2.a.2}
d F_{(1)} & =  0 & d \tilde F_{(3)} & =  H_{(3)} \wedge F_{(1)}
\no \\
d H_{(3)} & = 0 & d F_{(5)} & = H_{(3)} \wedge \tilde F_{(3)}
\end{align}
The field equations for the dilaton and axion are given by,
\bea
\label{2.d.1}
0 & = & \nabla ^M \nabla_M \phi 
- e^{2 \phi } F_{(1)} \cdot F_{(1)} + {1 \over 12} e^{-\phi} H_{(3)} \cdot H_{(3)} 
-{ 1 \over 12} e^{ \phi} \tilde F_{(3)} \cdot \tilde F_{(3)}
\no \\
0 & = & \nabla ^M \big (e^{2 \phi} F_{(1)M} \big ) + { 1 \over 6} e^\phi H_{(3)} \cdot \tilde F_{(3)}
\eea
while those for the 3-form field strengths are, 
\bea
\label{2.d.2}
0 & = & \nabla ^P \big ( e^{-\phi} H_{(3)MNP} \big )  - e^{ \phi}  F_{(1)}^P \tilde F_{(3)MNP}
-{ 1 \over 6}  \tilde F_{(3)}^{PQR} F_{(5) MNPQR} 
\no \\
0 & = & \nabla ^P \big ( e^\phi \tilde F_{(3)MNP} \big ) + { 1 \over 6} H_{(3)} ^{PQR} F_{5 MNPQR} 
\eea
and the ones for the metric are given by,
\bea
\label{2.d.3}
0 & = & 
R_{MN} - \half  \p_M \phi \, \p_N \phi - \half e^{2 \phi} F_{(1)M} F_{(1)N} -{1 \over 96} (F_{(5)}^2)_{MN} 
-{1 \over 4}  e^{-\phi} H_{(3)M}{}^{PQ} H_{(3)NPQ}
\no \\ &&
-{ 1 \over 4} e^\phi \tilde F_{(3)M}{}^{PQ} \tilde F_{(3)NPQ}
 + { 1 \over 48} G_{MN} \Big ( e^{ - \phi}  H_{(3)} \cdot H_{(3)} +  e^\phi \tilde F_{(3)} \cdot \tilde F_{(3)} \Big )
\eea
These equations agree with the complexification of the field equations in Einstein frame in terms of real fields given in \cite{Polchinski:2000uf}.

\subsection{Flux field equations as Bianchi identities}

Following the \textit{democratic formulation} of \cite{Bergshoeff:2001pv}, the field equations for $F_{(1)}, \tilde F_{(3)}, F_{(5)}$ and $H_{(3)}$ may be expressed as Bianchi identities of their Poincar\'e dual fields. To do so, we use the relation for the square of the Poincar\'e dual on an arbitrary form field $F_{(p)}$ of rank $p$ in spacetime dimension $d$ with a metric whose signature has $s$ negative entries, 
\bea
\label{2.d.4}
\star \big ( \star F_{(p)} \big ) = (-)^{s+p(d-p)} F_{(p)}
\eea
Since here we have $d=10$ with Minkowski signature $s=1$ the square of the duality $\star ^2$ equals the identity for all odd rank $p$. In particular, we have the familiar relation $\star F_{(5)}= F_{(5)}$ and define $F_{(7)}$ and $F_{(9)}$ as follows,
\begin{align}
\label{2.d.5}
F_{(7)} & = - \, e^\phi \star \tilde F_{(3)} & \star F_{(7)} & = - \, e^{\phi} \tilde F_{(3)}
\no \\
F_{(9)} & = e^{2 \phi} \star F_{(1)} & \star F_{(9)} & =  e^{2 \phi} F_{(1)}
\end{align} 
In terms of these dual fields, the field equations for $F_{(1)}, \tilde F_{(3)}$ and $H_{(3)} = dB_{(2)}$ may be recast in the form of Bianchi identities. Below we give the complete list of all such Bianchi identities for the RR field strengths and their solution in terms of corresponding RR potentials $\chi = C_{(0)}, C_{(2)}, C_{(4)}, C_{(6)}$ and $C_{(8)}$, 
\begin{align}
\label{2.d.7}
d F_{(1)} & = 0  & F_{(1)} & = d \chi 
\no \\
d \tilde F_{(3)} & = H_{(3)} \wedge F_{(1)} & \tilde F_{(3)} & = dC_{(2)} - H_{(3)} \, \chi
\no \\
d F_{(5)} & = H_{(3)} \wedge \tilde F_{(3)} & F_{(5)} & = d  C_{(4)} - H_{(3)} \wedge C_{(2)}
\no \\
d F_{(7)} & =H_{(3)} \wedge F_{(5)} & F_{(7)} & = dC_{(6)} - H_{(3)} \wedge  C_{(4)}
\no \\
d F_{(9)} & = H_{(3)} \wedge F_{(7)} & F_{(9)} & = dC_{(8)} - H_{(3)} \wedge C_{(6)}
\end{align}
In terms of the standard RR fields, the potentials $C_{(6)}$ and $C_{(8)}$ may be defined by,
\bea
\label{2.d.8}
dC_{(6)} & = & - e^\phi \star \tilde F_{(3)} + H_{(3)} \wedge  C_{(4)} 
\no \\
dC_{(8)} & = &  e^{2 \phi} \star F_{(1)} + H_{(3)} \wedge C_{(6)} 
\eea
The construction for the field $B_{(6)}$ which is dual to $B_{(2)}$ is more involved.  We begin by recasting the field equation for $H_{(3)}$ on the first line of (\ref{2.d.2}) in form notation, 
\bea
\label{2.d.9}
d \star \big ( e^{-  \phi} H_{(3)} \big ) & = &  -  F_{(1)} \wedge F_{(7)} + \tilde F_{(3)} \wedge F_{(5)} 
\eea
The right side is closed,  as may be verified by using the Bianchi identities  (\ref{2.d.7}),  and may be expressed as a total differential with the help of their potentials,  
\bea
\label{2.d.10}
d \star \big ( e^{-  \phi} H_{(3)} \big ) =
 d \Big ( - \chi \, F_{(7)} + C_{(2)} \wedge d  C_{(4)} 
 - \half C_{(2)} \wedge C_{(2)} \wedge H_{(3)} \Big ) 
\eea
Therefore a natural definition of the field $B_{(6)}$ as the dual of $B_{(2)}$, expressed in standard fields by eliminating $F_{(7)}$ in favor of $- e^\phi \star \tilde F_{(3)}$,  is as follows, 
\bea
\label{2.d.11}
d B_{(6)} =  e^{-  \phi} \star H_{(3)} - e^\phi \, \chi \star \tilde F_{(3)} - C_{(2)} \wedge F_{(5)}  
 - \half C_{(2)} \wedge C_{(2)} \wedge H_{(3)}
\eea

\subsection{Solutions to Type IIB$_\CC$ invariant under $F(4;\CC)$}

Invariance under the Lie superalgebra $F(4;\CC)$ requires invariance under its maximal bosonic subalgebra
 $\ms \mo (7;\CC) \oplus \ms \mo (3;\CC)$ which in turn requires a spacetime of the form,
\bea
\label{2.e.1}
\cM_{6  \CC}  \times \cM_{2  \CC} \times \Sigma _\CC
\eea
where the first two factors are  warped over $\Sigma_\CC$. The spaces $\cM_{6  \CC}$ and $\cM_{2  \CC}$ have complex dimensions 6 and 2 and their complex-valued metrics $ds^2_{\cM_6}$ and $ds^2 _{\cM_2}$ of unit radius  are invariant under  $\ms \mo (7;\CC) $ and $ \ms \mo (3;\CC)$, respectively. The full metric is then given by,\footnote{The metric in (\ref{3.a.1}) is expressed in Einstein frame. The superscript $E$ used for the components $f_2$ and $f_6$ in Einstein frame in \cite{DHoker:2025nid} will be omitted throughout.}
\bea
\label{3.a.1}
ds^2 = f_6^2 \, ds^2_{\cM_{6  \CC}} + f_2^2 \, ds^2_{\cM_{2  \CC}} + ds^2_{\Sigma_\CC}
\eea
where the frame fields and reduced metrics are related as follows,
\begin{align}
\label{3.a.2}
e^m&=f_6 \, \hat e^m & ds^2_{\cM_{6  \CC}} &= \eta _{mn} \, \hat e^m \, \hat e^n & m,n&=0,..,5
	\nonumber\\
e^i&=f_2\, \hat e^i & ds^2_{\cM_{2  \CC}} &= \delta_{ij} \, \hat e^i \, \hat e^j  &  i,j&=6,7
	\nonumber\\
e^a& & ds^2 _{\Sigma_\CC} &= \delta_{ab} \, e^a \, e^b  & a,b&=8,9
\end{align}
Here, $m,i,a$ stand for the frame indices and the frame metric $\eta$ has signature $(-+++++)$.  The flux fields 
$F_{(1)}, \tilde F_{(3)}, H_{(3)}, F_{(5)} $ and their respective potentials $C_{(0)}, C_{(2)}, B_{(2)}$ and $C_{(4)}$ decompose as follows,
\begin{align}
\label{3.a.3}
F_{(1)} & = d C_{(0)}
\no \\
\tilde F_{(3)} & = g_a \, e^a \wedge e^{67} &
C_{(2)} & = C_a \, \hat e^{67}
\no \\
H_{(3)} & = h_a \, e^a \wedge e^{67} &
B_{(2)} & = B_a \, \hat e^{67}
\no \\
F_{(5)} & =0 & C_{(4)} & =0 
\end{align}
where $e^{67} = e^6 \wedge e^7$ and $\hat e^{67} = \hat e^6 \wedge \hat e^7$. The functions $\chi, g_a,h_a, C_a, B_a$ depend only on $\Sigma _\CC$. For $F_{(5)}=0$ the Bianchi identities of (\ref{2.d.7}) imply $H_{(3)} \wedge \tilde F_{(3)}=0$ and $dC_{(4)}=H_{(3)} \wedge C_{(2)}$ which, in view of (\ref{3.a.3}), is realized by $C_{(4)}=0$ up to a gauge transformation that shifts $C_{(4)}$ by a closed form. Combining the relations $C_{(4)}= F_{(5)}=0$ with $C_{(2)} \wedge C_{(2)}=0$ as a result of (\ref{3.a.3}), the expressions of (\ref{2.d.8}) and (\ref{2.d.11}) simplify and we obtain, 
\bea
\label{3.a.4}
d C_{(6)} & =  & - e^\phi \star \tilde F_{(3)}
\no \\
dB_{(6)} & = & e^{- \phi} \star H_{(3)} - e^\phi \, \chi \star \tilde F_{(3)}  
\no \\
dC_{(8)} & = & e^{2 \phi} \star F_{(1)} + H_{(3)} \wedge C_{(6)}
\eea
Each equation is integrable thanks to the Bianchi identities. 

\subsection{Exact solutions}
\label{sec:exactsolution}

In this section we summarize the $F(4;\CC)$ invariant solutions to Type IIB$_\CC$ obtained in~\cite{DHoker:2025nid}, and extend those results by the expression for the flux fields and potentials. 

\sm

A system of local holomorphic coordinates $w$ and $\tw$ is introduced on $\Sigma _\CC$ in terms of which the complexified metric and frame  take the following form,\footnote{Their complex conjugates $w^*$ and $\tw^*$ will not enter the complex solutions until we consider real forms of the solutions.}
\bea
\label{2.g.1}
ds^2_{\Sigma _\CC} = 4 \rho^2 \, dw d\tilde w \hskip 1in e^z = \rho \, dw \hskip 0.5in e^\tz= \rho \, d\tw
\eea
The solutions of \cite{DHoker:2025nid} are expressed in terms of four independent functions $\cA_1(w)$ , $\cA_2(w)$, $\tilde \cA_1(\tw)$ and $\tilde \cA_2(\tw)$ through the following combinations,
\bea
\label{2.g.2}
\cG & =  & i \big ( \tilde \cA_1 \cA_2 - \cA_1 \tilde \cA_2 - \cB + \tilde \cB \big )
\no \\
\kappa ^2 & = & i \big ( \p_w \cA_1 \, \p_{\tilde w} \tilde \cA_2 - \p_w \cA_2 \p_{\tilde w} \tilde \cA_1 \big ) 
= - \p_w \p_{\tilde w} \cG
\eea
where $\cB$ and $\tilde \cB$ are defined, up to an additive constant, by the relations,
\begin{align}\label{2.g.2b}
\p_w \cB & =  \cA_2 \p_w \cA_1 - \cA_1 \p_w \cA_2
&
\p_{\tilde w} \tilde \cB & =  \tilde \cA_2 \p_{\tilde w} \tilde \cA_1 - \tilde \cA_1 \p_{ \tilde w} \tilde  \cA_2
\end{align}
and the composite $T$ is defined by,
\begin{align}
\label{2.g.3}
T^2 &=1 + { 2 \over 3} {\p_w \cG \, \p_{\tilde w} \cG \over \kappa ^2 \, \cG}
= 1 - { 2 \over 3} {\p_w \cG \, \p_{\tilde w} \cG \over  \cG \,  \p_\tw \p_w \cG}
\end{align}

\subsubsection{Solutions for the dilaton, axion and metric factors}

The dilaton and axion fields of the solution are conveniently expressed through $\tau_\pm$, 
\begin{align}
\label{2.f.1}
\tau _\pm &= \chi \pm i \, e^{-\phi}
\end{align}
and are given by,
\bea
\label{2.f.2}
\tau_\pm = \frac{
(T\pm 1) \partial_{\tilde w}\tilde\cA_2 \partial_w\cG
+ (T\mp 1) \partial_w\cA_2 \partial_{\tilde w}\cG}
{(T\pm 1)  \partial_{\tilde w}\tilde\cA_1\partial_w\cG
+ (T\mp 1) \partial_w\cA_1\partial_{\tilde w}\cG}
\eea
The Einstein-frame metric factors $\rho^2, f_6^2$ and $f_2^2$ take the form,
\begin{align}
\label{eq:metric}
\rho^2 & = -\frac{c_6}{k_1}\kappa^2\sqrt{-\frac{T}{3\cG}}
&
f_6^2 & =  2k_1 c_6 T^{- \eta} \sqrt{- {3 \cG \over T}}
& f_2^2 & =  -2 {k_1 c_2^2 \over c_6}  T^{\eta} \sqrt{- {3 \cG \over T}}
\end{align}
where $k_1^4=1$ and $c_2,c_6$ are arbitrary non-zero constants.  The exponent $\eta$ was determined to take the values $\eta = \pm 1$ in \cite{DHoker:2025nid} and labels different branches of the square roots involved in the solutions. In appendix \ref{sec:potentials} we show  that   only the branch, 
\begin{align}
    \eta &= -1
\end{align}
is consistent with integrability conditions following from the Bianchi identities for the 3-form flux fields $dH_{(3)}=dF_{(3)}=0$ and we shall henceforth use this value for $\eta$.

\subsubsection{Gauge potentials}
\label{sec:potentials-sum}

The components $g_z$ and $h_z$ of the 3-form field strengths were obtained in \cite{DHoker:2025nid}.  The full set of RR and NSNS potentials is determined in Appendix~\ref{sec:potentials} and given as follows. The 2-form potential $B_{(2)}$ is defined by $dB_{(2)} = H_{(3)}$ while $C_{(2)}$ is defined by the second equation of the right side of (\ref{2.d.7}). These equations are integrable thanks to the Bianchi identities and  we obtain the following expressions for the potentials,
\begin{align}
\label{eq:2-forms}
B_{(2)}&=\frac{6 i c_2^2k_1\nu_1}{c_6}
	\left ( \frac{\cU_1}{3T^2}+\cA_1-\tilde\cA_1\right ) \, \hat e^{\, 67}
\nonumber\\
	C_{(2)}&=\frac{6 i c_2^2k_1\nu_1}{c_6}
	\left ( \frac{\cU_2}{3T^2}+\cA_2-\tilde\cA_2\right ) \, \hat e^{\, 67}
\end{align}
where the composites $\cU_i$ for $i=1,2$ are defined by,
\bea
\kappa^2 \, \cU_{i} = \partial_\tw\tilde\cA_{i}\partial_w\cG-\partial_w\cA_{i}\partial_\tw\cG
\eea
The six-form potentials $B_{(6)}$ and $C_{(6)}$ are defined by the first two lines in equation (\ref{3.a.4}). These equations are again integrable thanks to the Bianchi identities,  and are found to take the following values on the solutions, 
\begin{align}
\label{eq:6-forms}
B_{(6)}&=40c_6^3k_1^3\nu  \left ( \frac{3i}{5}\cG \, \cU_2
	+4(\cW_2+\tilde\cW_2)
	-(\cA_2-\tilde\cA_2)(2\tilde\cA_1\cA_2-2\cA_1\tilde\cA_2+3i\cG)\right ) \hat e^{\, 0 \cdots 5}
\no \\	
C_{(6)}&=40c_6^3k_1^3\nu  \left ( \frac{3i}{5}\cG \, \cU_1
	+4(\cW_1+\tilde\cW_1)
	-(\cA_1-\tilde\cA_1)(2\tilde\cA_1\cA_2-2\cA_1\tilde\cA_2+3i\cG)\right ) \hat e^{\, 0 \cdots 5}
\end{align}
where the holomorphic functions $\cW_i$ and $\tilde\cW_i$ for $i=1,2$ are defined by,
\begin{align}
\label{eq:wi-def}
	\partial_w\cW_{i}&=\cA_{i}\partial_w\cB & \partial_\tw\tilde\cW_{i}&=\tilde\cA_{i}\partial_\tw\tilde\cB
\end{align}
Finally, the RR 8-form potential is defined by the last line in (\ref{3.a.4}). We give the modified 8-form potential,
\begin{equation}
\tilde C_{(8)}=C_{(8)}-B_{(2)}\wedge C_{(6)}
\end{equation}
with Bianchi identity $d\tilde C_{(8)}=e^{2\phi}\star F_{(1)}-B_{(2)}\wedge dC_{(6)}$, which evaluates as follows on the solutions,
\begin{align}
\label{eq:8-forms}
\tilde C_{(8)}
&= 48c_2^2c_6^2k_1^4 \bigg  (
	3\cG \cU_1(\cA_1-\tilde\cA_1)
	-\frac{6\cG^2\partial_w\cA_1\partial_\tw\tilde\cA_1}{\kappa^2}
	-15i (\cY_1-\tilde\cY_1)
	-\cV_1 \bigg ) \hat e^{\, 0 \cdots 7}
\end{align}
where the holomorphic functions $\cY_i$ and $\tilde\cY_i$ for $i=1,2$ are defined by,
\begin{align}
\label{eq:yi-def}
	\partial_w\cY_{i}&=\cA_{i}^2\partial_w\cB
	&
	\partial_\tw\tilde\cY_{i}&=\tilde\cA_{i}^2\partial_\tw\tilde\cB
\end{align}
and $\cV_i$ is given by,
\begin{align}
	\cV_i&=5i \big (\cA_i^2+\tilde\cA_i^2-\cA_i\tilde\cA_i \big ) \big (\cA_1\tilde\cA_2-\tilde\cA_1\cA_2-2 i\cG \big )
\end{align}

With the potentials in hand, we can identify a set of transformations of the functions $\cA_{1,2}$, $\tilde \cA_{1,2}$ whose only effect is to induce gauge transformations. To this end, we consider shifts of the form
\begin{align}
\label{gaugetrans}
\cA_{i} &\to \cA_i + \alpha_i & \tilde \cA_{i} &\to \tilde \cA_i - \alpha_i& \quad i&=1,2
\end{align}
with complex  constants $\alpha_{1,2}$. Under these shifts 
\begin{align}
\partial_w \cB& \to \partial_w \cB + \alpha_2 \partial_w \cA_1- \alpha_1 \partial_w \cA_2 & \partial_{\tilde w} \tilde \cB& \to \partial_{\tilde  w}\tilde  \cB - \alpha_2 \partial_ {\tilde w } \tilde \cA_1+ \alpha_1 \partial_{\tilde  w} \tilde  \cA_2
\end{align}
while $\cG$ and hence all metric functions, as well as  $\tau_\pm$, are invariant.  For the two form potentials $B_{(2)}$ and $C_{(2)}$ the expressions multiplying $\hat e^{67}$ transform by constant shifts, which constitute gauge transformations.  It is straightforward to show that $C_{(6)}$ and $B_{(6)}$ as well as $C_{(8)}$ are invariant under these shifts.

\subsection{Transformations under $SL(2,\CC)$ symmetry of Type IIB$_\CC$}
\label{sec:3.9}

Type IIB$_\CC$ supergravity is invariant under a global $SL(2,\CC)$. This symmetry commutes with $F(4;\CC)$ and transforms the $F(4;\CC)$ invariant solutions by transforming the functions $\cA_1, \cA_2, \tilde \cA_1, \tilde \cA_2$ as doublets under an arbitrary $M \in SL(2,\CC)$,
\bea
\label{3.AA}
\left ( \bma \cA_2 \cr \cA_1 \ema \right ) 
\to M \left ( \bma \cA_2 \cr \cA_1 \ema \right )
\hskip 0.8in 
\left ( \bma \tilde \cA_2 \cr \tilde \cA_1 \ema \right ) 
\to M \left ( \bma \tilde \cA_2 \cr \tilde \cA_1 \ema \right )
\hskip 0.8in 
M= \left ( \bma a & b \cr c & d \ema \right )
\eea
with $a,b,c,d \in \CC$ and $\det M=1$. It follows from (\ref{2.g.2}) and (\ref{2.g.3})  that $\kappa^2$, $\p_w \cB$ and $\p_\tw \tilde \cB$ are invariant. We shall choose suitable integration constants such that $\cB - \tilde \cB$ is invariant.  The combinations $\cG$ and $T$ are then also invariant which implies that the Einstein-frame metric functions $f_2^2, f_6^2$ and $ \rho^2$  in (\ref{eq:metric}) are invariant as well. The transformation of $\tau_\pm$ can be determined from their definition in (\ref{2.f.2}), which leads to
\bea
\label{3.tautr}
\tau _\pm \to { a \tau_\pm + b \over c \tau_\pm +d } 
\eea
Finally, since $\cU_i$, $\cW_i$ and $\tilde \cW_i$ transform as doublets, like $\cA_i$ and $\tilde \cA_i$, it follows that,
\bea
\label{3.BB}
\left ( \bma C_{(2)}  \cr B_{(2)}  \ema \right ) 
\to M \left ( \bma C_{(2)}  \cr B_{(2)}  \ema \right ) 
\hskip 0.8in 
\left ( \bma  B_{(6)} \cr C_{(6)} \ema \right ) 
\to M \left ( \bma B_{(6)} \cr C_{(6)} \ema \right ) 
\eea
Finally, we note that $C_{(8)}$ can be embedded into a triplet under $SL(2,\CC)$ following \cite{Bergshoeff:2005ac}. We will not discuss the full multiplet here and focus our attention on $C_{(8)}$, but note that the S-dual potential is obtained by exchanging $(\cA_1,\tilde\cA_1)\rightarrow(\cA_2,\tilde\cA_2)$.

\subsection{Collapsing $\cM_2$ or $\cM_6$ on $\partial\Sigma$}

For Riemann surfaces with boundary the  regularity conditions comprise conditions on the boundary $\partial\Sigma$ of $\Sigma$  to ensure that the ten-dimensional solutions close off into globally smooth geometries. These conditions may be formulated on the complex solutions to Type IIB$_\CC$ and will descend smoothly to any of the real forms. As  such, they  amount to conditions for the spaces $\cM_6$ or $\cM_2$ to collapse smoothly at the boundary of $\Sigma$. To analyze the behavior of the radii of these spaces, we shall examine the following combinations of the metric factors $f_2^2, f_6^2$ and $\rho^2$ of (\ref{eq:metric}) for $\eta=-1$, 
\begin{align}
\label{eq:comb-1}
 f_6^2 f_2^2&=-4c_2^2 \, { \kappa^4 \over \rho^4}
 &
 \frac{f_6^2}{f_2^2}&=-\frac{c_6^2}{c_2^2} \, T^2
 &
 f_6^6 f_2^2&=-144 \, c_2^2 c_6^2 \, \cG^2
\end{align}
The first and third relations in (\ref{eq:comb-1}) imply that collapsing $\cM_6$ or $\cM_2$ or both requires $\kappa^2/\rho^2 \to 0 $ and $\cG \to 0$ at the boundary $\p \Sigma$. The second relation in (\ref{eq:comb-1}) implies that the behavior of $T$ controls which one of the spaces $\cM_2$, $\cM_6$ collapses.
In summary, the conditions for collapsing either $\cM_2$ or $\cM_6$ while keeping the other finite are,
\begin{align}
\label{eq:collapseM2M6}
	\text{collapsing $\cM_2$:}& \qquad 
	\kappa^2/\rho^2 \to 0 , ~ \cG \rightarrow 0, ~  T  \to \infty
\nonumber\\
	\text{collapsing $\cM_6$:}& \qquad 
	\kappa^2/\rho^2 \to 0, ~\cG \rightarrow 0, ~ T  \to 0
\end{align}
It is natural to characterize the boundary of $\Sigma_\CC$ as the subspace of $\Sigma$ where the scalar functions $\cG$ and $\kappa^2/\rho^2 $ vanish. The possibility that both $\cM_2$ and $\cM_6$ collapse simultaneously, which is  allowed in principle by the form of the metric factors of the solutions in (\ref{eq:comb-1}), will not be considered any further  here.

\subsubsection{The behavior of $\cG$ and $T$ near $\p \Sigma$}

Choosing local complex coordinates $w$ and $\tilde w$ in terms of which  the boundary is specified by $\tilde w = w$,
we assume that the function $\cG$ has the following behavior near the boundary, 
\begin{align}
\label{eq:G-y-exp}
	\cG=(w - \tilde w)^\alpha g(w, \tilde w)
\end{align}
for a function $g(w, \tilde w)$ that is bounded away from 0 and $\infty$ at the boundary and an exponent $\a$ that remains to be determined. In terms of this parametrization, we readily evaluate the behavior of $T^2$ near $\p \Sigma$ using the second relation in (\ref{2.g.3}),  
\begin{align}
T^2 = 1 - { 2 \a \over 3 (\a-1)} + \cO(w - \tilde w)
\end{align}
Assuming that either $\cM_2$ or $\cM_6$ remains of finite size at $\p \Sigma$ while the other goes to zero size, we have the following two options, 
\begin{align}
\label{eq:alpha-T}
\text{collapsing $\cM_2$:}  \qquad T  \to \infty, \qquad & 	\alpha  =1
\no \\
\text{collapsing $\cM_6$:} \qquad T  \to \, 0 ~ ,  \qquad	&\alpha  =3
\end{align}
This fully specifies the behavior of $\cG$ at the boundary $\p \Sigma$, under the assumption that either $\cM_2$ or $\cM_6$, but not both simultaneously, collapse, and the resulting two cases are reminiscent of the analyses in \cite{DHoker:2017mds,Corbino:2018fwb}.

\subsubsection{The behavior of $\cA_1, \cA_2, \tilde \cA_1, \tilde \cA_2$ near $\p \Sigma$}
\label{abehavior}

To realize the fall-off behaviors for $\cG$ with $\alpha\in\{1,3\}$ in terms of the holomorphic functions $\cA_1, \cA_2, \tilde \cA_1$ and $ \tilde \cA_2$ we evaluate the first order derivatives of $\cG$,
\begin{align}
\partial_w\cG&= i \big (\cA_1+\tilde\cA_1\big ) \partial_w\cA_2-i \big ( \cA_2+\tilde\cA_2 \big ) \partial_w\cA_1
\no \\
\partial_\tw\cG&=-i \big ( \cA_1+\tilde\cA_1 \big ) \partial_\tw\tilde\cA_2+i \big (\cA_2+\tilde\cA_2 \big )\partial_\tw\tilde\cA_1
\end{align}
as well as its second derivatives, 
\begin{align}
\partial_w ^2 \cG&= i \big (\cA_1+\tilde\cA_1\big ) \partial_w^2 \cA_2-i \big ( \cA_2+\tilde\cA_2 \big ) \partial_w^2 \cA_1
\no \\
\partial_\tw ^2 \cG&=-i \big ( \cA_1+\tilde\cA_1 \big ) \partial_\tw ^2 \tilde\cA_2+i \big (\cA_2+\tilde\cA_2 \big )\partial_\tw ^2 \tilde\cA_1
\end{align}
while the second derivative $\p_w \p_\tw \cG = - \kappa^2$ is given by the first equation in (\ref{2.g.2}).

\sm

$\bullet$ For $\alpha=3$, which correspond to $\cM_6$ collapsing, the derivatives $\p_w \cG, \p _\tw \cG, \p_w^2 \cG, \p _\tw^2  \cG$ and $\p_w \p_\tw \cG = - \kappa^2$ need to vanish at $w=\tw$. While the matrix $(\p_w \cA_i, \p_\tw \tilde \cA_i)$ of first order derivatives  has vanishing determinant in view of $\kappa^2=0$ at $\p \Sigma$, the matrix $(\p_w^2 \cA_i, \p_\tw^2 \tilde \cA_i)$ of second order derivatives will be assumed to be invertible on $\p \Sigma$, so that the vanishing of $\p_w^2 \cG, \p _\tw^2  \cG$ requires the following conditions on the boundary $\tilde w=w$, 
\begin{align}
\label{eq:alpha3-rel}
	\tilde\cA_1(\tw)&=- \cA_1(w)
	& 
	\tilde\cA_2(\tw)&=-\cA_2(w)
\end{align}
It follows that the first derivatives $\p_w \cG$ and $\p_\tw \cG$ then automatically vanish. Finally, the vanishing of $\cG$ itself additionally requires $\cB(w) = \tilde \cB(\tw)$ for $\tw = w$ which, in particular, relates the additive integration constants in $\cB$ and $\tilde \cB$. 

\sm

$\bullet$ For $\alpha=1$, which corresponds to $\cM_2$ collapsing,  the derivative along the boundary should vanish, namely $\p_w \cG + \p_\tw \cG=0$, but the transverse derivative need not vanish. These conditions are realized as follows on the boundary $\tw = w$,
\begin{align}
\label{eq:alpha1-rel}
	\partial_\tw \tilde\cA_1(\tw)&= \partial_w \cA_1(w)
	& 
	\partial_\tw\tilde\cA_2(\tw)&= \partial_w\cA_2(w)
\end{align}
These conditions are not mutually independent on $\p \Sigma$, as $\kappa^2$ vanishes there. They imply that $\cG$ must be constant along any connected component of the boundary $\p \Sigma$. When $\cM_2$ collapses on a single boundary component, the condition $\cG=0$ may be realized by a suitable choice of integration constants for $\cB$ and $\tilde \cB$. When $\cM_2$ collapses along multiple boundary components there will be further conditions relating the integration constants of $\cB$ and $\tilde \cB$ connecting the different components. Unlike for $\alpha=3$, the case $\alpha=1$ only imposes relations on the differentials, not on the values of the functions on $\p \Sigma$.

\subsection{Cycles, charge densities and charges}
\label{sec:charges}

The flux fields produce charge distributions that are invariant under $\ms \mo (7;\CC) \oplus \ms \mo (3;\CC)$ and therefore constant on $\cM_2$ and $\cM_6$.  Upon restricting to a real form of the solutions, they account for charge densities associated with the types of branes present in the particular real form of Type IIB, and evaluate to physical charges whenever $\cM_2$ or $\cM_6$, or both, are compact, in which case they are $S^2$ or $S^6$, respectively. Here we shall provide the formulas for the complex charge densities which can encompass all of the above cases. 

\sm

To determine the complex charges of the solution we start by identifying the corresponding cycles. Depending on whether $\cM_2$ or $\cM_6$ collapses at the boundary $\p \Sigma$, we have 3 relevant types of cycles, each of which is  realized by an open curve $\cC$ (topologically an interval) on $\Sigma$:
\begin{itemize}
\itemsep =-0.02 in
	\item 3-cycles: curves with $\cM_2$ collapsing at both end points $\sim$ $(p,q)$ 5-brane charges
	\item 7-cycles: curves with $\cM_6$ collapsing at both end points $\sim$ $(p,q)$ string charges
	\item 9-cycles: $\cM_6$ collapsing on one end, $\cM_2$ on the other $\sim$ instanton charge
\end{itemize}
There are no relevant 5-cycles as the corresponding flux field $F_{(5)}$ vanishes for all our solutions. Furthermore, 1-cycles correspond to D7 branes, which will not be considered. 

\sm

The Page charge densities $\cQ_{\rm D5}$, $\cQ_{\rm NS5}$, $\cQ_{\rm D1}$, $\cQ_{\rm F1}$ and $\cQ_{\rm Din}$, corresponding to D5 and NS5 branes, D1 and F1 strings and the number of D-instantons, respectively, are obtained by integrating the corresponding potentials over the associated  curve $\cC \subset \Sigma$,
\begin{align}
	\cQ_{\rm D5}&=\int_\cC dC_{(2)}
	&
	\cQ_{\rm D1}&= \int_\cC dC_{(6)}
	&
	\cQ_{\rm Din}&= \int_\cC d\tilde C_{(8)}
	\nonumber\\
	\cQ_{\rm NS5}&= \int_\cC dB_{(2)}
	&
	\cQ_{\rm F1}&= \int_\cC dB_{(6)}
	&
\end{align}
with normalizations to be determined and expressions for  the potentials $C_{(2)}, B_{(2)}, C_{(6)}, B_{(6)}$ and $\tilde C_{(8)}$ given in section \ref{sec:potentials-sum}.   The integrals reduce to the differences between the potentials at the end points of the curve $\cC \subset \Sigma$. Upon restricting to a real form in which either $\cM_2$ is $S^2$ or $\cM_6$ is $S^6$ or both, the corresponding charges $Q_{\rm D5}$, $Q_{\rm NS5}$, $Q_{\rm D1}$, $Q_{\rm F1}$ and $Q_{\rm Din}$ are obtained by integrating over those unit spheres,
\bea
Q_{\rm D5/NS5} =\int_{\hat S^2} \cQ_{\rm D5/NS5}
\hskip 0.5in
Q_{\rm D1/F1} = \int_{\hat S^6} \cQ_{\rm D1/F1}
\hskip 0.5in
Q_{\rm Din} = \int_{\hat S^2 \times \hat S^6}  \cQ_{\rm Din}
\eea
To evaluate the charge densities we shall denote the endpoints of the curves $\cC$ by $w_0, w_1$.

\sm
  
For the 5-brane charges, the curves $\cC$ connect boundary segments where $\cM_2$ collapses. On each segment we have   $\kappa^2\rightarrow 0$ and $T^2\rightarrow\infty$ (see~(\ref{eq:collapseM2M6})). To keep $\cM_6$ from collapsing as well, we use the relation $\rho^2 f_6^2 = -2 c_6^2 \kappa^2 T$ deduced from (\ref{eq:metric}) with $\eta=-1$ to imply that $\kappa^2 T$ remains finite, which implies that $\kappa^2T^2\rightarrow\infty$ and $\cU_i/T^2\rightarrow 0$ in (\ref{eq:2-forms}). As a result, the 5-brane charge densities become,
\bea
\cQ_{\rm D5} & = & \frac{6ic_2^2k_1\nu_1}{c_6}  \left[\cA_2-\tilde\cA_2\right]_{w_0}^{w_1} \, \hat e^{67}
\no \\ 
\cQ_{\rm NS5} & = & \frac{6ic_2^2k_1\nu_1}{c_6}  \left[\cA_1-\tilde\cA_1\right]_{w_0}^{w_1} \, \hat e^{67}
\eea
The conditions in (\ref{eq:alpha1-rel}) guarantee that $\cA_1-\tilde\cA_1$ and $\cA_2-\tilde\cA_2$ remain constant along each segment where $\cM_2$ collapses, so that the charge densities do not depend on the specific point on a given boundary segment where the curve $\cC$ ends.

\sm

For the string charges, the curves $\cC$ connect segments where $\cM_6$ collapses. Thanks  to the boundary conditions of (\ref{eq:G-y-exp}) and (\ref{eq:alpha-T}) we have $\cG \to 0$ and $\cG/\kappa^2\rightarrow 0$, so that $\cG \, \cU_i \to 0$ in (\ref{eq:6-forms}) while the boundary conditions of (\ref{eq:alpha3-rel}) make the second term inside the large parentheses of (\ref{eq:6-forms}) vanish as well. As a result, we obtain, 
\bea
\label{eq:QD1F1}
\cQ_{\rm D1} & = & 160c_6^3k_1^3\nu_1 \left[\cW_1+\tilde \cW_1\right]_{w_0}^{w_1} \, \hat e^{0\cdots 5} 
\no \\
\cQ_{\rm F1} & = & -160c_6^3k_1^3\nu_1 \left[\cW_2+\tilde\cW_2\right]_{w_0}^{w_1} \, \hat e^{0\cdots 5} 
\eea
The boundary conditions in (\ref{eq:alpha3-rel}) ensure that these expressions do not depend on the specific point along a given boundary segment where the curve $\cC$ ends.

\sm

For the instanton charge, the curves $\cC$ connect a boundary segment where $\cM_2$ collapses at one endpoint $w_0$ of $\cC$ to a boundary segment where  $\cM_6$ collapses at the other endpoint $w_1$ of $\cC$. Where $\cM_6$ collapses, we again use that $\cG \, \cU_i$ vanishes in (\ref{eq:8-forms}). Due to the rapid decay of $\cG$ at the boundary, $\cG^2\partial_w \cA_i\partial_\tw\tilde\cA_i$ vanishes as well, and the conditions in (\ref{eq:alpha3-rel}) make the last term in (\ref{eq:8-forms}) vanish as well. For segments where $S^2$ collapses, $\cG^2/\kappa^2$ vanishes, and so does the second term in (\ref{eq:8-forms}). Putting all together, we obtain,
\begin{align}\label{eq:q-dinst}
\cQ_{\rm Din} &= 48 c_2^2c_6^2k_1^4 \bigg \{\cV_1 (w_0)
-  3\cG \, \cU_1 (\cA_1-\tilde\cA_1)(w_0) - 15i \Big [ \cY_1-\tilde\cY_1 \Big ]^{w_1}_{w_0}   \bigg \} \, \hat e^{0\cdots 7}
\end{align}
For charges which can be determined by choosing $\cC$ infinitesimally close to a branch point where the boundary conditions change from $\cM_2$ collapsing to $\cM_6$ collapsing, the relations between $\cA_i$ and $\tilde\cA_i$ from the $\cM_6$ segment hold up to terms that vanish as the branch point is approached. This makes the $\cG \, \cU_i$ and $\cV_i$ terms vanish. Overall, the value of the potentials should be constant along the boundary segments, as we will verify in examples.

\newpage

\section{$dS_{1,5}\times S^2\times\Sigma$ solutions to Type IIB$^\star$}
\label{sec:3}

A complete classification of the consistent real forms of Type IIB$_\CC$ supergravity was obtained in \cite{Bergshoeff:2007cg} and recently reviewed in \cite{DHoker:2025nid}, whose notation we shall follow throughout. For every real form, the coordinates of spacetime are restricted to real coordinates and the dilaton and the components of the spacetime metric are required to be real,
\bea
G^*_{MN} = G_{MN} \hskip 1in \phi^*=\phi
\eea
The reality conditions on the other bosonic fields specify the real forms of Type IIB:
\begin{itemize}
\itemsep=-0.04in
\item Type IIB$_\RR$ in Minkowski signature: real $B_{(2)}, B_{(6)}$ and real RR fields;
\item Type IIB$^\star$ in Minkowski signature: real $B_{(2)}, B_{(6)}$ and imaginary RR fields;  
\item Type IIB$^\prime$  in Minkowski signature: real $C_{(2)}, C_{(6)}$ and imaginary $\chi, B_{(2)}, B_{(6)} , F_{(5)}$;
\item real forms with multiple time-like directions, which will not be considered here. 
\end{itemize}
The corresponding reality conditions on the spinor fields given in \cite{Bergshoeff:2007cg,DHoker:2025nid} will not be needed. The Type IIB$^\star$ and Type IIB$^\prime$ supergravities are related to one another by $S$-duality, so we shall restrict here to considering only Type IIB$^\star$.  

\sm

The solutions to Type IIB$_\RR$ that are invariant under a real form of $F(4;\CC)$ have a Riemann surface of Euclidean signature and come in two varieties:
\begin{itemize}
\itemsep=-0.04in
\item $AdS_{1,5} \times S^2$ warped over $\Sigma$ with maximal bosonic subgroup $SO(2,5) \times SO(3)$  \cite{DHoker:2016ujz};
\item $AdS_{1,1} \times S^6$ warped over $\Sigma$ with maximal bosonic subgroup $SO(2,1) \times SO(7)$ 
 \cite{Corbino:2017tfl}.
\end{itemize}
The possible spacetimes for solutions to Type IIB$^\star$ with one time-like direction and a Riemann surface $\Sigma $ of Euclidean signature also come in two varieties:
\begin{itemize}
\itemsep=-0.04in
\item $dS_{1,5} \times S^2$ warped over $\Sigma$ with maximal bosonic subgroup $SO(1,6) \times SO(3)$;
\item $dS_{1,1} \times S^6$ warped over $\Sigma$ with maximal bosonic subgroup $SO(1,2) \times SO(7)$.
\end{itemize}
In the remainder of this paper we shall focus on the $dS_{1,5} \times S^2$ solutions.

\subsection{Implementing the conditions for a real form solution}

We shall now implement the restriction to the real form of the solutions with $dS_{1,5}\times S^2$ geometry in Type IIB$^\star$. Following the analysis of section 4 of \cite{DHoker:2025nid}, the reality conditions relate the holomorphic coordinates $w$ and $\tw$ to one another by complex conjugation (denoted throughout by a~${}^*$ superscript), and relate the  functions $\cA_i$ and $\tilde \cA_i$ to one another by the following relations, 
\begin{align}
\label{3.k.1}
	\tilde w &= w^* & \tilde \cA_1(\tw) & = - \cA_1(w) ^* & \tilde\cA_2 (\tw) & = \cA_2(w) ^*
\end{align}
These reality conditions reduce the $SL(2,\CC)$ invariance to $SL(2,\RR)$ and impose the condition $M^* = \sigma ^3 M \sigma ^3$ on the allowed transformations in (\ref{3.AA}).  Furthermore, the third line of Table 1 of \cite{DHoker:2025nid} together with equation (3.37) of \cite{DHoker:2025nid} imply the following relations between the constants $k_1, k_2, c_2, c_6$ and $\nu$, 
\begin{align}
\label{3.k1k2}
	k_1&=k_2=i
	&
	\nu c_6&=3 c_2
\end{align}
where  $c_2$, $c_6$ are real in view of the first line of (5.17) in \cite{DHoker:2025nid} and $\nu^2=1$. With these identifications $\cG$ and $\kappa^2$ are imaginary, while $T^2$ is real. It will be convenient to work with real functions instead, and we define $\hat \cG$, $\hat \kappa^2$ and $\hat T$ by,
\begin{align}
\label{3.k.2}
\hat\cG&=-i\cG= -\cA_1\cA^\star_2 - \cA_2 \cA_1^\star -\cB-\cB^\star
\no \\
\hat\kappa^2&=-i\kappa^2=\partial_w\cA_1(\partial_w \cA_2)^\star+ \partial_w\cA_2 (\partial_w\cA_1)^\star 
\no \\
\hat T & = -i T
\end{align}
The axion/dilaton fields $\tau_\pm$ become,
\bea
\label{2.f.2b}
\tau_\pm = \frac{
(\hat T \mp i ) (\partial_w \cA_2)^* \partial_w \hat \cG
+ (\hat T \pm i ) (\partial_w\cA_2) \partial_\sw \hat \cG}
{- (\hat T \mp i )  (\partial_w \cA_1)^* \partial_w \hat \cG
+ (\hat T \pm i ) (\partial_w\cA_1) \partial_\sw \hat \cG}
\eea
The reality of $\hat \cG$ and $ \hat T$ imply that the fields satisfy $(\tau_\pm)^* = - \tau_\pm$, so that both fields $\tau_\pm$ are purely imaginary. In view of (\ref{2.f.1}) this is consistent with the reality conditions of Type IIB$^\star$ supergravity for which the dilaton $\phi$ is real but the axion $\chi$ is imaginary. The complex metric in (\ref{3.a.1}) reduces for this real form to,
\begin{equation}
	ds^2 = f_6^2 \, ds^2_{S^6} + f_2^2 \, ds^2_{S^2} + ds^2_{\Sigma}
\end{equation}
The metric functions in (\ref{eq:metric}) for $\eta=-1$ in terms of the hatted quantities are given by,
\begin{align}
\label{eq:metric-2}
\rho^2 & = - c_6 \hat \kappa^2\sqrt{-\frac{\hat T}{3 \hat \cG}}
&
f_6^2 & =  - 2 c_6 \hat T  \sqrt{- {3 \hat \cG \over \hat T}}
& f_2^2 & =  - {2 c_6 \over 9 } { 1 \over \hat T}  \sqrt{- {3 \hat \cG \over \hat T}}
\end{align}
where we have used the relations of (\ref{3.k1k2}) to simplify the above expressions. The relations of (\ref{eq:comb-1}) simplify accordingly, 
\begin{align}
\label{eq:comb-10}
 f_6^2 f_2^2&={ 4 \over 9} c_6^2 \, { \hat \kappa^4 \over \rho^4}
 &
 \frac{f_6^2}{f_2^2}&= 9\, \hat T^2
 &
 f_6^6 f_2^2&= 16 \,c_6^4 \, \hat \cG^2
\end{align}

\subsection{Implementing reality and positivity conditions on the fields}
\label{sec:metric-reg}

In addition to closing off the geometry smoothly on the boundaries of $\Sigma$, we need the proper reality and  signature conditions on the metric, and reality conditions on the dilaton and other bosonic fields in the interior of $\Sigma$. We start with the metric functions. Demanding positivity of $\rho^2, f_6^2$ and $f_2^2$ imposes the following conditions:
\begin{enumerate}
\itemsep = -0.04in
\item Positivity of $f_6^2$ and $ f_2^2$  requires $\hat \cG$ and $\hat T$ to be real in view of  (\ref{eq:comb-10}). 
\item Reality of $\hat \kappa^2/\rho^2$, $f_6^2, f_2^2$ requires $\hat \cG$ and $\hat T$ to have opposite signs
in view of (\ref{eq:metric-2}).  
\item Positivity of $\rho^2$ requires $\hat \kappa^2$ and $\hat T$ to have the same sign in view of (\ref{eq:metric-2}).
\item The branch of the square root in (\ref{eq:metric-2}) is chosen so that $-c_6(-\hat T/\hat \cG)^\half  >0$. 
\end{enumerate}
Together, these conditions fix the relative signs of $\hat\kappa^2, \hat T$ and $\hat \cG$:
\begin{align}
\label{3.b.1}
	\hat\kappa^2, \, \hat T, \, -\hat\cG  \quad \hbox{ have the same sign}
\end{align}
guaranteeing that $\rho^2, f_6^2$ and $f_2^2$ are positive. The conditions also guarantee that $\tau_\pm$ are purely imaginary, so that $e^{-\phi}$ is real.

\sm

The above conditions are not sufficient to guarantee that $e^{- \phi}>0$ or, equivalently, that the dilaton $\phi$ is real.  In fact, unlike for standard Type IIB supergravity solutions where the sign of $e^{-\phi}$ is automatically preserved by its $SL(2,\RR)$ symmetry, for Type IIB$^\star$ solutions the sign of $e^{-\phi}$ is not preserved by $SL(2,\RR)$ transformations. In Type IIB$^\star$, as discussed after equation (\ref{3.k.1}), the field $\tau_\pm$ transforms as follows under  a transformation by $ M \in SL(2,\RR)$,
\bea
\tau_\pm \, \to \, \frac{a\tau+b}{c\tau+d} 
\hskip 1in  
M=\begin{pmatrix}a&b\\c&d\end{pmatrix} \quad \hbox{ with } \quad M^\star=\sigma^3M\sigma^3
\eea
so that $a,d, ib, ic \in \RR$. Given these reality conditions, $e^{-\phi}$ transforms as follows, 
\bea\label{eq:phi-transf}
e^{-\phi} \, \to \, \frac{e^{-\phi}}{|c\chi+d|^2-|c|^2e^{-2\phi}}
\eea
For sufficiently large $c$ the sign of $e^{-\phi}$ flips under this transformation, and there exist no real values of $e^{-\phi}$ and $i \chi$ such that their entire orbit under $SL(2,\RR)$ has $e ^{- \phi} >0$. 

\sm

To shed light on this behavior in Type IIB$^\star$ we start with a discussion of the scalar manifolds and their parametrization for the standard Type IIB$_\RR$ supergravity and for Type IIB$^\star$, building on the discussion in section 3.1 of \cite{Cremmer:1998em}. The scalar manifolds are $SL(2,\RR)/H$, with $H=SO(2)$ for Type IIB$_\RR$ and $H=SO(1,1)$ for Type IIB$^\star$ \cite{Hull:1998vg}.
In the following we denote a coset representative as $\mathcal V\in SL(2,\RR)$, to be acted on from the left by $H$ and from the right by global $SL(2,\RR)$ duality transformations.

\sm

In Type IIB$_\RR$, an $H$-invariant parametrization can be obtained in terms of a real matrix $\mathcal M$ defined by
\begin{equation}\label{eq:MR}
	\mathcal{M}=\mathcal V^T\mathcal V
\end{equation}
By construction, $\mathcal M$ is symmetric positive definite, with $\det\mathcal M=+1$. This guarantees the existence of a Cholesky decomposition, which reconstructs from given $\mathcal M$ an associated $\mathcal V$ in upper triangular form, or Borel gauge. Global $SL(2,\RR)$ duality transformations, parametrized by a real matrix $M\in SL(2,\RR)$, act on $\mathcal M$ as
\begin{equation}\label{eq:cM-SL2}
	\mathcal M \rightarrow M^T\cM M
\end{equation}
The usual axion-dilaton Lagrangian can be obtained from $\mathcal L=\frac{1}{4}e\,\tr\left(\partial_\mu \cM^{-1}\partial^\mu\cM\right)$
by parametrizing $\mathcal V$ and the associated $\mathcal M$ as
\begin{align}
	\mathcal V&=\begin{pmatrix}e^{\phi/2} & \chi e^{\phi/2}\\ 0 &e^{-\phi/2}\end{pmatrix}
	&
	\mathcal M&=\begin{pmatrix} e^\phi & \chi e^\phi\\ \chi e^\phi & \chi^2e^\phi+e^{-\phi}\end{pmatrix}
\end{align}
In Type IIB$_\RR$, this parametrization in terms of $(\phi,\chi)$ covers all matrices $\mathcal M$ with the properties resulting from the definition in (\ref{eq:MR}).\footnote{Given an arbitrary $\mathcal M$ as in (\ref{eq:MR}), positive definiteness guarantees $\mathcal M_{1,1}>0$, which defines real $\phi$ by $\cM_{1,1}=e^{\phi}$. The off-diagonal entries, which have no sign restriction, then define real $\chi$ by $\cM_{1,2}=\cM_{2,1}=\chi e^\phi$. The remaining entry $\mathcal M_{2,2}$ is then fixed by $\det \mathcal M=1$ to take the form in the parametrization.}
Therefore, the $SL(2,\RR)$ transformations (\ref{eq:cM-SL2}) can be expressed globally as transformations of $(\phi,\chi)$.

\sm

In Type IIB$^\star$, due to the mixed signature of $H=SO(1,1)$, a similar $H$-invariant parametrization can be obtained in terms of a matrix $\mathcal M_\star$ defined by
\begin{equation}\label{eq:Mstar}
		\mathcal{M}_\star=\mathcal V^T\eta\mathcal V
\end{equation}
where $\eta={\rm diag}(1,-1)$. This $\mathcal M_\star$ is indefinite with $\det \mathcal M_\star=-1$. The axion-dilaton action can be obtained by choosing $\mathcal V$ and the associated $\mathcal M_\star$ as
\begin{align}\label{eq:Vstar-1}
	\mathcal V&=\begin{pmatrix}e^{\phi/2} & \chi_\star e^{\phi/2}\\ 0 &e^{-\phi/2}\end{pmatrix}
	&
	\mathcal M_\star&=\begin{pmatrix} e^\phi & \chi_\star e^\phi\\ \chi_\star e^\phi & \chi_\star^2e^\phi-e^{-\phi}\end{pmatrix}
\end{align}
In this expression $\chi_\star$ is real (with negative definite kinetic term) and related to the imaginary $\chi$ in our Type IIB$^\star$ formulation by $\chi=i\chi_\star$. The crucial difference compared to Type IIB$_\RR$ is that the coordinates  $(\phi,\chi_\star)$ do not cover arbitrary $\mathcal M_\star$ defined in (\ref{eq:Mstar}), as can be seen from the fact that $\cM_{\star\,1,1}$ is not necessarily positive for an indefinite matrix. The root is that $SO(1,1)$ transformations are not sufficient to bring a generic matrix $\mathcal V\in SL(2,\RR)$ to upper triangular form, in contrast to the case of $SO(2)$.\footnote{Starting with a $\cV\in SL(2,\RR)$ with $\cV_{12}\neq 0$ and parametrizing $h\in SO(2)$ as $h=\cos\alpha\,\mathds{1}+i\sin\alpha\,\sigma_2$, forcing $h\cV$ to be upper triangular leads to $\cot\alpha=\cV_{1,1}/\cV_{1,2}$, which can be solved for $\alpha$ without restriction. Parametrizing $h\in SO(1,1)$ as $\cosh\alpha\,\mathds{1}+\sinh\alpha\,\sigma_2$, on the other hand, leads to $\coth\alpha=-\cV_{1,1}/\cV_{1,2}$, which requires $|\cV_{1,1}/\cV_{1,2}|>1$. If $|\cV_{1,1}/\cV_{1,2}|<1$, we can instead set $\cV_{1,1}=0$, by choosing $\tanh\alpha=-\cV_{1,1}/\cV_{1,2}$.\label{foot:gf}}
As a result, the $SL(2,\RR)$ transformations (\ref{eq:cM-SL2}) can map elements from the patch of $SL(2,\RR)/SO(1,1)$ covered by the $(\phi,\chi_\star)$ coordinates to elements outside this patch, e.g.\ with $\mathcal M_{1,1}<0$. This would appear to turn $e^\phi$ negative and $\phi$ complex, as noted above.

\sm

Instead of accepting a complex dilaton, however, one should define a second coordinate patch for $SL(2,\RR)/SO(1,1)$, with a different parametrization for $\cV$ which covers the necessary part. Following footnote \ref{foot:gf}, we choose a form with first diagonal entry zero, as
\begin{align}\label{eq:Vstar-2}
	\tilde \cV&=\begin{pmatrix}0 & -e^{-\phi/2}\\ e^{-\phi/2} & e^{\phi/2}\chi_\star\end{pmatrix} & 
	\tilde \cM_\star=\begin{pmatrix} -e^{\phi} & -\chi_\star e^\phi\\ -\chi_\star e^\phi &e^{-\phi}-e^\phi\chi_\star^2 \end{pmatrix}
\end{align}
The parametrization (\ref{eq:Vstar-2}) accommodates $\tilde \cM_{\star\,1,1}<0$ with real $\phi$ and in particular covers the $SL(2,\RR)$ images with apparent $e^{-\phi}<0$ pointed out below (\ref{eq:phi-transf}).
Instead of implying complex $\phi$, they are covered by a different coordinate patch on $SL(2,\RR)/SO(1,1)$.\footnote{The transformation from one coordinate patch to the other is a field redefinition rather than an active $SL(2,\RR)$ transformation, in the sense that the matrix $\cM_\star$ is unchanged instead of transforming by (\ref{eq:cM-SL2}) and the other supergravity fields which would transform under $SL(2,\RR)$ also do not change.}
The axion-dilaton and two-form Lagrangian, $\mathcal L=\frac{1}{4}e\,\tr\left(\partial_\mu \cM_\star^{-1}\partial^\mu\cM_\star\right) + \frac{1}{12}e\mathcal H_{\mu\nu\rho}^T\cM_\star\mathcal H^{\mu\nu\rho}$ with $\mathcal H=(idC_{(2)},H_{(3)})^T$, leads in components to the same kinetic terms for $\phi$ and $\chi_\star$ with the parametrizations (\ref{eq:Vstar-1}) and (\ref{eq:Vstar-2}), while the combinations among the two-form fields with positive/negative kinetic terms are swapped between (\ref{eq:Vstar-1}) and (\ref{eq:Vstar-2}). The coordinate patch (\ref{eq:Vstar-2}) thus realizes the component field properties of the Type IIB$^\prime$ real form.

\sm

In summary, rather than implying complex $\phi$, the transformations in (\ref{eq:phi-transf}) which seemingly lead to $e^\phi<0$ signal the necessity to change coordinates on $SL(2,\RR)/SO(1,1)$. This maintains a real dilaton with standard action for the axion-dilaton system. Similar to solutions with 7-branes in Type IIB$_\RR$, solutions in Type IIB$^\star$ do not in general have globally defined axion and dilaton fields, but they nevertheless exist locally.
At an operational level, one can allow $e^\phi<0$ with the understanding that, in regions where $e^\phi<0$, one should construct the associated matrix $\cM_\star$ with $e^\phi<0$ taken at face value and then transition to a different parametrization of $\cM_\star$ where $e^\phi>0$. Alternatively, one can work with $\cM_\star$ directly.
In the following we therefore only insist on $e^\phi$ being real, or $\tau_\pm$ both being imaginary, which is already guaranteed by the conditions in (\ref{3.b.1}). The resulting solutions are regular at the level of Type IIB supergravity. The fate of string and brane excitations, e.g.\ when crossing transition regions, will be left for future investigation.

\newpage

\section{Solving for $S^6 \times S^2 \times \Sigma$ solutions}
\label{sec:4}

As advocated in the Introduction, our main interest in the $dS_{1,5}\times S^2\times\Sigma$ solutions is for their straightforward analytic continuation to the geometry $S^6\times S^2\times\Sigma$ relevant to the Euclidean polarized IKKT model. The transition is accomplished by continuing the six-dimensional symmetric space in the metric as $dS_{1,5}\rightarrow S^6$. The warp factors and the expressions for the remaining basic supergravity fields are unaffected, and the reality and regularity conditions discussed in the previous section therefore carry over unchanged.

\sm
	
While the $dS_{1,5}\times S^2\times\Sigma$ solutions were understood within the Type IIB$^\star$ real form of Type~IIB$_\CC$, the $S^6\times S^2\times\Sigma$ solutions cannot be understood within any of the Type~IIB real forms we discussed, since none of them has signature $(10,0)$. Correspondingly, the Killing spinors inherited from the $dS_{1,5}\times S^2\times\Sigma$ solutions fail to satisfy appropriate reality conditions for Euclidean signature, though the $S^6\times S^2\times\Sigma$ solutions are nevertheless invariant under the associated linearized supersymmetry transformations.

\sm

We now discuss the concrete implementation of the regularity and positivity conditions for solutions with geometry of the form $S^6 \times S^2$ warped over a Riemann surface $\Sigma$ with Euclidean signature. For such solutions, we allow both the $S^6$ and the $S^2$ components to collapse at the boundary of $\Sigma$ and to close off into $S^7$ and $S^3$ spheres. We shall investigate solutions here under the following assumptions (in the notations of section \ref{sec:3}):
\begin{enumerate}
\itemsep=-0.04in
\item $\Sigma$ is the upper half complex plane;
\item all functions and differentials are single-valued in the interior of $\Sigma$;
\item $\hat\kappa^2$ is positive,  without loss of generality.
\end{enumerate}

\subsection{Implementing positivity of $\hat\kappa^2$}

As a first step we solve the positivity constraint for $\hat\kappa^2$. 
The construction proceeds in two steps. The first parallels a discussion in section 3.1 of \cite{DHoker:2017mds}. To bring out the parallels between the discussion here and in \cite{DHoker:2017mds}, we change basis for the holomorphic functions,
\begin{align}
\label{4.a.1}
	\cA_\pm&=\frac{1}{\sqrt{2}}\left(\cA_1\mp\cA_2\right)
\end{align}
The expressions for $\hat \cG$, $\hat\kappa^2$ and $\hat T$ in terms of $\cA_\pm$ are then given by, 
\begin{align}
\label{4.a.2}
\hat\cG&=|\cA_+|^2-|\cA_-|^2-\cB-\cB^\star
&
\partial_w\cB&=-\cA_+\partial_w\cA_-+\cA_-\partial_w\cA_+
\no \\
\hat\kappa^2&=-\left|\partial_w\cA_+\right|^2+\left|\partial_w\cA_-\right|^2
&
\hat T^2 & = -1 -{2 \over 3} \, { \p_w \hat \cG \, \pbw \hat \cG \over \hat \kappa ^2 \, \hat \cG}
\end{align}
Using the reality conditions (\ref{3.k.1}), the closing off conditions (\ref{eq:alpha3-rel}) and (\ref{eq:alpha1-rel})  for $\cM_6=S^6$ and $\cM_2 = S^2$  become,
\begin{align}
\label{eq:bndy-cond-cApm}
	S^6\rightarrow 0:&&
	\cA_\pm(w)&= \big (\cA_\mp(w^\star) \big )^\star  &&
	\nonumber\\
	S^2\rightarrow 0:&& \partial_w\cA_\pm(w)&=- \big ( \partial_{w^\star}\cA_\mp(w^\star) \big )^\star &&
\end{align}
The positivity requirements for $\hat\kappa^2$ are then equivalent to demanding $|\p_w \cA_+|^2 < | \p_w \cA_-|^2$ throughout the interior of the upper half complex plane $\Sigma$, with equality on the boundary $\p \Sigma$. The constraint is conveniently expressed in terms of an electrostatics potential $\Phi$ which is defined by,
\bea
\label{4.a.3}
\Phi(w,\sw) =-\ln\left |\frac{\p_w\cA_+}{\p_w\cA_-}\right|^2
\eea
It satisfies $ \Phi >0$ and $\p_w\p_\sw\Phi=0$ in the interior of $\Sigma$ and vanishes on the boundary thanks to the  
reflection conditions for $\partial\cA_\pm$ which imply, 
\begin{align}
\label{4.a.4}
	 \Phi(w,\sw)&=-\Phi(\sw,w)^\star
\end{align}
As is familiar from electrostatics, such a potential may be realized in terms of an arbitrary set of positive charges at points $s_a$ with $\Im(s_a) >0$ in the interior of the upper half plane and corresponding mirror negative charges at points $s_a^*$ in the lower half plane. Taking $N$ positive charges to be all equal to unity, we obtain the following realization of $\Phi$, 
\begin{align}
\label{4.a.5}
	\Phi&=-\sum_{a=1}^N \ln\left|\frac{w-s_a}{w-s_a^\star}\right|^2
\end{align}
The differentials $\partial \cA_\pm$ are then given by,
\begin{align}
\label{4.a.6}
	\partial_w\cA_+&=a_+f(w)\prod_{a=1}^N(w-s_a)
	&
	\partial_w\cA_-&=a_-f(w)\prod_{a=1}^N(w-s_a^\star)
\end{align}
in terms of an arbitrary non-vanishing holomorphic function $f(w)$. The complex constants $a_\pm$ are constrained by $|a_+/a_-|^2=1$ and can be reduced to phases by redefining $f$.

\subsection{Singularities on $\partial\Sigma$}
\label{sec:singularities}

The behavior of the functions $\cA_\pm$ near the boundary $\p \Sigma$ governs the collapsing of the spheres $S^6$ and $S^2$ according to (\ref{eq:bndy-cond-cApm}). We now classify admissible isolated singularities at the boundary of $\Sigma$.  This classification involves the positivity conditions of (\ref{3.b.1}) in the interior of $\Sigma$, and hence depends on the real form of Type IIB supergravity and the particular real form solution.

\sm

To analyze the behavior near a point on $\p \Sigma $, we use local complex coordinates, $w, w^*$ that vanish at this point and are real on $\p \Sigma$. To either side of $w=0$ on the real line, either $S^2$ or $S^6$ may collapse.  Consistently with the general form of the differentials in (\ref{4.a.6}), the differentials near $w=0$ are assumed to take the following form,
\begin{align}
	\partial_w\cA_\pm &=a_\pm w^p + b_\pm w^{p+1}+ \cO(w^{p+2}) \hskip 1in a_\pm \not=0
\end{align}
for a  real exponent $p$. The functions $\cA_\pm$ then behave as follows,
\begin{align}\label{eq:Apm-sing}
	\cA_\pm &=\cA_\pm^0+ \begin{cases}a_\pm \ln w + b_\pm w  + \cO(w^2) & p=-1
		\\
		-\frac{a_\pm}{w}+b_\pm\ln w + \cO(w) & p=-2
		\\
		a_\pm\frac{w^{p+1}}{p+1} + b_\pm \frac{w^{p+2}}{p+2} + \cO(w^{p+3}) & p\neq -2,-1
	\end{cases}
\end{align}
where $\cA^0_\pm$ are integration constants. For $p \not = -1, -2$ (the cases $p=-1,-2$ will be discussed separately below), the combination $\cB$ evaluates to, 
\begin{align}\label{eq:cB-exp}
	\cB&=\cB_0-\cA_+^0\cA_-+\cA_-^0\cA_+-(a_+b_--a_-b_+)\begin{cases}\frac{1}{4}\ln w & p=-\frac{3}{2} \\ \frac{w^{2p+3}}{(p+1)(p+2)(3+2p)} & p\neq -\frac{3}{2}\end{cases}
\end{align}
We discuss separately the cases where $S^6$ collapses for $w>0$ and where $S^2$ collapses for $w>0$. They are realized, respectively, by the following conditions relating the parameters,
\begin{align}\label{eq:bc-rel}
	\text{$S^6\rightarrow 0$ for $w>0$:}&&
	(\cA_-^0)^\star&=\cA_+^0
	&
	(a_-)^\star&=+a_+ & (b_-)^\star&=+b_+&&
	\nonumber\\
	\text{$S^2\rightarrow 0$ for $w>0$:}&&
	&&
	(a_-)^\star&=-a_+ & (b_-)^\star&=-b_+
\end{align}
We note that the conditions of (\ref{eq:bndy-cond-cApm}) for the collapse of $S^6$ imply a relation  between the integration constants $\cA_\pm^0$ as well as relations between $a_\pm$ and $b_\pm$.

\sm

We shall now give a more detailed analysis of the allowed exponents $p$ for the cases of either $S^6$ or $S^2$ collapsing for $w>0$. For integer $p$, the boundary conditions do not change across $w=0$. For half integer $p$, the conditions relating $a_\pm$ and $b_\pm$ transition across $w=0$ to those appropriate for the other sphere to collapse, and the solution can be regular to both sides of $w=0$. For  $p$ neither integer nor half integer, the boundary conditions for $w<0$ and $w >0$ cannot be matched, so that all such values are excluded. Below, we shall analyze in detail the remaining cases where $p \in \ZZ/2$, including the cases $p=-1$ and $p=-2$ that were left unanalyzed earlier.

\subsubsection{$S^6\rightarrow 0$ for $\Re(w)>0$ on $\p \Sigma$}

We now analyze the conditions for collapsing $S^6$ for $\Re(w)>0$ of the boundary $\p \Sigma$. 

$\bullet$ 
For $p=-1$ the monodromy associated with the logarithmic branch cuts in $\cA_\pm$ is inconsistent with the reflection relations: $\cA_\pm$ both shift by an imaginary constant, which is inconsistent with $(\cA_-^0)^\star=\cA_+^0$. This situation is analogous to the one encountered in section 2.3 of  \cite{Corbino:2018fwb}. 
Hence $p=-1$ is ruled out in this case.

\sm

$\bullet$  For $p=-2$, the logarithmic singularity is again inconsistent with the reflection relation, as was the case for $p=-1$. Cancelling the logarithmic singularity by requiring $b_\pm=0$ produces single-valued $\cA_\pm$; then demanding $\cG$ to have no monodromy forces the next order to vanish as well and, if this is assumed, then $\hat T^2$ cannot be non-negative. Hence, the case  $p=-2$ is ruled out in this case.

\sm

$\bullet$  For $p=-\frac{3}{2}$, $\cB$ in (\ref{eq:cB-exp}) shifts by a real constant at $w=0$ due to the logarithmic term. Since $\cA_\pm$ do not shift, this translates to a shift in $\cG$. The condition $\cG\big\vert_{\partial\Sigma}=0$ therefore fails upon crossing $w=0$. Hence $p=-{ 3 \over 2}$ is ruled out in this case.

\sm

$\bullet$  For $p\neq -2,-\frac{3}{2},-1$, and with the conjugation relations for $S^6$ collapsing, we find
\begin{align}
\hat\cG&=\frac{a_+b_--a_-b_+}{(p+1) (p+2)}\left(\frac{w^{3+2p}-\bar w^{3+2p}}{2 p+3}-(w\bar w)^{p+1}(w-\bar w)\right)
+ \cO(|w|^{2p+4})
\end{align}
Using (\ref{2.g.3}) to compute $T^2$ and (\ref{3.k.2}) to evaluate $\hat T$ using $\hat \cG$, we see that the overall multiplicative constant in $\hat \cG$ drops out and we obtain the following behavior of $\hat T^2$ in polar coordinates $w = r \, e^{i \theta}$ for small $r$ and $0 < \theta < \pi$,
\bea
\hat T^2 & = & -1 +{ (2p+3)N
\over 3(p+1)(p+2)\sin \theta \big ((2p+3) \sin \theta - \sin((2p+3) \theta) \big )} + \cO(r)
\eea
where the numerator $N$ is given by,
\bea
N & = & p^2+3p+3+(p+1) \cos \big ((2p +4)\theta \big ) - (p+2) \cos\big ( (2p +2) \theta \big )
\no \\ && 
 -(p+1)(p+2) \cos(2 \theta)
\eea
A necessary condition for positivity of $\hat T^2$ is positivity  for $\theta={\pi \over 2}$ for which we have, 
\begin{align}
	\hat T^2=\frac{1}{3} \left(1+\frac{1}{p+1}-\frac{1}{p+2}-\frac{4 p+6}{2 p+\cos (\pi  p)+3}\right) + \cO(r)
\end{align}
Recall that only half-integers $p \not = -1,-2, -{3 \over 2}$ are allowed from our earlier analysis so that $\cos(\pi p)$ may be either $1, 0$, or $ -1$, in which cases the above formula simplifies,
\bea
-{p \over 3(p+1)} \hskip 1in -{ p^2+3p+1 \over 3(p+1)(p+2)} \hskip 1in -{p+3 \over 3(p+2)} 
\eea
Clearly $p >0$ and $p< -3$ are ruled out, leaving the following allowed exponents,
\begin{align}
	p\in \Big \{-3,-\frac{5}{2},-\frac{1}{2} \Big \}
\end{align}
For the cases $p=-{5 \over 2}$ and $p = - \half$, one  verifies $\hat T^2 >0$ for all $0 < \theta < \pi$.  For $p=-3$, the leading order in $y$ computed above vanishes for all $\theta$.  Note that zeros in the differentials correspond to $p=1$ and are excluded on boundary segments where $S^6$ collapses. For $p=-3$ compatibility with the condition on $\cA_\pm^0$ requires $\Res_{w=0}(\partial_w\cA_\pm)=0$ and an analysis of the sign of the order $\cO(r)$ contribution to $\hat T^2$.

\subsubsection{$S^2\rightarrow 0$ for $\Re(w)>0$}

We now analyze the conditions for collapsing $S^2$ for $\Re(w)>0$ of the boundary $\p \Sigma$. Contrarily to the case of $S^6$ collapsing, the cases $p=-1,-2$ are not ruled out right away on boundary segments where $S^2$ collapses, since the condition (\ref{eq:bndy-cond-cApm}) only constrains the differentials, while shifts in $\cA_\pm$ are allowed.

\sm

$\bullet $ 
For $p=-1$, keeping $\cG\vert_{\partial\Sigma}=0$ on both sides of $w=0$ requires the residue of $\partial_w\cB$ to be real, as in \cite{DHoker:2017mds,Kaidi:2018zkx}. With this reality condition, however, the near-pole expansion reproduces equations (3.32) and (3.34) of \cite{DHoker:2017mds}, so that $\hat\kappa^2$ and $\hat\cG$ have the same sign. This was desired for $AdS_6\times S^2\times\Sigma$ but is not allowed here, so $p=-1$ is ruled out.

\sm

$\bullet$ 
For $p=-2$, we find near $w=0$
\begin{align}
	\hat\cG&\sim(a_+b_+^\star-a_+^\star b_+)\frac{w-\bar w}{w\bar w}\ln(w \bar w)
\end{align}
The leading terms in $\hat\kappa^2$ and $\hat\cG$ have opposite signs, as required, so that $\hat T^2$ is positive and  $p=-2$ is an admissible singularity on boundary segments with $S^2$ collapsing.

\sm

$\bullet$ For $p \in \ZZ+ \half$ the boundary conditions change across the singularity. These cases are covered by the previous analysis for $S^6$ collapsing (related by the transformation $w\rightarrow -\bar w$). We conclude that $p\in\{-\frac{5}{2},-\frac{1}{2}\}$ are the allowed half-integer values.

\sm

$\bullet$ The remaining cases are integer $p \not= -1,-2$. The expression for $\hat\cG$ near $w=0$ is,
\begin{align}
	\hat\cG&=-2|\cA_+^0|^2+ 2|\cA_-^0|^2 +(\cA_+^0-\bar\cA_-^0)(\bar\cA_++\cA_-)+(\bar\cA_+^0-\cA_-^0)(\cA_++\bar\cA_-)
	\nonumber\\&\hphantom{=}
	+\frac{a_+b_--a_-b_+}{(p+1) (p+2)}\left(\frac{w^{3+2p}-\bar w^{3+2p}}{2 p+3} - (w\bar w)^{p+1}(w-\bar w)\right)
\end{align}
The constants $\cA_\pm^0$ in the expansion around $w=0$ are not constrained.\footnote{$\bar\cA_++\cA_-$ is constant along boundary segments with $S^2$ collapsing, and $\bar\cA_+-\cA_-$ is not constrained.}
The first line dominates the second unless $\bar \cA_\pm^0=\cA_\mp^0$ so that the first line vanishes, or $p< -2$.  

\sm

If the second line is dominant, an argument based on $\hat T^2$ analogous to the one for segments with $S^6$ collapsing before rules out $p\geq 0$ and $p\leq -3$. Since $p=0$ corresponds to regular boundary points, we conclude that $\cA_+^0-(\cA_-^0)^\star$ has to be non-zero at regular boundary points. For $\cA_+^0-(\cA_-^0)^\star\neq 0$ we find $\hat\cG \sim r^{1+p}\sin((2+p)\varphi)$ with $w=re^{i\varphi}$, which fails positivity for $\Im(w)>0$ if $p>0$. So $p\leq - 3$ and $p>0$ are ruled out.

\sm

In summary, the discussion so far leaves as potential options
\begin{align}
	p\in\left\{-\frac{5}{2},-2,-\frac{1}{2}\right\}
\end{align}
where we again do not list $p=0$, which corresponds to regular boundary points. Zeros in the differentials on the boundary are again ruled out by this discussion.

\subsection{No poles in the interior of $\Sigma$}

We shall now show that the presence of a pole in the interior of $\Sigma$ is incompatible with the reality and positivity conditions of (\ref{3.b.1}). Let us assume that the differentials $\p_w \cA_\pm$ have a pole in $w$ at a point $u$ in the upper half plane, namely $\Im(u) >0$. We shall denote the order of the leading pole by $q \geq 1$. If $q \geq 2$, then there may also be contributions from sub-leading poles of order $q-1$ and so on. Here, we shall need only the leading pole to argue that poles cannot occur consistently with (\ref{3.b.1}).

\sm

Since the conditions (\ref{3.b.1}) depend only on the functions $\hat \cG$ and $\hat \kappa ^2$, which are both  invariant under $SU(1,1)$ transformations on the functions $\cA_\pm$, we may use an $SU(1,1)$ transformation to rotate the leading pole in $ \cA_\pm$ to $\cA_+$ for $\hat \cG >0$  or to $\cA_-$ when $\hat \cG <0$.  Furthermore, swapping $\cA_+$ and $\cA_-$ amounts to a simultaneous change of sign in both $\hat \cG$ and $\hat \kappa ^2$ which leaves $\hat T^2$ unchanged and leaves the conditions (\ref{3.b.1}) unchanged.  Thus, we may rotate the leading pole in $w$ at $u$ to occur only in $\cA_+$ while $\cA_-$ may have poles at $u$ of lower order strictly less than $q$. 
We have the following leading behavior for $q \geq 1$,
\bea
\cA_+ \sim { a_+ \over (w-u)^q}
\hskip 1in
\hat \cG \sim { |a_+|^2 \over |w-u|^{2q}}
\hskip 0.8in
\hat \kappa ^2 \sim - { q^2 |\a_+|^2 \over |w-u|^{2q+2}}
\eea
The signs of $\hat \cG$ and $\hat \kappa^2$ are  opposite to one another as indeed required by (\ref{3.b.1}). Proceeding to analyze the sign of $\hat T^2$ using the last formula in (\ref{4.a.2}) we find,
\bea
\hat T^2 \sim - { 1 \over 3}
\eea
independently of the order $q$ of the pole. It is easy to check that the same value for $\hat T^2$ is obtained when $\cA_+ \sim \ln (w)$ when the leading pole of $\p_w \cA_+$ is simple. Since we have $\hat T^2<0$ in all cases, we find an inconsistency with the conditions of (\ref{3.b.1}) which precludes having any poles  in the interior of $\Sigma$, namely, in the upper half plane.

\subsection{The differentials $\p \cA_1$ and $ \p \cA_2$}
\label{sec:differentials}

Our analysis of the regularity and positivity conditions on the supergravity fields has determined  the general structure of  the differentials $\partial_w\cA_\pm$, which take the following form,  
\begin{align}
\label{4.d.1}
	\partial_w\cA_+&=a_+ f(w)\prod_{a=1}^N(w-s_a)
	&
	\partial_w\cA_-&=a_-f(w)\prod_{a=1}^N(w-s_a^\star)
\end{align}
for a common  function $f(w)$ that we shall now determine from the allowed singularities identified in section \ref{sec:singularities}. The allowed singularities on $\partial\Sigma$ are square root branch points with the  behavior $\p_w \cA_\pm \sim (w-b_\ell) ^{-1/2}$ or $\p_w \cA_\pm \sim (w-c_k)^{-5/2}$, as well as poles with $\p_w \cA_\pm \sim  (w-d_p) ^{-2}$ on boundary segments where $S^2$ collapses or $\p_w \cA_\pm \sim  (w-e_q)^{-3}$ on boundary segments where $S^6$ collapses. Finally, no singularities or branch points are allowed in the interior of $\Sigma$. This leads to the following general form of the function $f(w)$, 
\begin{align}
\label{4.d.2}
	f(w)&= \prod_{\ell=1}^L\frac{1}{\sqrt{w-b_\ell}}\prod_{k=1}^K\frac{1}{\sqrt{(w-c_k)^5}}
	\prod_{p=1}^P\frac{1}{(w-d_p)^{2}}\prod_{q=1}^Q\frac{1}{(w-e_q)^{3}}
\end{align}
where $b_\ell$, $c_k$, $d_p$ and $e_q$ are distinct points on the real axis, namely on $\p \Sigma$. The values $K,L,P,Q$ specify the multiplicity of each singularity while $N$ is the number of zeros in the interior of $\Sigma$. 
Assuming that the differentials $dw \, \p_w \cA_\pm$ are regular as $w \to \infty$  and non-vanishing, the orders  $K,L, P, Q \geq 0$ of the singularities are related to one another by,
\bea
\label{4.d.4}
N +2 = \half L +{5 \over 2} K + 2P + 3Q
\eea
where the addition of 2 on the left arises from the factor $dw$ in the 1-forms.  If a branch point or a pole is moved to $w=\infty$, then the formula is valid with the proviso that the singularity is properly included in the corresponding number $K, L, P$ or $Q$. 

\sm

Equation (\ref{4.d.4}) implies that $K+L$ must be an even integer. Also, for $N=0$, the form $\hat\kappa^2$ vanishes everywhere, which does not support a viable solution. Henceforth, we shall assume $N\geq 1$ so that we have at least one zero $s_a$ in the interior of $\Sigma$.

\sm

\begin{figure}[htb]
	\begin{tikzpicture}[xscale=1.2, yscale=1.6]
		\draw[red] (-0.5,0) -- (2.5,0);
		\draw[blue] (2.5,0) -- (5,0);
		\draw[red] (5,0) -- (7.5,0);
		\draw[blue] (7.5,0) -- (10,0);
		\node at (1.25,-0.3){\footnotesize $S^2\rightarrow 0$};
		\node at (3.75,-0.3){\footnotesize $S^6\rightarrow 0$};
		\node at (6.25,-0.3){\footnotesize $S^2\rightarrow 0$};
		\node at (8.75,-0.3){\footnotesize $S^6\rightarrow 0$};
		\node at (11.5,0) {\footnotesize $\partial\Sigma$};
		\node[blue] at (10.5,0) {$\cdots$};
		\node [red] at (-1,0) {$\cdots$};
		
		\node at (2.5,-0.25) {\footnotesize $b_{k-1}$};
		\node at (5,-0.25) {\footnotesize $c_{\ell}$};
		\node at (7.5,-0.25) {\footnotesize $b_{k}$};
		\draw (8,0) arc (0:180:2cm and 1cm);
		\node at (7.5,1) {\footnotesize $\cC_{m,n}$};
		
		\draw[thick] (2.5,-0.1) -- (2.5,0.1);
		\draw[thick] (5,-0.1) -- (5,0.1);
		\draw[thick] (7.5,-0.1) -- (7.5,0.1);

		\node at (0.5,0) {\textcolor{red}{$\times$}}; \node at (0.5,0.3) {\footnotesize $d_p$};
		\node at (9.5,0) {\textcolor{blue}{$\times$}}; \node at (9.5,0.3) {\footnotesize $e_q$};
	\end{tikzpicture}
	\caption{Boundary segments with $S^2$ and $S^6$ collapsing alternatingly, with square root branch points $b_k$ and $c_\ell$, as well as poles $d_p$ on boundary segments where $S^2$ collapses and $e_q$ on boundary segments where $S^6$ collapses.
	Also shown are contours $\cC_{m,n}$ connecting segments along which  $S^6$ collapses.\label{fig:segments}}
\end{figure}
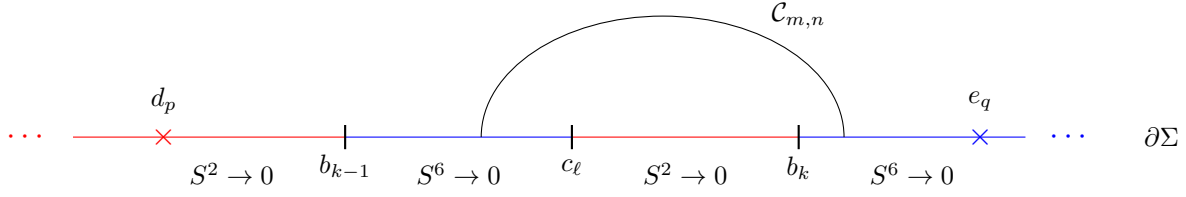

Inspection of (\ref{4.d.1}) and (\ref{4.d.2}) reveals that the arguments of the square roots are real on the real axis, so that the square roots take  either real or imaginary values. As a result, the function $f(w)$ alternates between real and imaginary values on the real line.  With a suitable choice of $a_\pm$ the differentials $\partial\cA_\pm$ satisfy the boundary conditions (\ref{eq:bndy-cond-cApm}) alternatingly on segments along $\partial\Sigma$, so that segments with $S^2$ collapsing and segments with $S^6$ collapsing take turns, with transitions at $w=b_\ell$ and $w=c_k$ (fig.~\ref{fig:segments}).

\subsection{Matching conditions for boundary segments}
\label{sec:4.5}

The local conditions for collapsing $\cM_2$ or $\cM_6$ at the boundary of $\Sigma$  were derived in section \ref{abehavior}. For the real form solution they  are  expressed in terms of reflection conditions on $ \cA_{\pm}$ for the collapse of $S^6$ and $\partial_w \cA_{\pm}$  for the collapse of $S^2$, respectively in (\ref{eq:bndy-cond-cApm}). For the collapse of $S^2$ no additional conditions have to be imposed. For the segments along which $S^6$ collapses, one first imposes the differential condition $\partial_w\cA_+- (\partial_w \cA_-)^\star=0$, on the boundary, which ensures that $\cA_+-\cA_-^\star$ is constant along the boundary segment. With a single segment where $S^6$ collapses, the integration constants in $\cA_\pm$ can then be chosen such that $\cA_+-\cA_-^\star=0$ on this segment and (\ref{eq:bndy-cond-cApm}) is satisfied. For multiple segments on which $S^6$ collapses, however, non-trivial conditions arise.  They take the form,
\begin{align}
\label{eq:lift}
	\int_{\cC_{m,n}} \Big ( dw \, \partial_w \cA_+ - \big (dw \, \partial_w  \cA_- \big )^\star \Big ) =0
\end{align}
where $\cC_{m,n}$ is a contour connecting the $m^{\rm th}$ and $n^{\rm th}$ boundary segments with $S^6$ collapsing (see fig.~\ref{fig:segments}). If these conditions are satisfied, $\cA_+-\cA_-^\star$ takes the same value on all boundary segments where $S^6$ collapses, and we can make this value zero by a judicious choice of integration constants in $\cA_\pm$.

\sm

For $p=-3$ poles, which may be present along boundary segments on which $S^6$ collapses, additional conditions arise from the fact that logarithmic branch cuts in $\cA_\pm$ are incompatible with the boundary conditions (see \cite[section 2.3]{Corbino:2018fwb}), and cannot occur. This leads to the following condition for each triple pole,
\begin{align}
\label{eq:reg-res-dAp}
	\Res_{w=e_q}(\partial_w \cA_\pm)&=0
\end{align}
For $p=-2$ poles, which may be present along segments where $S^2$ collapses, no analogous condition arises, since the boundary conditions only involve the differentials.

\sm

The next step is to realize the boundary condition $\hat\cG\vert_{\partial\Sigma}=0$ globally. Locally, along a given segment, this can be accomplished by a suitable choice of the integration constant for  $\cB$.
Moreover, with the above conditions we have $(\partial_w+\partial_\sw)\hat\cG=0$ along the entire boundary of $\Sigma$. The remaining conditions therefore amount to the absence of monodromy in $\cB+\cB^\star$ at singular points.
At $p=-1/2$ branch points, the scaling is such that there is no monodromy in $\cB$, so no additional conditions arise. For $p=-5/2$ branch points and $p=-2,-3$ poles, however, matching leads to further  conditions 
\begin{align}
\label{eq:reg-dB}
	\int_{\cC_\epsilon(c_k)}dw \, \partial_w\cB+{\rm c.c.}&= 0&
	\int_{\cC_\epsilon(d_p)} dw \, \partial_w\cB+{\rm c.c.}&= 0&
	\int_{\cC_\epsilon(e_q)}dw \, \partial_w\cB+{\rm c.c.}&= 0
\end{align}
where $\cC_\epsilon(\cdot)$ denotes a contour connecting boundary segments to either side of the branch point or pole.
The discussion so far solves the regularity and positivity conditions for $\hat\kappa^2$ ($\hat\kappa^2>0$ in the interior of $\Sigma$ and $\hat\kappa^2\vert_{\partial\Sigma}=0$) and the boundary conditions for $\hat\cG$.

\subsection{Behavior at singular points}\label{sec:sing-points}

Here we discuss the behavior of the supergravity solutions near the singular points.
We use the notation of section \ref{sec:singularities} for the expansion of $\partial\cA_\pm$, extended to the following form,
\begin{align}\label{dAexp-2}
	\partial_w\cA_\pm &=a_\pm w^p + b_\pm w^{p+1}+ c_\pm w^{p+2}+d_\pm w^{p+3} + \cO(w^{p+4})
\end{align}
Regularity conditions may constrain the expansion coefficients, with the form of the constraints depending on the choice of $p$, as discussed in sections \ref{sec:singularities} and \ref{sec:4.5}.

\subsubsection{$p=-1/2$ transition points}\label{sec:p-1-half}
For $p=-1/2$ the boundary conditions transition from $S^6$ collapsing to $S^2$ collapsing. Using the relations in the first line of (\ref{eq:bc-rel}) with $w=re^{i\varphi}$, we obtain, 
\begin{align}
	\hat\kappa^2&\approx 3 \alpha  \sin \varphi 
	&
	\hat\cG&\approx-8 \alpha  r^2\sin\varphi\sin^2\left(\frac{\varphi}{2}\right)
	&
	\hat T&\approx\frac{1}{3} \tan \left(\frac{\varphi }{2}\right)
\end{align}
where
\begin{align}\label{eq:alpha-exp-def}
	\alpha&=\frac{2}{3i}(a_+^\star b_+-a_+b_+^\star)
\end{align}
This leads to the following expressions for the metric factors, 
\begin{align}
	\rho^2&\approx -\frac{c_6}{2}\frac{\sqrt{\alpha}}{r}
&
	f_2^2&\approx -8c_6\sqrt{\alpha } r \cos^2 \left(\frac{\varphi }{2}\right)
&	f_6^2&\sim-8 c_6 \sqrt{\alpha } r \sin^2 \left(\frac{\varphi }{2}\right)
\end{align}
For $c_6<0$ the metric functions are non-negative. They exhibit the following behavior,  $f_2\rightarrow 0$ for $\varphi \rightarrow \pi$ and $f_6\rightarrow 0$ for $\varphi\rightarrow 0$, as desired.
For the metric on $\Sigma$ we note that $4\rho^2|dw|^2\approx -2c_6\sqrt{\alpha}|dw/\sqrt{w}|^2$. This can be uniformized by setting $w=z^2$, so that this singularity maps to a corner. The dilaton can be determined from (\ref{eq:expphi}), and we find
\begin{align}
	e^\phi&\approx \frac{i (a_++a_+^\star) (b_++b_+^\star)}{a_+^\star b_+-a_+b_+^\star}
	&
	\chi&=\frac{a_+^\star-a_+}{a_+^\star+a_+}
\end{align}
Transition points produce cycles which could in principle lead to instanton charge. But with the scalings for $p=-1/2$, $\cY$ does not shift at the transition point and following the discussion in section \ref{sec:charges} there is no instanton charge.

\subsubsection{$p=-5/2$ transition points}
\label{sec:p-5-half}

For $p=-5/2$ the boundary conditions also transition from $S^6$ collapsing to $S^2$ collapsing and we again use the relations in the first line in (\ref{eq:bc-rel}).  A non-trivial constraint on the expansion coefficients arises from the requirement that $\hat\cG$ should not shift at the singularity. This requirement constrains the subleading coefficients $d_\pm$ in (\ref{dAexp-2}) to satisfy,
\begin{equation}
	a_+d_+^\star-a_+^\star d_+ +3 b_+ c_+^\star-3 b_+^\star c_+=0
\end{equation}
With $w=re^{i\varphi}$ and the definition of $\alpha$ as before the expansions lead to
\begin{align}
	\hat\kappa^2&\approx\frac{3 \alpha  \sin\varphi}{r^4}
	&
	\hat\cG&\approx-\frac{8\alpha}{r^2}\sin\varphi\sin^2\left(\frac{\varphi}{2}\right)
	&
	\hat T&\approx\frac{1}{3} \tan \left(\frac{\varphi }{2}\right)
\end{align}
with $\alpha$ defined in (\ref{eq:alpha-exp-def}). Regularity requires $\alpha>0$.
The resulting metric functions are
\begin{align}
	\rho^2&\approx-\frac{c_6}{2}\frac{\sqrt{\alpha}}{r^3}
&
	f_2^2&\approx-\frac{8 c_6 \sqrt{\alpha }}{r} \cos^2 \left(\frac{\varphi }{2}\right)
&	f_6^2&\approx-\frac{8 c_6 \sqrt{\alpha }}{r}\sin^2 \left(\frac{\varphi }{2}\right)
\end{align}
They are non-negative for $c_6<0$.
As $r \to 0$, the radii of both $S^2$ and $S^6$ blow up, as does $\rho^2$, and the entire geometry decompactifies. The dilaton and the axion behave as follows,
\begin{align}
\label{4.3a}
	e^\phi&\approx 
	-\frac{4}{9 \alpha  }(a_++a_+^\star)^2
	\frac{2+\cos\varphi}{r}
	&
	\chi&=\frac{a_+^\star-a_+}{a_+^\star+a_+}
\end{align}
The exponentiated dilaton thus becomes negative if $a_++a_+^\star\neq 0$. The expansion changes qualitatively when $a_+$ is imaginary, $a_+^\star=-a_+$, so that the leading term in $e^\phi$  as $r \to 0$ vanishes. The expansions for $\tau_\pm$ in this case are
\begin{align}\label{eq:p5half-tau}
	\tau_+&\approx \frac{10}{3Cr^4} &
	\tau_-&\approx \frac{4a_+}{3 (b_++b_+^\star)}\frac{2+\cos\phi}{r}
\end{align}
with $\tau_-$ subleading compared to $\tau_+$ and the imaginary constant $C$ given by
\begin{equation}\label{eq:p5half-C-def}
	C=\frac{3 (a\cdot e)+5(b\cdot d)}{a_+^2}+\frac{15(a\cdot c)(b\cdot c)}{a_+^2 (a\cdot b)}+\frac{5(a\cdot c)^3}{a_+^2 (a\cdot b)^2}
\end{equation}
where the dot products are defined as $(a\cdot b)=a_+b_+^\star-a_+^\star b_+$ and likewise for the other expansion coefficients.
This leads to the behavior for $e^\phi$ and $\chi$ as
\begin{align}
	\label{eq:p5half-dilaton-asympt}
	e^\phi&\approx \frac{3i}{5}Cr^4
	&
	\chi&\approx \frac{5}{3Cr^4}
\end{align}
Now the exponentiated dilaton tends to zero and the sign of the leading contribution depends on $C$. The axion, however, now diverges for $r\rightarrow 0$. The constants $a_{\pm}$ transform under $SL(2,\RR)$, following the discussion in sections \ref{sec:3.9} and \ref{sec:metric-reg}. If the condition $a_+=-a_+^\star=-a_-$ is satisfied, the $SL(2,\RR)$ transformations maintaining it have $c=0$ and $d=1/a$.
The two-form fields  behave as follows for general non-zero $a_+$,  
\begin{align}
	B_{(2)}&\approx +\frac{8}{9} \sqrt{2} c_6 \nu (a_++a_+^\star)\left(\frac{1}{\sqrt{w}}+\frac{1}{\sqrt{\sw}}\right)^3 \hat e^{67}
	\nonumber\\
	C_{(2)}&\approx-\frac{8}{9}\sqrt{2}c_6\nu(a_+-a_+^\star)\left(\frac{1}{\sqrt{w}}+\frac{1}{\sqrt{\sw}}\right)^3  \hat e^{67}
\end{align}
For the $p=-5/2$ transition point the scalings of $\cA_\pm$ can give D-instanton charge. Following the discussion in section \ref{sec:charges}, for a contour $\cC_\epsilon$ infinitesimally close to the branch point, the D-instanton charge is given by,
\begin{equation}
	Q_{\rm D(-1)}= -720c_2^2c_6^2 \, V_{S^6}V_{S^2}\; 
	\lim_{\epsilon\rightarrow 0}
	\int_{\cC_\epsilon} dw \, \p_w (i\cY_1-i\tilde\cY_1)
\end{equation}
where $V_{S^6}$ and $V_{S^2} $ are the volumes of the spheres $S^6$ and $S^2$ of unit radii, respectively. For the $dS_{1,5}\times S^2\times\Sigma$ real form $\cY_i$ and $\tilde\cY_i$ are related by $\tilde\cY_i=-\cY_i^\star$, so the D-instanton charge is imaginary. It can be expressed as,
\begin{equation}
	Q_{\rm D(-1)}= +720\pi c_2^2c_6^2 \, V_{S^6}V_{S^2} \Res_{w=0}(\partial_w\cY_1)- {\rm c.c.}
\end{equation}
We explicitly evaluate this charge for the case $a_++a_+^\star=0$, where, with $C$ given in (\ref{eq:p5half-C-def}),
\begin{equation}\label{eq:p5half-QDm1}
	Q_{\rm D(-1)}=1536\pi c_2^2c_6^2 \, V_{S^6}V_{S^2}a_+^2  (b_++b_+^\star)^2C
\end{equation}
In order to exhibit the form of the solution near the $p=-5/2$ branch point more clearly, we change coordinates on $\Sigma$ to real non-negative coordinates $(R_1,R_2)$ as follows,
\begin{align}
	r&=\frac{1}{R_1^2+R_2^2} & \varphi&=2\tan^{-1}\left(\frac{R_1}{R_2}\right)
\end{align}
The $p=-5/2$ transition point is at $R_1^2+R_2^2\rightarrow\infty$. The metric becomes manifestly flat,
\begin{equation}
	ds^2\approx-8\sqrt{\alpha}c_6\left(R_1^2ds^2_{S^6}+R_2^2ds^2_{S^2}+dR_1^2+dR_2^2\right)
\end{equation}
and the 3-form field strengths then take the form,
\begin{align}
	dB_{(2)}&\approx \frac{64}{3}\sqrt{2}\nu c_6(a_++a_+^\star)R_2^2 dR_2 \wedge  \vol_{S^2}
	\nonumber\\
	dC_{(2)}&\approx -\frac{64}{3}\sqrt{2}\nu c_6(a_+-a_+^\star)R_2^2 dR_2 \wedge \vol_{S^2}
\end{align}
We note that, in terms of the $\RR^3$ formed by the radial coordinate $R_2$ and the $S^2$ angular directions, $R_2^2dR_2\wedge \vol_{S^2}=\vol_{\RR^3}$.
The dilaton and axion for $a_++a_+^\star\neq 0$ become,
\begin{align}
	e^\phi&\approx -\frac{4}{9\alpha}(a_++a_+^\star)^2(R_1^2+3R_2^2) & \chi&\approx\frac{a_+^\star-a_+}{a_+^\star+a_+}
\end{align}
while for $a_++a_+^\star= 0$ they are given by,
\begin{align}
	e^{-\phi}&\approx \frac{5}{3iC}(R_1^2+R_2^2)^4
	& 
	\chi&\approx \frac{5}{3C}(R_1^2+R_2^2)^4
\end{align}
At leading order $ie^{-\phi}=\chi$, in line with $\tau_-$ being subleading compared to $\tau_+$ in (\ref{eq:p5half-tau}). The subleading $\tau_-$ becomes,
\begin{equation}
	\tau_-\approx \frac{4a_+}{3 (b_++b_+^\star)}\left(R_1^2+3 R_2^2\right)
\end{equation}
In line with the general presence of D-instanton charge, the flat Einstein-frame metric matches the D-instanton solution \cite{Gibbons:1995vg}. The 2-form fields and the behavior of the axion-dilaton are deformed compared to the original D-instanton solution. For $a_++a_+^\star=0$ the emerging asymptotic region is weakly-coupled and matches the original D-instanton asymptotics.  Moreover, the non-vanishing (but subleading) $\tau_-$ and the non-vanishing $C_{(2)}$ may be seen as deformation of the original instanton solution, which has $\tau_-=0$ and vanishing 2-forms. The discussion here, however, covers full $SL(2,\RR)$ orbits of solutions, and for the regularity discussion for $e^\phi$ we refer to section \ref{sec:metric-reg}.

\subsubsection{$p=-2$}

For $p=-2$ we again use the parametrization of $\cA_\pm$ as in (\ref{eq:Apm-sing}) with relations among the coefficients such that $S^2\rightarrow 0$. That is,
\begin{align}
	\cA_\pm &=\cA_\pm^0-\frac{a_\pm}{w}+b_\pm\ln w & p=-2
\end{align}
with $a_\pm^\star=-a_\mp$ and $b_\pm^\star=-b_\mp$. The condition for $\hat\cG$ to have no monodromy leads to $(\cA_\pm^0)^\star=+\cA_\mp^0$.
Then
\begin{align}
	\hat\kappa^2&\approx -a\cdot b\frac{w-\sw}{|w|^4} 
	& 
	\hat\cG&\approx -a\cdot b\frac{w-\sw}{|w|^2}\log |w|^2
	&
	\hat T^2&\approx \frac{2|w|^2 \log|w|^2}{3 (w-\sw)^2}
\end{align}
where $a\cdot b=a_+^\star b_+ - a_+ b_+^\star$ is imaginary. 
$\hat T^2$ is positive and $\hat\kappa^2$ and $\hat\cG$ have opposite signs.
The metric functions become
\begin{align}
	\rho^2&\approx c_6\left(\frac{(a\cdot b)^2}{27|w|^{10}\log|w|} \right)^{\frac{1}{4}}
&
	f_6^2&\approx -6|w|^2\log|w|^2\rho^2
\no \\
&&	f_2^2&\approx -(w-\sw)^2\rho^2
\end{align}
All warp factors blow up at $|w|\rightarrow 0$ and the geometry decompactifies, while the exponentiated 
dilaton tends to zero,
\begin{align}
	e^\phi&\approx -(b_+-b_+^\star)^2\sqrt{\frac{|w|^2\log|w|}{12(a\cdot b)^2}}
\end{align}

\subsubsection{$p=-3$}

For a $p=-3$ singular point on a boundary segment where $S^6\to 0$, the analysis parallels the one given in section 3.2 of \cite{Corbino:2018fwb}  and we will be brief.
The expansion of  $\cA_{\pm}$ in (\ref{eq:Apm-sing})  near the singular point $w=0$  takes the form
\begin{align}
\cA_{\pm} = A^0_\pm + {a_{\pm}\over w^2} + {b_\pm \over w} + c_{\pm} \ln w+ d_{\pm} w+ \cdots 
\end{align}
where the vanishing condition (\ref{eq:bndy-cond-cApm}) for $S^6\to 0$  implies $a_{\pm}=a^*_{\mp}$,    $ b_\pm =b^*_{\mp}$, $ d_\pm =d^*_{\mp}$ and $ A^0_\pm =(\cA^0_\mp)^*$. The  condition for the absence of a monodromy in $\hat\cG$ implies that $c_{\pm}=c_{\pm}^*=0$. With $w=r e ^{i \phi}$  the 
expansion becomes 
\begin{align}
\hat\kappa^2& \approx  2i(b_-a_-^*-a_-b_-^*) {\sin \phi \over r^5}+ 2i (d_- a_-^*-d_-^*a_-) {\sin 3 \phi \over r^3} +\cdots\nonumber \\
\hat\cG & \approx -{4\over 3} i (b_-a_-^*-a_-b_-^*) {\sin^3 \phi \over  r^3}- 4i (d_- a_-^*-d_-^*a_-){\sin^3 \phi \over r} + \cdots
\end{align}
If $i(b_-a_-^*-a_-b_-^*) $ is real and positive the regularity conditions $\hat\kappa^2>0, \hat\cG<0$  are satisfied in a neighborhood of the origin.

\subsection{The regularity condition for $\hat\cG$}

So far, the construction implements the regularity requirements for $\hat \kappa^2$, namely $\hat \kappa^2>0$ in the interior of $\Sigma$ and $\hat \kappa^2=0$ on $\partial\Sigma$ (away from isolated singular points).   In the following we show that the regularity condition on the function $\hat \cG$, namely that $\hat \cG<0$ in the interior of  $\Sigma$  and $\hat \cG = 0$ on the boundary $\partial \Sigma$,  forces  the presence of non-integrable  singular points in $\partial_w \cA_{\pm}$ on the boundary $\partial \Sigma$.    The relation $\partial_w\partial_{\bar w} \hat \cG = - \hat \kappa ^2 $, deduced from the first two lines in (\ref{4.a.2}),  can be solved to determine $\hat \cG $ in terms of $\hat \kappa^2$ using a Green function,\footnote{For notational ease we suppress the dependence on the complex conjugate argument in the following.}
\begin{align}
\label{greensfun}
\hat \cG(z)  = {1\over \pi}  \int d^2 w\;  G(z,w) \hat \kappa^2(w)
\end{align}
where the Green function satisfies,
\begin{align}
\partial_z \partial_{\bar z} G(z,w) = -\pi \delta^2(z-w) 
\hskip 1in 
G(z,w) \Big |_{z\in \partial \Sigma} =0
\end{align}
and the rate of vanishing on the segments of the boundary is determined by (\ref{eq:G-y-exp}).  By a general argument (see e.g. section 2.3 in \cite{DHoker:2017mds}) the Green function $G(z,w)$ is strictly positive inside $\Sigma$. Hence,  if the  integral (\ref{greensfun}) exists, then it follows from the positivity of $G$ and $\hat \kappa^2$ that  $\hat \cG$ must be positive in the interior of $\Sigma$. This  violates the regularity conditions, and makes the existence of regular solutions seemingly impossible.  

\sm

As was already pointed out, the above argument assumes that the integral in (\ref{greensfun}) exists. This loophole in the argument was  pointed out in \cite{Corbino:2018fwb}   for the   $AdS_{2}\times S_{6}  \times \Sigma_2$ Type IIB solutions.   If $\hat \kappa^2$ has a singularity at the boundary which is not integrable with respect to the Green function then the integral representation (\ref{greensfun}) does not exist and the argument presented above fails.  It is straightforward to check from the expressions in section \ref{sec:sing-points} for $\hat \kappa^2$ and $\hat \cG$ that for $p=-1/2$ the singular point is integrable whereas for $p=-5/2, p=-2$ and $p=-3$ points the singular points are indeed non-integrable. In sections \ref{sec:L1-sol} and \ref{sec:sol} we give explicit examples of regular solutions with such singularities.

\sm

We shall now show that, under certain conditions, the regularity condition for $\hat \cG$, namely $\hat \cG<0$ in the interior of $\Sigma$ and $\hat \cG=0$ on $\partial\Sigma$, is implied, without relying on a construction in terms of a Green function.

\sm

We start with a compact $\Sigma$ and will discuss the modifications brought about by singularities and non-compact regions, as they typically appear in physically relevant solutions, afterwards. We start with the following characterization of critical points of $\hat \cG$:

\medskip

\textbf{Lemma:} If $\hat \kappa^2=-\partial_w\partial_{\bar w} \hat \cG$ is positive in the interior of $\Sigma$, any critical points of $\hat \cG$ in the interior of $\Sigma$ are necessarily maxima.

\medskip

Proof: The proof proceeds in two steps. The first is a characterization of the critical points. Assume we have a critical point, $\partial \hat \cG=\bar\partial \hat \cG=0$. This means
\begin{align}
	\partial_w \hat \cG=\left(\overline{\cA}_+-\cA_-\right)\partial_w\cA_+
	+\left(\cA_+-\overline{\cA}_-\right)\partial_w\cA_-&=0
\end{align}
If we had $\overline{\cA}_+-\cA_-\neq 0$, we would conclude
\begin{align}
	\partial\cA_+ + e^{i\varphi} \partial\cA_-&=0
	&
	e^{i\varphi} &= \frac{\cA_+-\overline{\cA}_-}{\overline{\cA}_+-\cA_-}
\end{align}
where $e^{i\varphi}$ is a phase. This would, however, imply that $\hat \kappa^2=-|\partial\cA_+|^2+|\partial\cA_-|^2$ would vanish. By the assumptions, this cannot happen in the interior of $\Sigma$.
We conclude that critical points in the interior of $\Sigma$ have 
\begin{align}\label{eq:crit}
	\overline{\cA}_+-\cA_-&=0
\end{align}
The second step is to use this characterization. From the definition of $\cG$, we have
\begin{align}
	\partial_w^2 \hat \cG&=
	\left(\overline{\cA}_+-\cA_-\right)\partial_w^2\cA_+
	+\left(\cA_+-\overline{\cA}_-\right)\partial_w^2\cA_-
\end{align}
At interior critical points, where (\ref{eq:crit}) holds, we therefore also have 
\begin{align}
	\partial_w^2 \hat \cG&=\partial_{\bar w}^2 \hat \cG=0
\end{align}
The complex Hessian thus becomes
\begin{align}
	\nabla^2 \hat \cG&=\begin{pmatrix} \partial_w^2 \hat \cG & \partial_w\partial_{\bar w} \hat \cG
	\\ 
	\partial_w\partial_{\bar w} \hat \cG &\partial_{\bar w}^2 \hat \cG
	\end{pmatrix}=-\begin{pmatrix} 0 & \hat \kappa^2\\  \hat \kappa^2& 0\end{pmatrix}
\end{align}
For $\hat \kappa^2>0$ the quadratic term in the Taylor expansion of $\hat \cG$ is negative and the critical point thus is a maximum.\hfill $\square$

\sm

With the help of this result, we will show that $\hat \cG$ is negative in the interior of $\Sigma$ provided that the behavior of $\hat \cG$ near the boundary of $\Sigma$ satisfies certain assumptions, whose validity in physically relevant solutions will be investigated afterwards:

\sm

\textbf{Corollary}: If $\hat \kappa^2$ satisfies the regularity conditions, and in addition every boundary point on $\partial\Sigma$ has an inward neighborhood in which $\hat \cG<0$, then $\hat \cG<0$ in the interior of $\Sigma$.

\sm

Proof by contradiction: Assume that $\hat \cG$ is positive somewhere in the interior of $\Sigma$ at a point $p\in\text{int}(\Sigma)$. By the assumptions, this point is separated from the boundary by a region in which $\hat \cG<0$. The mountain pass  theorem \cite{jabri2003mountain} then guarantees the existence of a saddle point: Any continuous path connecting $p$, where $\hat \cG>0$, to a boundary point $q$, where $\hat \cG=0$, has to go through a region where $\hat \cG<0$. Therefore, along any fixed path connecting $p$ and $q$ there is a minimum at which $\hat \cG<0$. Varying the path connecting $p$ and $q$, so as to maximize this minimum, leads to a saddle point. By the previous Lemma, however, all critical points are maxima. This is the desired contradiction. \hfill $\square$

\sm

In order to apply these results to the solutions of interest, we have to incorporate singularities on the boundary of $\Sigma$. These do not change the conclusion unless they invalidate the mountain ridge argument, e.g.\ by providing a run-off direction for the maximization procedure that produces the saddle point. For the singularities of interest, we have $\hat \cG \rightarrow -\infty$, e.g.\ at  $p=-5/2$ singularities, or $\hat \cG$ bounded, e.g.\ for $p=-1/2$. With this behavior no run-off directions arise and the argument remains intact.

\sm

The remaining task is to discuss the assumption regarding a neighborhood with negative $\cG$ for each boundary point. Along boundary segments where $S^6$ collapses, this is immediate: With  $w=x+iy$ and the boundary at $y=0$, we have the near-boundary behavior $\hat \cG = \frac{1}{6}\partial_y^3 \hat \cG\big\vert_{y=0}y^3 + \cO(y^4)$. At $y=0$ we further have $\partial_y^3 \hat \cG = -4\partial_y \hat \kappa^2$. Since $\hat \kappa^2$ is positive in the interior of $\Sigma$, we have $\partial_y \hat \kappa^2\vert_{y=0}>0$. The leading term in the expansion of $\hat \cG$ near the boundary is thus negative and there is a neighborhood in $\text{int}(\Sigma)$ with $\hat \cG<0$.

\sm

For boundary segments where $S^2$ collapses, $\hat \cG$ vanishes linearly towards the boundary, and we directly investigate
\begin{align}
	\partial_w \hat \cG&=(\bar \cA_+-\cA_-)\partial_w\cA_+ + (\cA_+-\bar\cA_-)\partial_w\cA_-
\end{align}
With the condition for the differentials in (\ref{eq:alpha1-rel}), $\partial_w \hat \cG$ is imaginary along the boundary segment and, moreover, $\bar\cA_+-\cA_-$ is constant. The requirement $\partial_y \hat \cG\vert_{y=0}<0$ thus reduces to constraints on the constant $\bar\cA_+-\cA_-$ and on the differentials.

\sm

In summary, the global requirement $\hat \cG\vert_{\text{int}(\Sigma)}<0$ reduces to conditions on the boundary behavior of the holomorphic functions along segments where $S^2$ collapses.

\subsection{Counting parameters}

In this section we will count the number of independent parameters of the solutions with general $L$.  In  section \ref{sec:differentials} we argued that the differentials $\partial_w \cA_{\pm}$ are characterized by  the numbers $L$ and $K$ of $p=-1/2$ and $p=-5/2$ branch points, respectively, the  numbers $P$ and $Q$ of  $p=-2,p=-3$ poles on the boundary, and the number $N$ of the bulk zeros. Solutions that are regular at $\infty$  have even values of $L+K$.  

\sm

The locations of the branch points and poles on the boundary  as well as of the bulk zeros account for a total of $2N+L+K+P+Q$ real parameters, of which three can be fixed using the action of $SL(2,R)$ on the upper half plane.  In addition, we have two  complex  integration constants $\cA_\pm^0$, two complex factors $a_{\pm}$ and one real integration constant for~$\cB$. Putting this together, the number of free parameters of the Ansatz is given by,
\begin{align}
	\label{N-Ansatz}
	2N+L+K+P+Q+6   
\end{align}
The regularity conditions impose the following constraints on the free parameters of the Ansatz: 
\begin{itemize}
	\item The matching condition (\ref{eq:bndy-cond-cApm})  on the collapse of $S^2$ relates the complex parameters $a_+$ and $a_-$ to one another, thereby fixing 2 real parameters.
	\item The matching condition (\ref{eq:lift})  on the collapse of  $S^6\to 0$  fixes $L+K$  real parameters, since there are $(L+K)/2$ segments where $S^6$ must vanish.
	\item The  condition (\ref{eq:reg-res-dAp}) on the absence of logarithmic branch cuts for $\cA_{\pm}$  at the $p=-3$ poles imposes $2Q$ real conditions.
	\item The monodromy conditions for $\cB$ at $p=-5/2$ branch points and $p=-2,-3$ poles (\ref{eq:reg-dB}) impose $K+P +Q$ real conditions. 
	\item The  gauge transformations on $B_{(2)}$ and $C_{(2)}$  given in (\ref{gaugetrans})  which are compatible with the real form fix two real parameters.
\end{itemize}
Subtracting all these conditions from the expression given in (\ref{N-Ansatz}) for the number of free parameters of the Ansatz gives the number of independent parameters $\cN$ of the solutions,
\begin{align}\label{eq:numind}
	{\cal N} &= 2N-K-2Q+2
\end{align}
Using the condition of regularity of the differentials at infinity, stated in  (\ref{4.d.4}), we may eliminate $N$ in terms of the other parameters to obtain the final count of the number of free parameters of the solutions,
\begin{align}\label{eq:calN}
	{\cal N} & = L+4K + 4P +4 Q -2
\end{align}
We note that a global regularity analysis for the case of $p=-2$ and $p=-3$ poles on the boundary has not been performed and the result above only imposes necessary local constraints near the poles. In  section  \ref{ref:sec.5.2} we give explicit examples of regular solutions with such poles, but  leave a complete analysis of these singularities for future work.

\clearpage

\section{Explicit solutions for $L=1$ and $L=2$}
\label{sec:L1-sol}

In this section we produce explicit solutions in the simplest cases where $L=1$ and $L=2$. For $L=1$, we recover the ``cavity" solution, which served as background for the probe D-string in \cite{Hartnoll:2024csr}. For $L=2$, we produce new solutions. Our analysis for $L=2$ will not be exhaustive, but it will show that regular solutions with $p=-2$ and $p=-3$ poles exist.

\subsection{Minimal $L=1$ solutions}
\label{sec:L1-min}

We start from the differentials constructed in section \ref{sec:differentials} with $L=1$. Since $K+L$ must be even, the simplest case corresponds to having $K=1$ and $P=Q=0$, namely one $p=-{5 \over 2}$ singularity on $\p \Sigma$ and  no poles. In view of (\ref{4.d.4}) we must then have $N=1$.  We can use the $SL(2,\RR)$ transformations on the upper half plane to set $b_1=0$ and $c_1=\infty$ and obtain the following differentials from (\ref{4.d.1}) and (\ref{4.d.2}), 
\begin{align}
	\partial_w\cA_+&=a_+\frac{w-s_1}{\sqrt{w}}
	&
	\partial_w\cA_-&=a_-\frac{w-s_1^\star}{\sqrt{w}}
\end{align}
Upon integration, we obtain, 
\begin{align}\label{Apm-L1}
	\cA_+&=\cA_+^0+\frac{2}{3}a_+\sqrt{w}(w-3s_1)
	&
	\cA_-&=\cA_-^0+\frac{2}{3}a_-\sqrt{w}(w-3s_1^\star)
\end{align}
Taking $(a_\pm)^\star=a_\mp$, the sphere $S^2$ collapses for $w<0$, while $S^6$ collapses for $w>0$. The latter requires that the integration constants  be related by $(\cA_+^0)^\star=\cA_-^0$.
The metric factors of the solution are then given by, 
\begin{align}
	\rho^2&=\frac{C}{4|w|}
&
	f_6^2&=-C\left(\sqrt{w}-\sqrt{w^\star}\right)^2
&	f_2^2&=
	C \left(\sqrt{w}+\sqrt{w^\star}\right)^2
\end{align}
where the constant $C$ is given by
\begin{align}
	C&=-4c_6|a_+|\sqrt{\frac{s_1-s_1^\star}{6i}}
\end{align}
The axion and dilaton are
\begin{align}\label{eq:dilaton-L1}
	e^\phi&=\frac{3(a_++a_+^\star)(a_+s_1+a_+^\star s_1^\star)+(a_++a_+^\star)^2 \left(w+ w^\star+4|w|\right)}{3|a_+|^2i(s_1-s_1^\star)}
	\nonumber\\
	\chi&=-\frac{a_+-a_+^\star}{a_++a_+^\star}+ie^{-\phi}
\end{align}
and the 2-forms determined from (\ref{eq:2-forms}) are,
\begin{align}
	B_{(2)}&=-\frac{2\sqrt{2}c_6}{9}\nu_1\left(4 (a_++a_+^\star) \left(\sqrt{w}+\sqrt{\bar w}\right)^3+3(\cA_+^0+(\cA_+^0)^\star)\right) \, \hat e^{67}
	\nonumber\\
	C_{(2)}&=+\frac{2\sqrt{2}c_6}{9}\nu_1\left(4 (a_+-a_+^\star) \left(\sqrt{w}+\sqrt{\bar w}\right)^3+3(\cA_+^0-(\cA_+^0)^\star)\right) \, \hat e^{67}
\end{align}
As expected in Type IIB$^\star$ supergravity, $B_{(2)}$ is real while the axion and  $C_{(2)}$ fields are imaginary. There is one 9-cycle, given by $S^6$ and $S^2$ fibered over a curve in $\Sigma$ connecting the boundary segment with $S^6$ collapsing to the segment with $S^2$ collapsing. There is no instanton charge, since the segments are connected by a $p=-1/2$ transition point.

\sm

The parameters $\cA_+^0$, $a_+$ and $s_1$ are $3$ complex degrees of freedom. We have used an $SL(2,\RR)$ transformation of the upper half plane to set $b_1=0$ and fix the $p=-\frac{5}{2}$ singularity at infinity. This leaves $1$ real redundancy from $SL(2,\RR)$ transformations of the upper half plane. So altogether we have 5 real parameters. Two parameters, encoded in $\cA_+^0$, correspond to gauge transformations. The number of remaining parameters, three, matches the number of parameters in Type IIB $SL(2,\RR)$ duality transformations.

\sm

The axion and dilaton deserve discussion. From the expression in (\ref{eq:dilaton-L1}), $e^\phi$ turns negative for large $|w|$ if $a_++a_+^\star\neq 0$, while the expression for the axion diverges everywhere if $a_++a_+^\star=0$. This could be taken to suggest that the solution is inevitably singular. However, following the discussion in section \ref{sec:metric-reg} we instead conclude that care is needed in parametrizing the scalar manifold $SL(2,\RR)/SO(1,1)$. The matrix $\mathcal M_\star$ used to fully parametrize the coset, with $\chi_\star=-i\chi$, and the parametrization in (\ref{eq:Vstar-1}), takes the form 
\begin{equation}\label{eq:Ms-L1}
	\mathcal M_\star=\frac{i}{3|a_+|^2(s_1-s_1^\star)}
	\begin{pmatrix} 
		-(a_++a_+^\star) (a_+ X+a_+^\star X^\star) & -i \left(a_+^2X-{a_+^\star}^2X^\star\right) \\
		-i \left(a_+^2X-{a_+^\star}^2X^\star\right) &
		(a_+-a_+^\star) (a_+X-a_+^\star X^\star)
		\end{pmatrix}
\end{equation}
where 
\begin{equation}
	X=3s_1+w+w^\star+4|w|
\end{equation}
This $\cM_\star$ is a symmetric matrix with $\det \cM_\star=-1$ and finite entries for all finite values of $a_+$ and $w$. In particular, it is finite for $a_+^\star=-a_+$. For $\cM_{\star\,1,1}>0$, $\cM_\star$ can be translated to dilaton and axion fields in a local field coordinate patch using the parametrization of $SL(2,\RR)/SO(1,1)$ in (\ref{eq:Vstar-1}), while for $\cM_{\star\,1,1}<0$ it can be translated to dilaton and axion fields using the parametrization in (\ref{eq:Vstar-2}), with a third patch needed to cover the transition region. This leads to finite and real axion and dilaton fields everywhere.

\subsection{Recovering the cavity solution of Hartnoll-Liu  in \cite{Hartnoll:2024csr}}
\label{sec:cavity}

The cavity background in equations (44-46) of \cite{Hartnoll:2024csr} was shown to preserve 16 supersymmetries and is expected to arise as a special case of the general complex $F(4)$-invariant supergravity solutions. Below, we will review the expressions for the various fields of this solution and then show how to recover it from the $L=1$ solution discussed above.

\sm

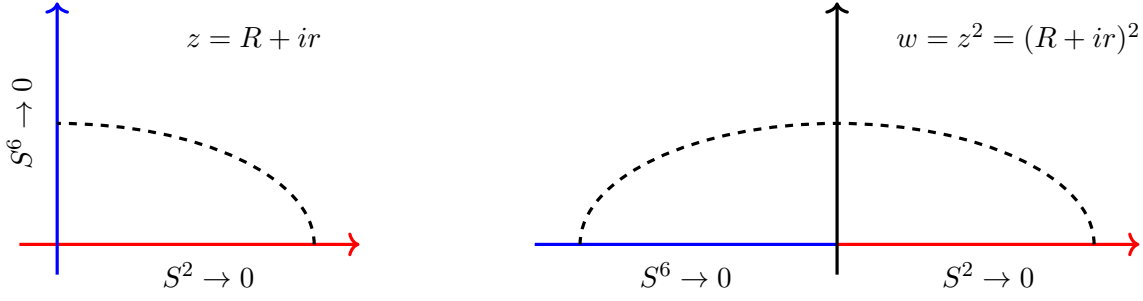
\begin{figure}
\begin{center}
	{\label{fig:cavity-quadrant}
		\begin{tikzpicture}[xscale=2, yscale=1.6]
			\draw [very thick, red,->] (-0.25,0) -- (2,0);
			\draw [very thick, blue,->] (0,-0.25) -- (0,2);
			\node at (1,-0.25) {\small $S^2\rightarrow 0$};
			\node [rotate=90] at (-0.25,1) {\small $S^6\rightarrow 0$};		
			\node at (1.3,1.7) {\small $z=R+ir$};
			\draw[very thick, dashed] (1.7,0) arc (0:90:1.7cm and 1cm);
		\end{tikzpicture}
	}
	\hskip 20mm
	\begin{tikzpicture}[xscale=2, yscale=1.6]
		\draw [very thick, blue] (-2,0) -- (0,0);
		\draw [very thick, red,->] (0,0) -- (2,0);
		\draw [very thick, ->] (0,-0.25) -- (0,2);
		\node at (1,-0.25) {\small $S^2\rightarrow 0$};
		\node at (-1,-0.25) {\small $S^6\rightarrow 0$};
		\draw[very thick, dashed] (1.7,0) arc (0:180:1.7cm and 1cm);
		\node at (1.2,1.7) {\small $w=z^2=(R+ir)^2$};
	\end{tikzpicture}
	\caption{Left: the cavity solution as part of the upper right quadrant. Right: The same solution mapped to the upper half plane via $w=z^2$.}\label{fig:cavity}
\end{center}
\end{figure}

The cavity solution of \cite{Hartnoll:2024csr} in Euclidean Type IIB has a flat Euclidean metric. We shall decompose this metric as the sum of the flat Euclidean metrics on $\RR^7$ and $\RR^3$, parametrized  in terms of radial coordinates $R$ and $r$ and unit-radius spheres $S^6$ and $S^2$, respectively. 
The metric may then be cast in the following form, 
\bea
\label{5.b.1}
ds^2 =R^2 ds^2_{S^6} + dR^2 +  r^2 ds^2 _{S^2} + dr^2 
\eea
where $ds^2_{S^6} $ and $ds^2 _{S^2}$ are the round metrics for spheres of unit radius and $R,r \geq 0$. 
At $R=0$ the $S^6$ collapses while at $r=0$ it is the $S^2$ that collapses (see figure \ref{fig:cavity}).
The dilaton, axion and NS 3-form field strength are given by,
\begin{align}
\label{5.b.2}
e^\phi&=-\frac{1}{\chi }=1-\frac{\mu^2}{32}\left(R^2 + 3r^2 \right)
&
H_{(3)} &=\mu \, r^2 dr \wedge \hat e^{67}
\end{align}
where $\mu$ is an arbitrary real constant.  The exponentiated dilaton $e^\phi$ of the solution is positive for $R^2+3r^2< 32/\mu^{2}$,  negative for $R^2+3r^2 > 32/\mu^{2}$ and vanishes at $R^2+3r^2 = 32/\mu^{2}$. This motivated restricting the solution in \cite{Hartnoll:2024csr} to the domain,
\bea
\label{5.b.3}
R^2+3r^2< \frac{32}{\mu^{2}}
\eea
and hence the term cavity. We will now show how to recover this solution from the general $L=1$ solutions discussed in the previous section, and clarify the interpretation of the axion and dilaton in accordance with the comments there.

\sm

The above cavity solution is a special case of the $L=1$ solutions. This can be made explicit by starting with $\cA_{\pm}$ in (\ref{Apm-L1}) and specializing them to
\begin{align}\label{eq:A12-cavity-UHP}
	\cA_1&=\frac{i}{c_6}\sqrt{w}\left(-\frac{g}{\mu}+\frac{\mu}{64}w\right) & \cA_2&=\frac{2}{c_6\mu}\sqrt{w}
\end{align}
with $\cA_\pm=(\cA_1\mp \cA_2)/\sqrt{2}$ as defined in (\ref{4.a.1}). This realizes the cavity solution on the upper half plane. Upon transforming the upper half plane to the upper right quadrant, by setting $w=z^2$, and further parametrizing $z$ as $z=R+ir$, the expressions in section~\ref{sec:L1-min} reproduce those quoted in equations (\ref{5.b.1}) and (\ref{5.b.2}).

\sm

With the cavity solution realized as a special case of the $L=1$ solutions in section~\ref{sec:L1-min}, the discussion around equation (\ref{eq:Ms-L1}) also carries over immediately. For the cavity solution with the relation in (\ref{5.b.2}), the matrix $\cM_\star$ parametrizing $SL(2,\RR)/SO(1,1)$ via (\ref{eq:Vstar-1}) takes the form
\begin{equation}
	\cM_\star=\begin{pmatrix} 1-\frac{\mu^2}{32}\left(R^2 + 3r^2 \right) & -1 \\ -1 & 0\end{pmatrix}
\end{equation}
This is symmetric with $\det \cM_\star=-1$ and finite entries for all finite $r$ and $R$. Following the discussion below (\ref{eq:Ms-L1}), regular axion and dilaton fields can be obtained by using multiple coordinate patches on $SL(2,\RR)/SO(1,1)$. Alternatively, the active $SL(2,\RR)$ transformation 
$\cM_\star\rightarrow \cM_\star^\prime= M^T\cM_\star M$ with
\begin{align}
M&=\begin{pmatrix} 1&0\\ 1& 1\end{pmatrix}
&
	\cM_\star^\prime=\begin{pmatrix} -1-\frac{\mu^2}{32}\left(R^2 + 3r^2 \right) & -1 \\ -1 & 0\end{pmatrix}
\end{align}
produces $\cM_\star^\prime$ which can be captured by the parametrization (\ref{eq:Vstar-2}) for all values of $r$ and $R$, i.e.\ using a single coordinate patch on $SL(2,\RR)/SO(1,1)$.
At the level of Type IIB supergravity, either description provides a continuation of the solution beyond the cavity domain, and we leave a full string theory analysis for the future.

\subsection{Minimal $L=2$ solutions}
\label{ref:sec.5.2}

The next case we consider is $L=2$. We will not provide an exhaustive discussion, but show that there are regular solutions. The minimal options that lead to non-trivial solutions have one pole, which can be either on a segment where $S^2$ collapses for a $p=-2$ pole or on a segment where $S^6$ collapses for a $p=-3$ pole. We will discuss these two cases in turn.

\subsubsection{$L=2$ solutions with $(p,q)$ string charge}

The differentials can be taken from the constructions in section~\ref{sec:differentials} for $L=2$, $K=P=0$ and $Q=1$, so that $N=2$ in view of (\ref{4.d.4}). Using $SL(2,\RR)$ transformations of the upper half plane we can set $b_1=1$, $b_2=-1$ and $e_1=\infty$ and  find the following differentials
\begin{align}
	\partial_w\cA_+&=a_+\frac{(w-s_1)(w-s_2)}{\sqrt{(w-1)(w+1)}}
	&
	\partial_w\cA_-&=a_-\frac{(w-s_1^\star)(w-s_2^\star)}{\sqrt{(w-1)(w+1)}}
\end{align}
For large $w$ the differentials $\partial_w\cA_\pm$ are linear in $w$, corresponding to a pole of degree 3, which properly accounts for $Q=1$. We choose $a_-=a_+^\star$ so that $S^2$ collapses for $w\in (-1,1)$ and $S^6$ collapses for $w \in \RR \smallsetminus (-1,1)$ and the pole at $w=\infty$ resides on a segment where $S^6$ collapses. The holomorphic functions $\cA_\pm$ and $\cB$ can all be expressed in terms of elementary functions, though we do not give the expressions here.

\sm

With only one segment on which $S^6$ collapses, there are no regularity constraints from the matching conditions of section \ref{sec:4.5}. The remaining regularity conditions include (\ref{eq:reg-res-dAp}) with $e_1=\infty$. Satisfying it requires
\begin{align}
	2 s_1 s_2&=-1
\end{align}
If $s_1$ is in the upper half plane, so is $s_2$.
The condition (\ref{eq:reg-dB}) at $e_1=\infty$ is then satisfied.

\sm

\begin{figure}
	\centering
	\subfigure[][]{\label{fig:L2string-metric}
		\includegraphics[width=0.32\linewidth]{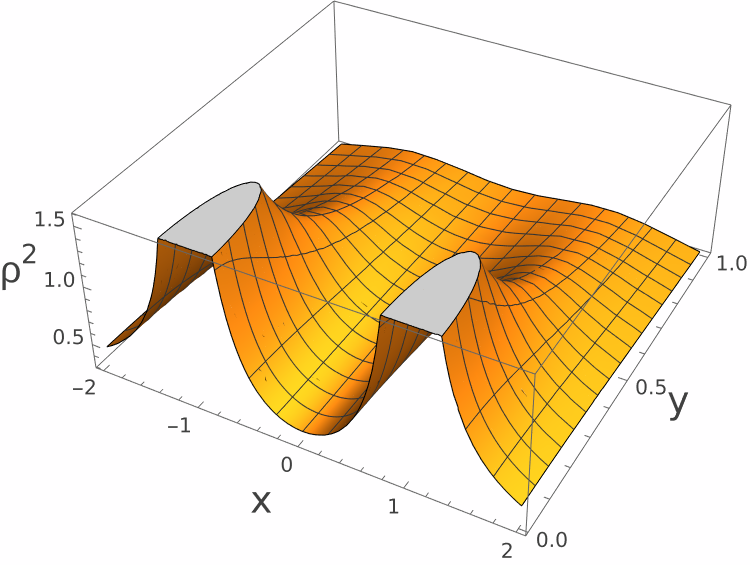}
		\includegraphics[width=0.32\linewidth]{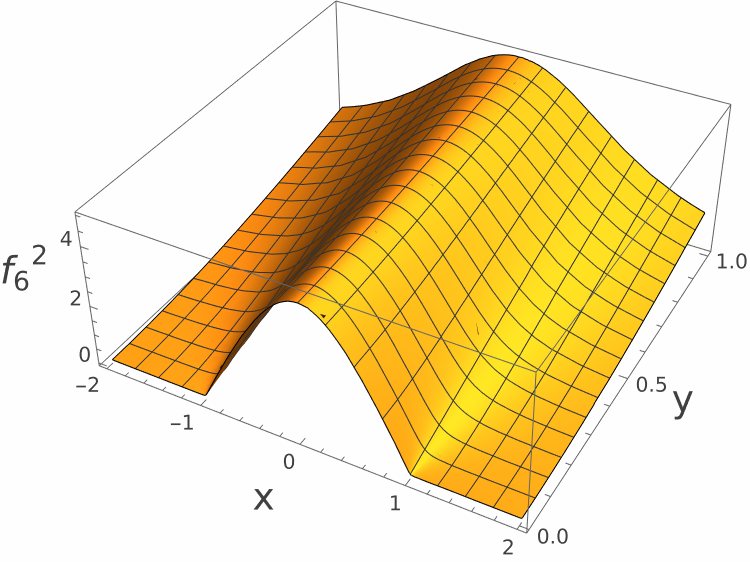}
		\includegraphics[width=0.32\linewidth]{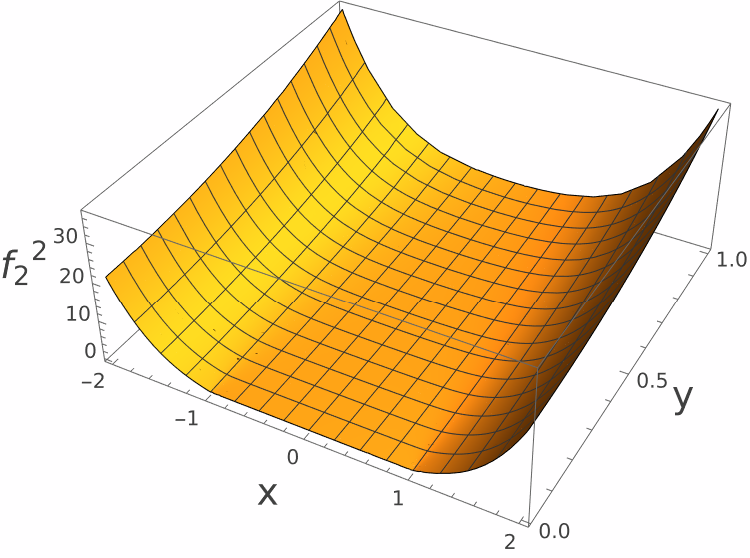}
	}
	\subfigure[][]{\label{fig:L2string-phi-chi}
		\includegraphics[width=0.32\linewidth]{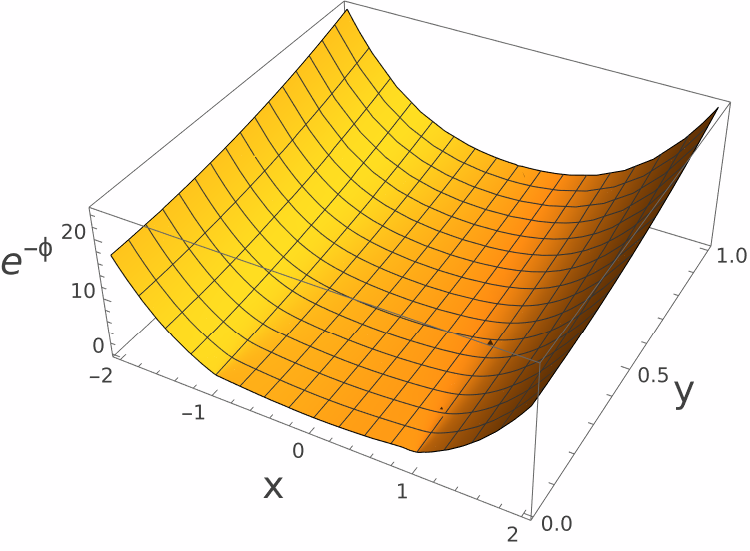}
		\qquad
		\includegraphics[width=0.32\linewidth]{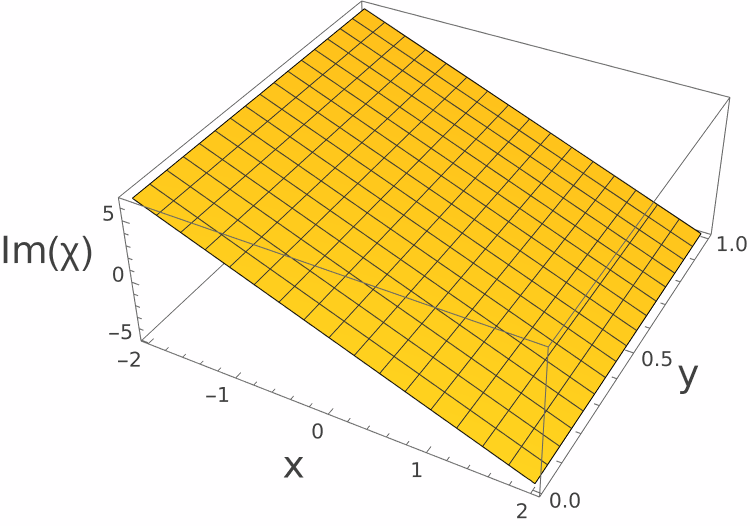}
	}
	\subfigure[][]{\label{fig:L2string-B6C6}
		\includegraphics[width=0.32\linewidth]{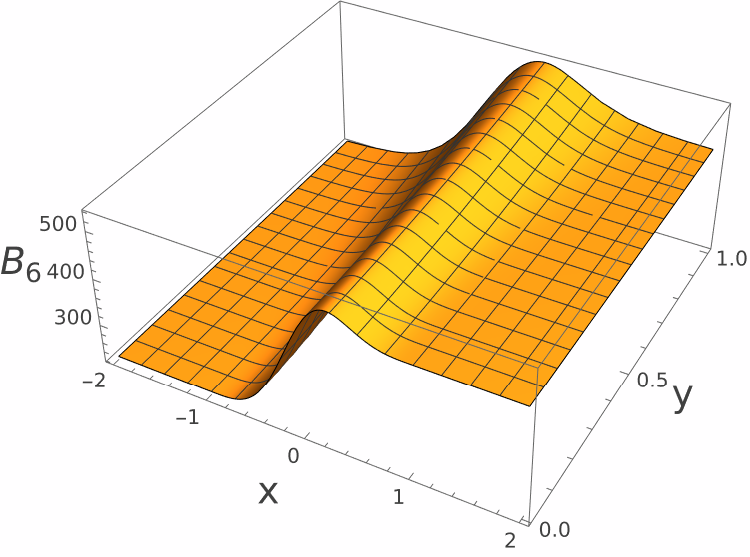}
		\qquad
		\includegraphics[width=0.32\linewidth]{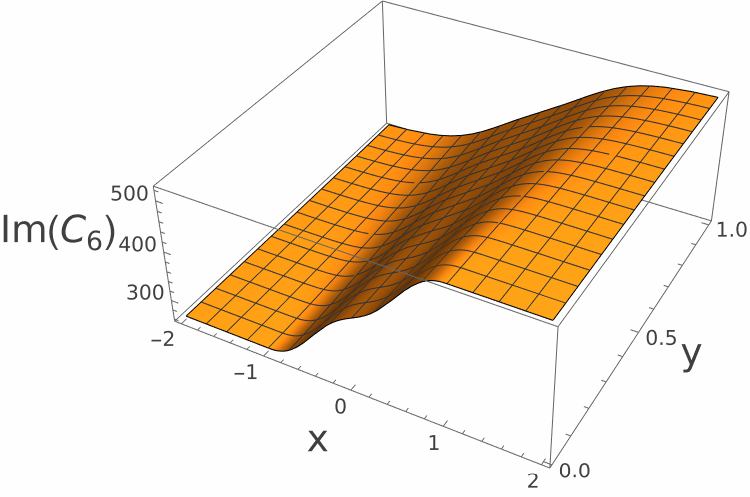}
	}
	\caption{Example $L=2$ solution with $(p,q)$ string charge, with parameters given in (\ref{eq:L2string-params}). The metric functions are in the top row, dilaton and axion in the second row, and the six-form potentials in the bottom row.}
\end{figure}
The remaining free parameters are  the location of one complex zero, which we take as $s_1$, the overall coefficient $a_+=a_-^\star$, and the integration constants $\cA_\pm^0$ which only produce gauge transformations. To show that fully regular solutions can be obtained, we show plots for a generic parameter choice, namely,
\begin{align}\label{eq:L2string-params}
a_+&=i & s_1&=1+\frac{i}{2} & \cA_\pm^0&=0
\end{align}
and we further choose $c_6=-1$. The non-trivial regularity requirements arise from the metric functions, which are shown in figure \ref{fig:L2string-metric}. They are non-negative throughout $\Sigma$, with the desired behavior at the branch points $w=\pm1$. The exponentiated dilaton and axion are shown in figure \ref{fig:L2string-phi-chi}. The figure shows that $e^{-\phi}$ is non-negative while the axion shows simple linear behavior as function of $x$. We found that choosing $a_+$ away from the imaginary axis spoils positivity of $e^{-\phi}$ while rendering the axion finite at the pole, similar to the behavior near $p=-5/2$ singularities discussed in section \ref{sec:p-5-half}. This would call for a regularity discussion similar to that for the $L=1$ solutions in section \ref{sec:L1-min}. The flux potentials do not impose regularity conditions; we find the RR potentials imaginary and the NSNS potentials real, as expected. But we briefly discuss the charges. With only one segment where $S^2$ collapses there are no 5-brane charges, and although there are 9-cycles the $p=-1/2$ transition points do not give rise to instanton charge, as discussed in section \ref{sec:p-1-half}. The only charge carried by the solutions is the string charge associated with the $p=-3$ pole at $w=\infty$. It can be identified in the plots in figure \ref{fig:L2string-B6C6} as the difference between the values of $(B_6,C_6)$ on the boundary segments $w>1$ and $w<-1$.

\subsubsection{$L=2$ solutions with $(p,q)$ 5-brane charge}

A similar analysis can be performed for $L=2$ solutions with $(p,q)$ 5-brane charge. The differentials are again taken from the constructions in section~\ref{sec:differentials} for $L=2$, $K=Q=0$ and $P=1$, so that $N=1$ in view of (\ref{4.d.4}). Using $SL(2,\RR)$ transformations of the upper half plane we can set $b_1=1$, $b_2=-1$ and $d_1=\infty$ and  find the following differentials
\begin{align}
	\partial_w\cA_+&=a_+\frac{w-s_1}{\sqrt{(w-1)(w+1)}}
	&
	\partial_w\cA_-&=a_-\frac{w-s_1^\star}{\sqrt{(w-1)(w+1)}}
\end{align}
For large $w$ the differentials $\partial_w\cA_\pm$ are constant in $w$, corresponding to a pole of degree 2, which properly accounts for $P=1$. For these solutions we choose $a_-=-a_+^\star$, so that $S^6$ collapses for $w\in (-1,1)$ and $S^2$ collapses for $w \in \RR \smallsetminus (-1,1)$, and the pole at $w=\infty$ resides on a segment where $S^2$ collapses.

\sm

\begin{figure}
	\centering
	\subfigure[][]{\label{fig:L2-5brane-metric}
		\includegraphics[width=0.32\linewidth]{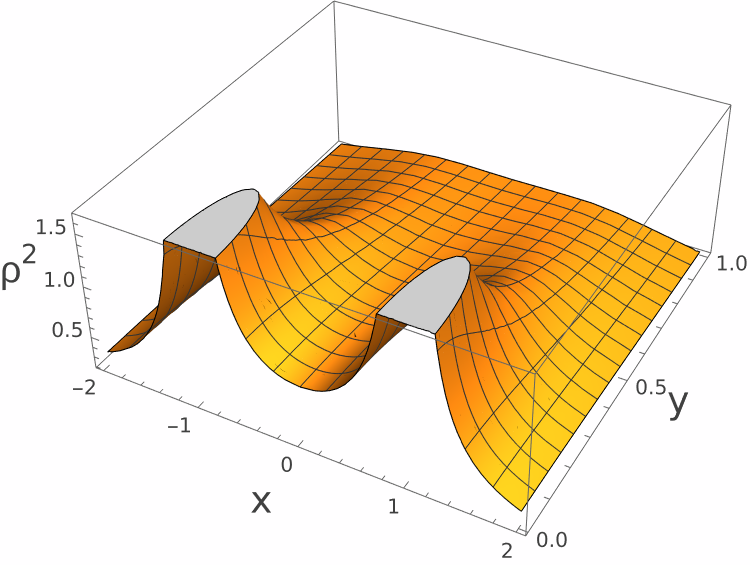}
		\includegraphics[width=0.32\linewidth]{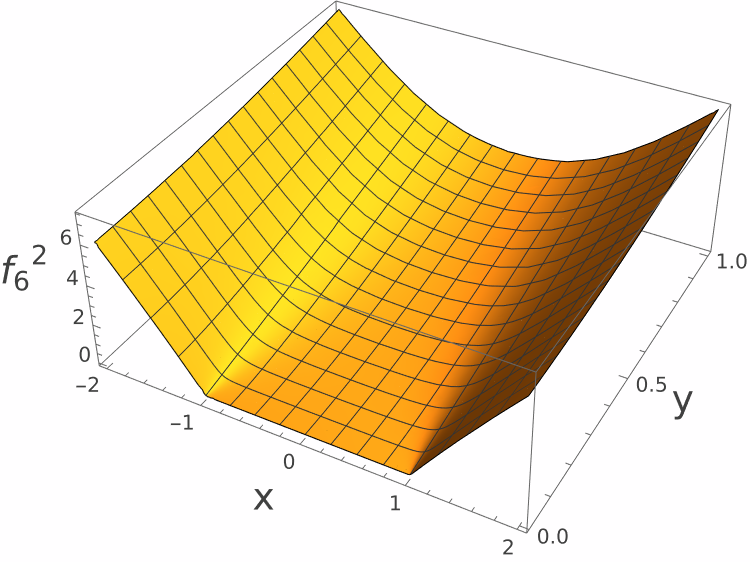}
		\includegraphics[width=0.32\linewidth]{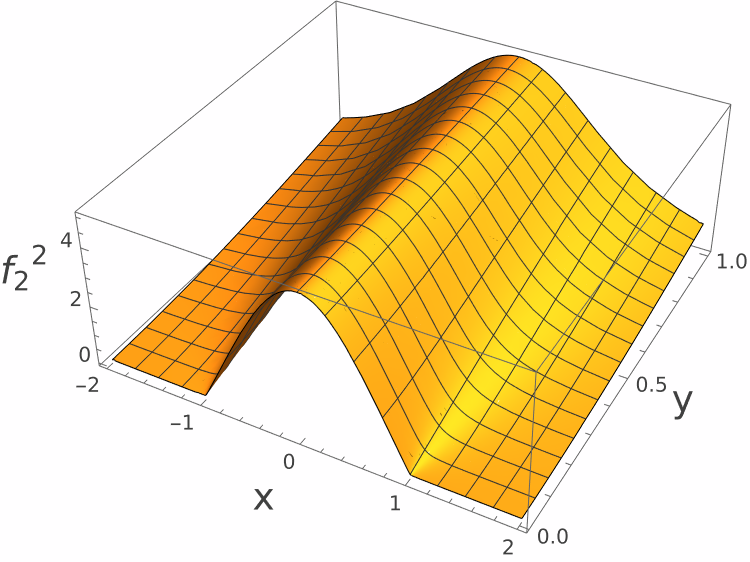}
	}
	\subfigure[][]{\label{fig:L2-5brane-phi-chi}
		\includegraphics[width=0.32\linewidth]{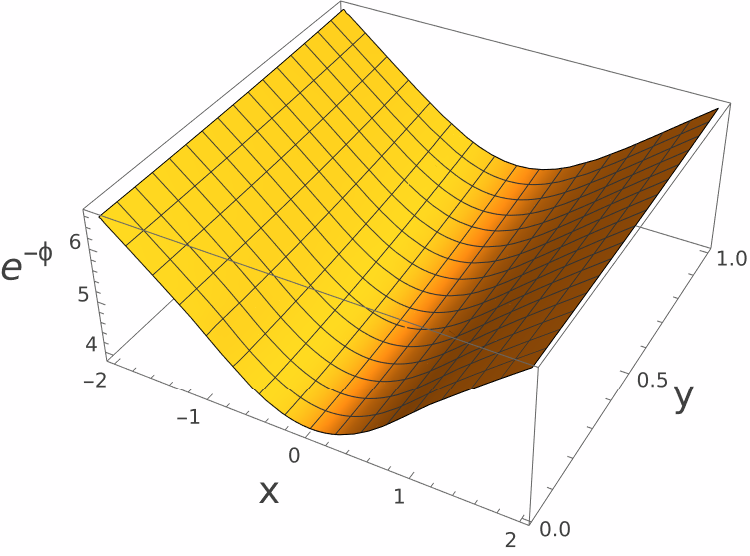}
		\qquad
		\includegraphics[width=0.32\linewidth]{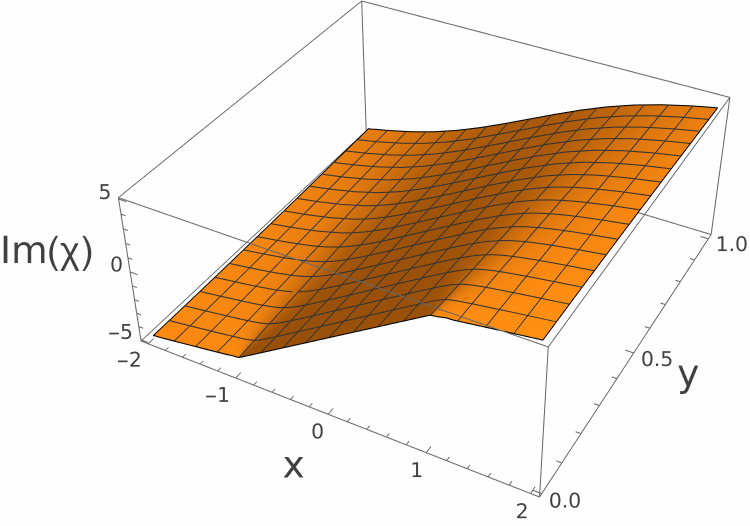}
	}
	\subfigure[][]{\label{fig:L2-5brane-B2C2}
		\includegraphics[width=0.32\linewidth]{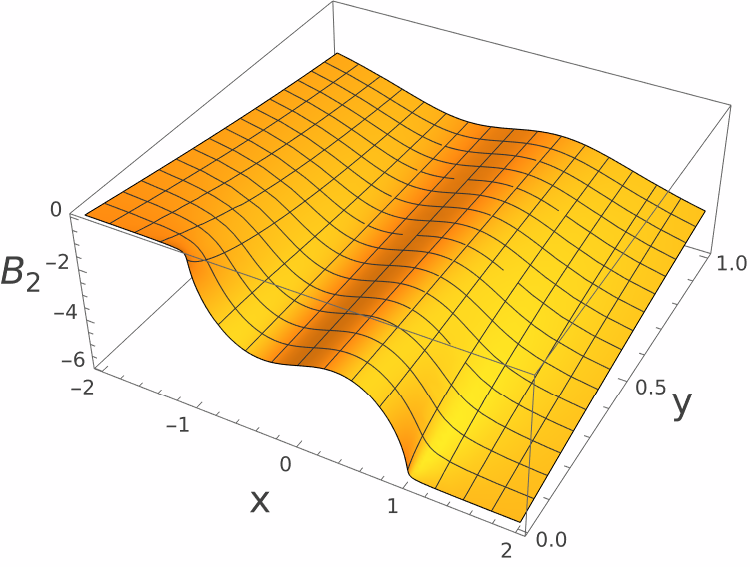}
		\qquad
		\includegraphics[width=0.32\linewidth]{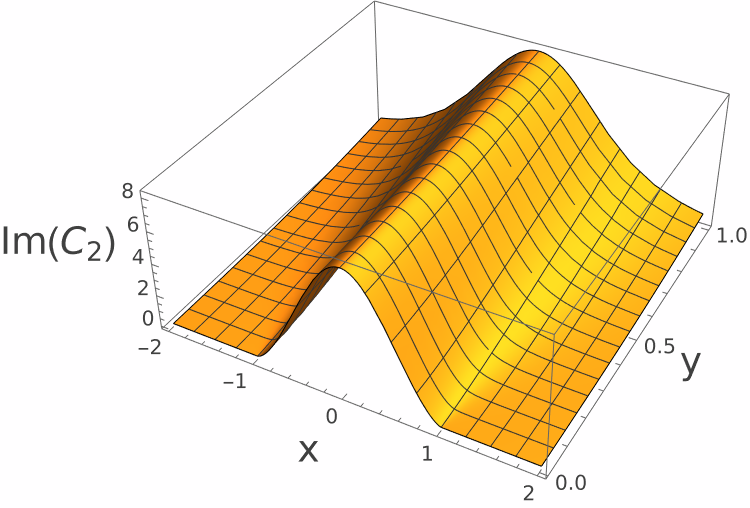}
	}
	\caption{Example $L=2$ solution with $(p,q)$ 5-brane charge, with parameters given in (\ref{eq:L2-5brane-params}). The metric functions are in the top row, dilaton and axion in the second row, and the two-form potentials in the bottom row.}
\end{figure}

Compared to the $(p,q)$ string case we have only one complex zero as compared to two, but also do not need the regularity condition in (\ref{eq:reg-res-dAp}). So we retain the same number of remaining parameters. We again content ourselves with showing that regular solutions exist. The parameters for the example solution are
\begin{align}\label{eq:L2-5brane-params}
a_+&=1 & s_1&=i & \cA_\pm^0&=0	
\end{align}
Plots of the metric functions are shown in figure \ref{fig:L2-5brane-metric} and again show that the metric functions are positive with the geometry closing off on the boundary segments of $\Sigma$ as desired. The exponentiated dilaton $e^{-\phi}$ is also non-negative throughout. The solution has no 7-cycles and carries no string charge, and also does not carry instanton charge for the same reason as in the $L=2$ string solution. The $(p,q)$ 5-brane charge is visible in the two-form potentials, shown in figure \ref{fig:L2-5brane-B2C2}.

\clearpage

\section{Solutions with D-instanton asymptotics and $L\geq 3$}
\label{sec:sol}

In this section we construct solutions with no poles, $P=Q=0$, and a single asymptotic region corresponding to a $p=-5/2$ transition point,  $K=1$. At this $p=-5/2$ transition point, the supergravity fields approach a configuration in the $SL(2,\RR)$ orbit of a deformed D-instanton solution. The boundary of $\Sigma$ is partitioned into $L+1$ segments, where $L$ is the number of $p=-1/2$ transition points, on which $S^6$ and $S^2$ collapse alternatingly. The number of zeros is $N=(L+1)/2$ in view of (\ref{4.d.4}), so that $L$ must be odd. The minimal case, $L=N=1$, was discussed in section \ref{sec:L1-min} and recovered the cavity background of \cite{Hartnoll:2024csr}. Here we focus on odd $L$ with $L\geq 3$. We produce fully explicit solutions for $L=3$ and the differentials with a count of the free parameters for $L>3$. The features of these solutions are qualitatively consistent with those of the solutions found in \cite{Komatsu:2024bop}, though we clarify the brane charges.

\subsection{The functions $\cA_\pm$ for elliptic solutions with $L=3$}

For $L=3$ the holomorphic functions $\cA_\pm$ may be expressed in terms of elliptic functions. The corresponding differentials $\p_w \cA_\pm$ are given by evaluating (\ref{4.d.1}) and (\ref{4.d.2}) in section~\ref{sec:differentials} for $K=1$, $L=3$, $N=2$ and $P=Q=0$.  
Using an $SL(2,\RR)$ transformation on the upper half $w$-plane, the branch points can be mapped to $b_1=0$, $b_2=1$ and $c_1=\infty$. This leaves the location of the third $p=-\half$ transition point, $b_3 = 1 /k^2\in \RR$, as a free parameter. Henceforth, we shall assume $0 < k <1$ without loss of generality. The resulting differentials are then given as follows, 
\begin{align}
	\label{eq:L3-dApm}
	\partial_w\cA_+&=k a_+ \frac{(w-s_1)(w-s_2)}{\sqrt{w(1-w)(1-k^2w)}}
	&
	\partial_w\cA_-&=k a_+^\star \frac{(w-s_1^\star)(w-s_2^\star)}{\sqrt{w(1-w)(1-k^2w)}}
\end{align}
where $a_+, s_1, s_2$ and $k$ remain to be related by the regularity conditions of section \ref{sec:4.5}. 

\sm

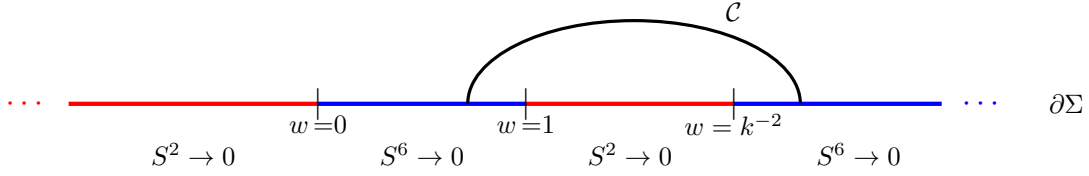
\begin{figure}[htb]
	\centering
	\begin{tikzpicture}[scale=1.1]
		\draw[ultra thick, red] (-0.5,0) -- (2.5,0);
		\draw[ultra thick, blue] (2.5,0) -- (5,0);
		\draw[ultra thick, red] (5,0) -- (7.5,0);
		\draw[ultra thick, blue] (7.5,0) -- (10,0);
		\draw (2.5,0) node{$|$};
		\draw (5,0) node{$|$};
		\draw (7.5,0) node{$|$};
		\draw [very thick] (8.3,0) arc (0:180:2cm and 1cm);
		\node at (7.5,1.1) {\footnotesize $\cC$};
		\node at (1,-0.6){\footnotesize $S^2\rightarrow 0$};
		\node at (3.75,-0.6){\footnotesize $S^6\rightarrow 0$};
		\node at (6.25,-0.6){\footnotesize $S^2\rightarrow 0$};
		\node at (9,-0.6){\footnotesize $S^6\rightarrow 0$};
		\node at (11.5,0) {\footnotesize $\partial\Sigma$};
		\node[blue] at (10.5,0) {$\cdots$};
		\node [red] at (-1,0) {$\cdots$};
		
		\node at (2.5,-0.25) {\footnotesize $w\,{=}0$};	
		\node at (5,-0.25) {\footnotesize $w\,{=}1$};
		\node at (7.5,-0.25) {\footnotesize $w=k^{-2}$};
	\end{tikzpicture}
	\caption{Boundary conditions and branch points for $L=3$. \label{fig:single-disc-bc}}
\end{figure}

As $w \to \infty$, the differentials behave as $\sqrt{w}$ which indeed corresponds to a $p=-{5 \over 2}$ branch point. Furthermore, the square roots behave as follows for $w \in \p \Sigma = \RR$, 
\begin{align}
	\frac{1}{\sqrt{w(1-w)(1-k^2w)}}\Bigg\vert_{\partial\Sigma}&=\begin{cases}\text{real} & \text{$ w>k^{-2}$ or $w\in(0,1)$}\\ \text{imaginary} & \text{$w<0$ or $w\in(1,k^{-2})$}\end{cases}
\end{align}
The differentials $\partial\cA_\pm$ realize reflection conditions making the $S^6$ collapse on the intervals $(0,1)$ and $(k^{-2},\infty)$, and reflection conditions  making the $S^2$ collapse on the intervals $(-\infty,0)$ and $(1,k^{-2})$.  This is illustrated in figure \ref{fig:single-disc-bc}.

The differentials can be integrated conveniently by decomposing them as follows,
\begin{align}
	\label{eq:dApm-L3-decomp}
	\partial_w\cA_+&=\partial_w\left(\frac{2 a_+}{3 k } \sqrt{w(1-w)(1- k^2w)} \right)
	+\frac{A(1-k^2 w) + B }{2\sqrt{w(1-w)(1-k^2 w)}}
	\nonumber\\
	\partial_w\cA_-&=\partial_w\left(\frac{2 a_+^\star }{3k } \sqrt{w(1-w)(1-k^2w)} \right)
	+\frac{A^* (1-k^2 w) + B^*}{2 \sqrt{w(1-w)(1-k^2w)} }
\end{align}
where the constants $A$ and $B$ are given by, 
\begin{align}
	\label{eq:L3-ABdef}
	A&=\frac{2a_+}{3k} \left( 3s_1+ 3s_2 - 2 - \frac{2}{k^2}  \right)
	&
	B&=\frac{2 k a_+}{3} \Big (3s_1s_2-{ 1 \over k^2} \Big )-A
\end{align}
The first term in each one of the equations for $\partial_w\cA_\pm$ may be integrated in terms of elementary functions. The integrals of the remaining terms are elliptic integrals and can be brought into the standard form for Jacobi elliptic integrals by the coordinate transformation $w=t^2$. The corresponding Abelian integrals (also referred to as incomplete elliptic integrals) of first and second kind are respectively defined by,
\bea
F(x;k) & = & \int _0^x dt \, {1 \over \sqrt{(1-t^2) ( 1 - k^2 t^2)}}
\no \\
E(x;k) & = & \int _0^x dt { 1-k^2 t^2 \over \sqrt{(1-t^2) ( 1 - k^2 t^2)}}
\eea
In terms of these functions, $\cA_\pm$ are given by, 
\begin{align}
	\label{eq:Apm-L3}
	\cA_+(w) &=\cA_+^0+\frac{2 a_+}{3k}\sqrt{w(1-w)(1-k^2w)}
	+A\, E(\sqrt{w},k) +B\,F(\sqrt{w},k)
	\nonumber\\
	\cA_-(w) &=\cA_-^0+\frac{2a_+^*}{3k} \sqrt{w(1-w)(1-k^2w)}
	+A^\star E(\sqrt{w},k)+B^\star F(\sqrt{w},k)
\end{align}
where $\cA_\pm^0$ are integration constants.

\sm

The regularity conditions in (\ref{eq:lift}), which ensure that the reflection condition on $\partial_w\cA_\pm$ can be lifted to $\cA_\pm$ for all segments with $S^6$ collapsing, give a single constraint,
\bea
\int _\cC \Big ( d w \, \p _w \cA_+ - \big ( dw \, \p_w \cA_- \big )^* \Big ) =0
\eea 
where $\cC$ is the contour of figure \ref{fig:single-disc-bc} that links the two intervals on which $S^6$ collapses. Taking the contour $\cC$, which connects the $(0,1)$ interval to the $(1/k^2,\infty)$ interval, as being infinitesimally above the real axis, the  resulting condition can be expressed as,
\begin{align}
	\int_1^{1/k^2}   \left( dw \frac{(w-s_1)(w-s_2)}{\sqrt{w(1-w)(1-k^2 w)}}-\left( dw \frac{(w-s_1^\star)(w-s_2^\star)}{\sqrt{w(1-w)(1-k^2 w)}}\right)^\star\right) =0
\end{align}
Along the segment $(1,1/k^2)+i\epsilon$ the two terms combine, and the condition becomes, 
\begin{align}
	\label{eq:L3-constr}
	\int_1^{1/k^2} dw \frac{(w -s_1)(w -s_2)}{\sqrt{w(w-1)(k^2 w-1)}} =0
\end{align}
With this constraint, the integration constants $\cA_\pm^0$ can be chosen so that the functions $\cA_\pm$ satisfy the desired boundary conditions everywhere.
In order for the boundary condition that the $S^6$ collapses, $(\cA_\pm)^\star=+\cA_\mp$, to be satisfied for $w\in(0,1)$, the integration constants have to satisfy, 
\begin{align}
	\label{eq:L3-Apm0-rel}
	(\cA_\pm^0)^\star=+\cA_\mp^0
\end{align}
The relation (\ref{eq:L3-constr}) ensures that the boundary conditions are satisfied on the second segment with $S^6\rightarrow 0$ as well. With the help of (\ref{eq:L3-Apm0-rel}) the conditions that $\cA_+- \cA_-^*$ take the same value on all segments where $S^6$ collapses then require $\cA_\pm(1/k^2)=\cA_\pm(1)$ or in terms of complete elliptic integrals,
\bea
A\, E(1/k,k) +B\,F(1/k,k) & = & A\, E(1,k) +B\,F(1,k)
\eea
along with its complex conjugate. 

\sm

The constraint (\ref{eq:L3-constr}) can be solved for one of the zeros, say $s_2$. Regularity requires that $s_2$ be in the upper half plane, which could lead to a secondary constraint on the remaining parameters. We now show that this is not the case: if $s_1$ is chosen in the upper half plane, the solution for $s_2$ resulting from (\ref{eq:L3-constr}) is in the upper half plane as well. We write (\ref{eq:L3-constr}) as
\begin{align}
	C_2-(s_1+s_2)C_1+s_1s_2 C_0&=0 & \qquad C_n&=\int_1^{1/k^2} dw \frac{w^n}{\sqrt{w(w-1)(k^2 w-1)}}
\end{align}
The solution for $s_2$ is given in terms of $s_1$ by, 
\begin{align}\label{eq:s2det}
	s_2&=\frac{C_2-C_1s_1}{C_1-C_0s_1}
\end{align}
Using this relation, we may compare the imaginary parts of $s_1$ and $s_2$ as follows, 
\begin{align}
	\frac{s_2-s_2^\star}{s_1-s_1^\star}&=\frac{C_0 C_2 - C_1^2}{|C_1-C_0s_1|^2}
\end{align}
To show that the right hand side is non-negative, the integrals can be combined as follows,
\begin{align}
	C_0C_2 - C_1^2 &=\int_{[1,1/k^{2}]\times[1,1/k^{2}]}dw dw'\frac{w^2- ww'}{\sqrt{w(w-1)(k^2 w-1)w'(w'-1)(k^2 w'-1)}}
	\nonumber\\
	&=\frac{1}{2}\int_{[1,1/k^{2}]\times[1,1/k^{2}]}dw dw'\frac{(w-w')^2}{\sqrt{w(w-1)(k^2 w-1) w'(w'-1)(k^2 w'-1)}}
\end{align}
The second line follows by symmetrizing the integrand in $w$ and $w'$, and the resulting integrand is manifestly non-negative, completing the argument.

\sm

We now discuss the asymptotic behavior of the solution at the $p=-5/2$ singularity. The large-$w$ behavior of $\partial_w \cA_{\pm}$ in (\ref{eq:L3-dApm}), given by,
\begin{align}
	\partial_w \cA_{+}  &= a_+ w^{1\over 2} + {a_+\over 2} (1+ {1\over k^2}-2s_1-2s_2) w^{-{1\over 2}}+\cO( w^{-{3\over 2}}) \nonumber\\
	\partial_w \cA_{-}  &= a^*_+ w^{1\over 2} + {a^*_+\over 2} (1+ {1\over k^2}-2s^*_1-2s^*_2) w^{-{1\over 2}}+\cO( w^{-{3\over 2}}) \nonumber\\
\end{align}
confirms that we have a  $p=-5/2$ branch point at infinity. Following the discussion in section \ref{sec:p-5-half}, this realizes asymptotic behavior in the $SL(2,\RR)$ orbit of a deformation of the D-instanton solution with flat Einstein-frame metric. The parameter choice for which the asymptotic behavior matches that of the original D-instanton solution is,
\begin{align}
	\label{eq:L3-ap-rel}
	a_+&=-a_+^\star
\end{align}
The asymptotic behavior then also qualitatively matches the behavior realized in \cite{Komatsu:2024bop}.

\subsection{Parameters and charges for elliptic solutions}
\label{sec:6.2}

The number of free parameters of the $L=3$ solutions can be obtained from equation (\ref{eq:calN}) with $P=Q=0$, $L=3$ and $K=1$, which yields 5 parameters. Fixing the asymptotic behavior at the $p=-5/2$ transition point via (\ref{eq:L3-ap-rel}) reduces this to 4 free real parameters. These can be taken as the location of one complex zero in the upper half plane, say $s_1$, the location of the third $p=-1/2$ branch point, $1/k^2$, and the imaginary constant $a_+$. The condition in (\ref{eq:L3-ap-rel}) partially fixes an $SL(2,\RR)$ frame, and only a subset of the $SL(2,\RR)$  transformation which are compatible with the $dS_{1,5}\times S^2\times\Sigma$ real form act on the parameters by mapping different solutions into each other.

\sm

We can verify this counting by direct inspection. The $L=3$ solution has twelve real parameters in the integration constants $\cA_{\pm}^0$, the complex zeros $s_{1,2}$, the parameter $a_+$, the real location $k$, and a real integration constant in $\cB+\cB^\star$ (which is the only form in which $\cB$ enters the supergravity solutions). The condition  in (\ref{eq:L3-Apm0-rel}) fixes $\cA_-^0$ in terms of $\cA_+^0$ while the relation (\ref{eq:s2det}) can be used to determine $s_2$.  The condition  that $\cG$ vanishes on the boundary fixes the integration constant for $\cB+\cB^\star$. Lastly, the condition in (\ref{eq:L3-ap-rel}) sets the real part of $a_+$ to zero. Imposing these  conditions leaves six real parameters.
The gauge transformations (\ref{gaugetrans}) of $C_{(2)}$, $B_{(2)}$ for  the complex solution  which are compatible with the reality condition  (\ref{3.k.1})  as well as with  (\ref{eq:L3-Apm0-rel})  are given by 
\begin{align}
	\label{gauge}
	\cA_+&\rightarrow \cA_+ + \alpha & \cA_-&\rightarrow \cA_- + \alpha ^\star
\end{align}
These shifts can be used to set $\cA_\pm^0=0$, leaving four real parameters in the $L=3$ solution.

\sm

The parameters of the solution are related to the brane charges as follows:  The solution has a 3-cycle   
${\cal C}\times S^2$, where ${\cal C}$ is a path connecting the two boundary segments $[-\infty,0]$ and $[1, 1/k^2]$ where the two-sphere collapses. This 3-cycle carries  $(p,q)$ 5-brane charges which were determined in  section \ref{sec:charges},
\begin{align}
	Q_{\rm D5}&=-\frac{2}{3}c_6\nu_1 V_{\hat S^2}\left[\cA_2 -\cA_2^\star\right]_{w_0}^{w_1}
	&
	Q_{\rm NS5}&=-\frac{2}{3}c_6\nu_1 V_{\hat S^2}\left[\cA_1+\cA_1^\star\right]_{w_0}^{w_1}
\end{align}
We can choose the endpoints of ${\cal C}$ on the boundary at $w_0=0$ and $w_1=1$. Thus,
\begin{align}
	Q_{\rm D5}&=+\frac{8}{3} \sqrt{2} \pi c_6\nu_1 \Big((A-A^\star) E(1,k) + (B-B^\star)F(1,k)\Big)
	\nonumber\\
	Q_{\rm NS5}&=-\frac{8}{3} \sqrt{2} \pi  c_6\nu_1\Big((A+A^\star)E(1,k)+(B+B^\star)F(1,k)\Big)
\end{align}
The solution also has a 7-cycle   
${\cal C}\times S^6$, where ${\cal C}$ is a path connecting the two boundary segments $[0,1]$ and $[1/k^2, \infty]$ where the six-sphere collapses.  This cycle supports $(p,q)$ 1-brane charges given by (\ref{eq:QD1F1}). Choosing $w_0=1$ and $w_1=1/k^2$ for the endpoints of ${\cal C}$,
\begin{align}
	Q_{\rm D1}&=-i {2^9 \over 3}  \pi^3 c_6^3\nu_1 \left[\cW_1+ \cW_1^\star\right]_{1}^{1/k^2}  &Q_{\rm F1}&=i {2^9 \over 3}  \pi^3 c_6^3\nu_1 \left[\cW_2- \cW_2^\star\right]_{1}^{1/k^2}
\end{align}
where $\cW_{1,2}$ is  obtained by integrating the holomorphic differential (\ref{eq:wi-def}).  We note that for generic choices of parameters the fundamental string charge $Q_{\rm F1}$ is nonzero, as will be illustrated explicitly in the next subsection.

\sm

Finally, there is a nine cycle $\cC\times S^2\times S^6$, where the path $\cC$ connects boundary segments where $S^2$ and $S^6$ collapse respectively. Following the discussion in section \ref{sec:p-5-half}, we can choose $\cC$ as arc around $w=\infty$.  This nine cycle supports D-instanton charge (\ref{eq:q-dinst}),
\begin{equation}
	Q_{\rm D(-1)}= \frac{2^{10}}{3} \pi^5 c_6^4 \Big( \Res_{w=\infty}(\partial_w\cY_1)- {\rm c.c.}\Big)
\end{equation}
where $\partial_w \cY_i, i=1,2$ were defined in (\ref{eq:yi-def}). An explicit expression can be extracted from the expansions of $\partial\cA_\pm$ using (\ref{eq:p5half-QDm1}).

\subsection{Supergravity fields for an example elliptic solution}\label{sec:L3plots}

In this section we illustrate an example elliptic solution with a generic choice of parameters through plots of the supergravity fields. With the functions $\cA_\pm$ derived in explicit form, the only remaining task is to integrate the holomorphic differential  $\partial_w \cB$  given in (\ref{2.g.2b}), to obtain the function $\cB$ which appears in the definition of $\cG$. A closed form for $\cB$ in terms of elliptic polylogarithms is derived in appendix \ref{app:cB-polylogs}. Practically, we perform the straightforward contour integral determining $\cB$ in the upper half plane numerically.

\sm

The free parameters for the solution we plot include one of the complex zeros in the auxiliary electrostatics problem, $s_1$, the location of the third branch point, $1/k^2$, the overall normalization coefficient $a_+$, and the constants $\cA_\pm^0$ which fix a gauge for the two-form potentials. We choose,
\begin{align}
	\label{eq:plot-param}
	a_+&=i & s_1&=\frac{3}{2}+\frac{i}{3} & k&={1\over \sqrt{2}} & \cA_\pm^0&=0
\end{align}
All remaining parameters can be derived from this set through the regularity conditions. The (unnormalized) brane charges can be found from the formulas of section \ref{sec:6.2},
\begin{align}
	(Q_{\rm D5},Q_{\rm NS5})&\approx (51.641i\,,\,39.792)
	&
	Q_{\rm D(-1)}&\approx 12337i
	\no\\
	(Q_{\rm F1},Q_{\rm D1})&\approx (27.369\,,\,-310.02i)
\end{align}
where the signs depend on the orientation of the cycles.

\sm

\begin{figure}
	\centering
	\includegraphics[width=0.32\linewidth]{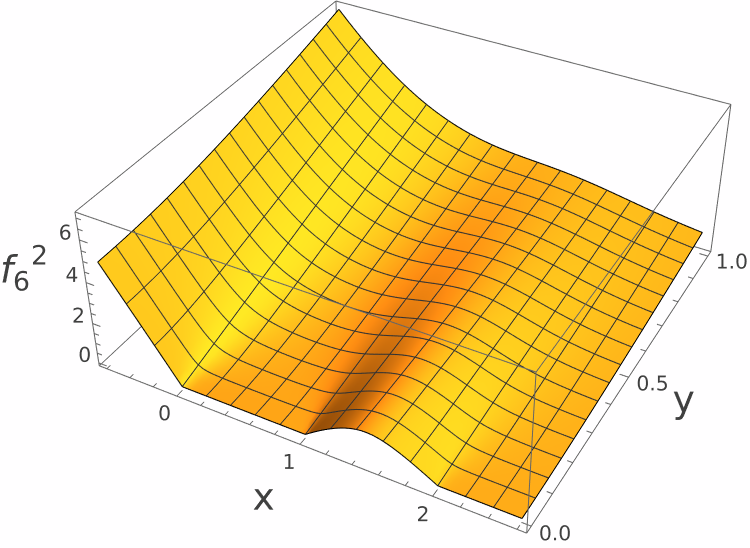}
	\includegraphics[width=0.32\linewidth]{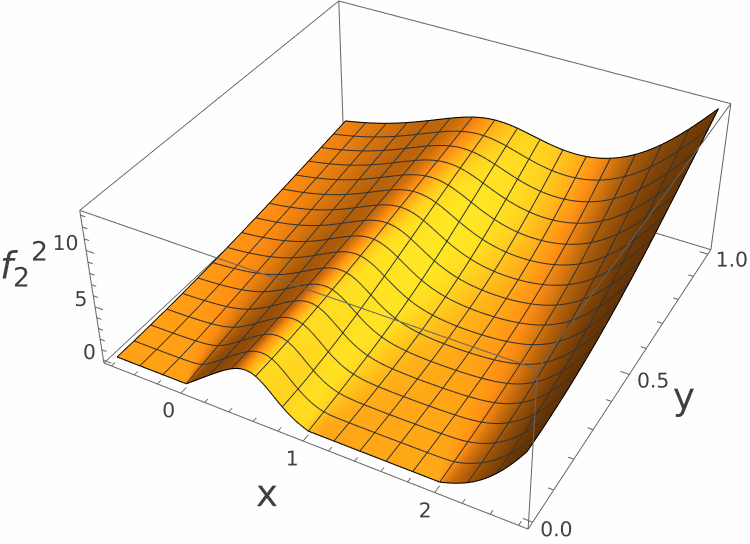}
	\includegraphics[width=0.32\linewidth]{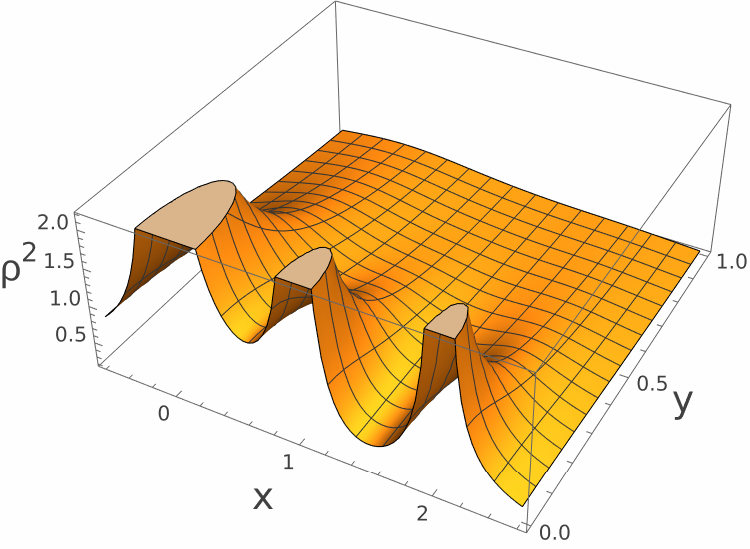}
	\caption{Metric functions for the elliptic solution with parameters in (\ref{eq:plot-param}).\label{fig:metric-plot}}
\end{figure}

For the plots we introduce real coordinates on the upper half plane, $w=x+iy$.
The metric functions are shown in figure \ref{fig:metric-plot}. The plots show that $f_6^2$, $f_2^2$ and $\rho^2$ are positive in the interior of $\Sigma$, with $f_2$ and $f_6$ alternatingly vanishing on segments of the real axis, with $S^2$ collapsing on the $x<0$ and $x\in (1,2)$ segments and the $S^6$ collapsing on the $x\in (0,1)$ and $x>2$ segments.
For generic $|w|\gg 1$, $f_6^2$ and $f_2^2$ both grow unboundedly. This is consistent with the geometry decompactifying and the metric approaching that of flat space, as expected based on the discussion of section \ref{sec:p-5-half}.

\sm

\begin{figure}
	\centering
	\includegraphics[width=0.32\linewidth]{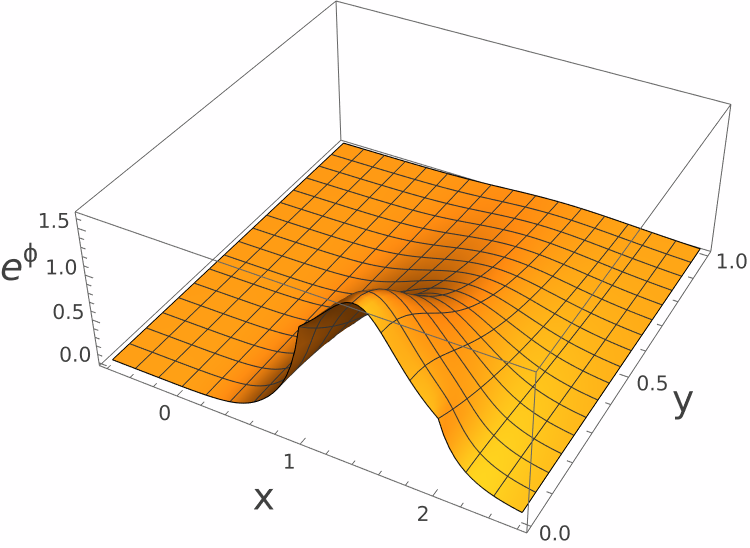}
	\qquad
	\includegraphics[width=0.32\linewidth]{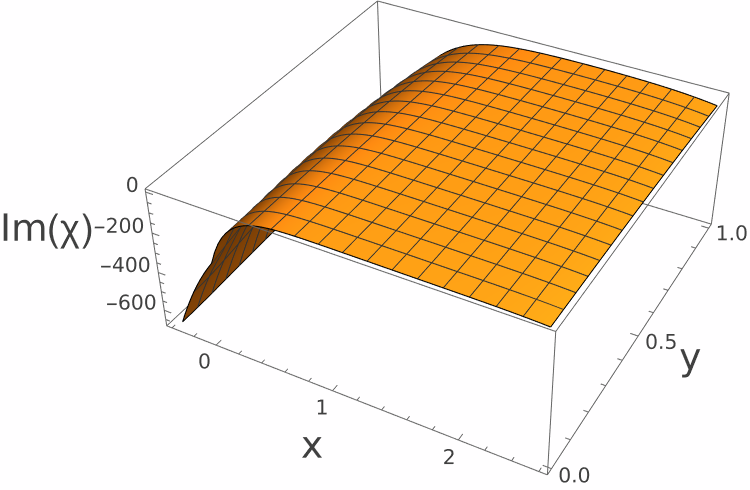}
	\caption{Dilaton and axion for the elliptic solution with parameters in (\ref{eq:plot-param}).\label{fig:phi-chi-plot}}
\end{figure}

The exponentiated dilaton and the imaginary part of the axion are shown in figure \ref{fig:phi-chi-plot}. We note that $e^\phi$ is non-negative throughout $\Sigma$ and vanishes for large $|w|$. The axion is imaginary, satisfying the same reality condition as in Type IIB$^\star$, and it grows unboundedly when $x$ becomes large and negative. This is the expected behavior based on the analysis of the asymptotic form in section \ref{sec:p-5-half}.

\sm

\begin{figure}
	\includegraphics[width=0.32\linewidth]{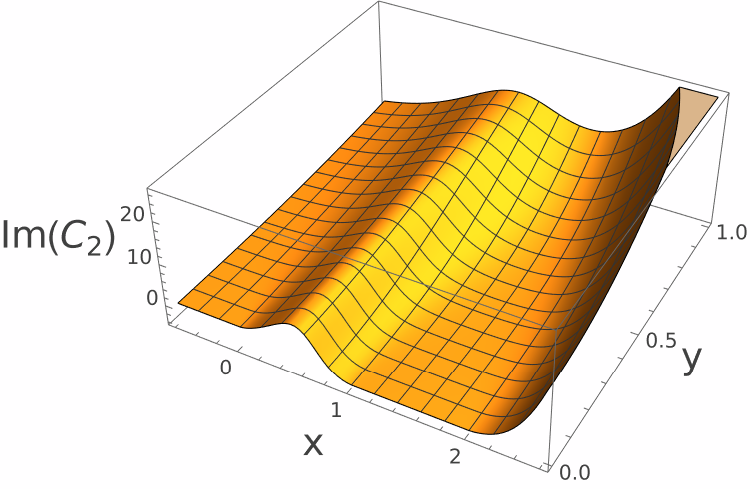}
	\includegraphics[width=0.32\linewidth]{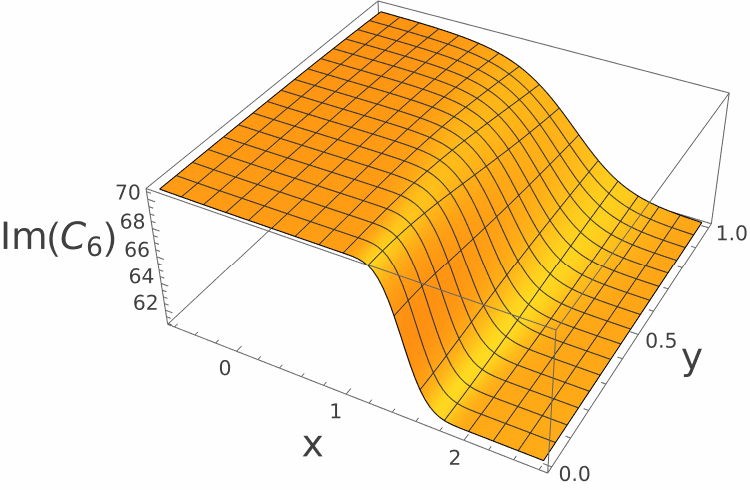}
	\includegraphics[width=0.32\linewidth]{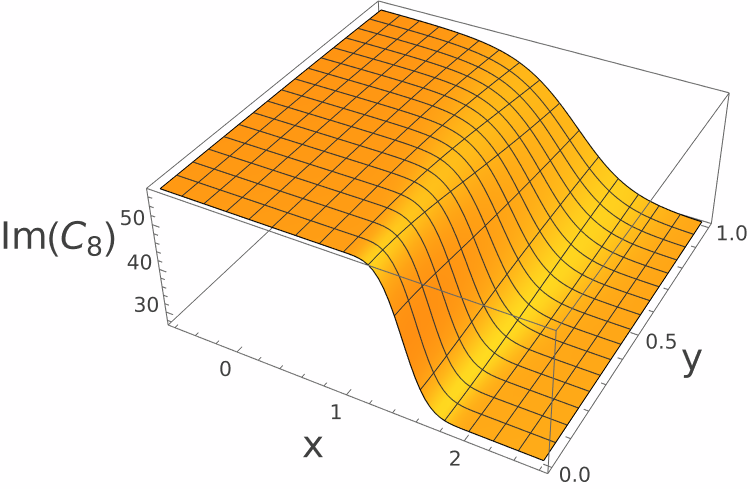}\\
	\includegraphics[width=0.32\linewidth]{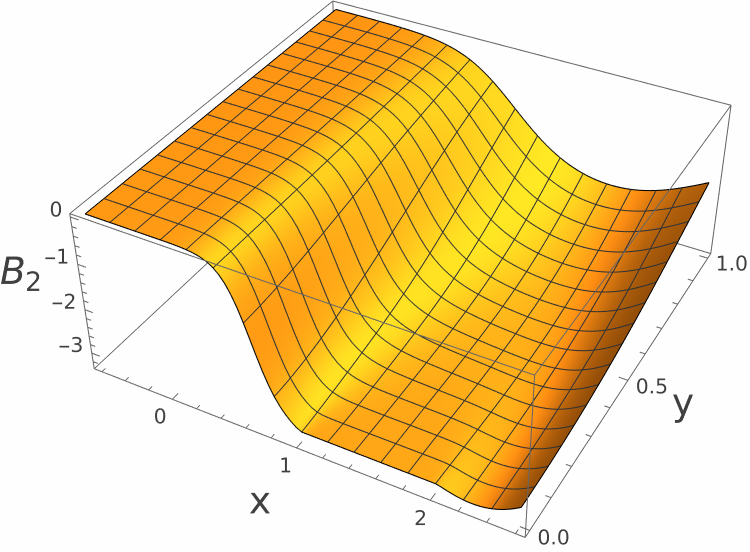}
	\includegraphics[width=0.32\linewidth]{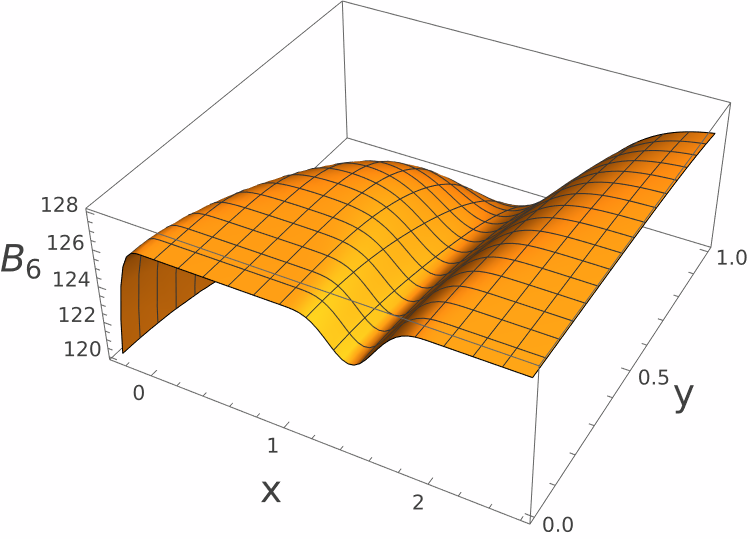}
	\caption{Flux potential functions for the elliptic solution with parameters in (\ref{eq:plot-param}).\label{fig:potentials-plot}}
\end{figure}

The remaining flux potentials are shown in figure \ref{fig:potentials-plot}. The RR potentials are imaginary and the NS potentials real, again as in Type IIB$^\star$. Along boundary segments where the $S^2$ collapses the two-forms are constant, and along boundary segments where $S^6$ collapses the six-forms are constant. The 5-brane charges of the solution are manifest in the plots as difference in the boundary values of the two-forms between the $x<0$ segment and the $x\in (1,2)$ segment. The plots clearly show that the solution carries both D5 and NS5 charge. The string charges are manifest in the difference in boundary values of the six-forms between the $x\in(0,1)$ and $x>2$ segments. Again, both the F1 and D1 charges are non-vanishing. The 8-form is constant on the two boundary segments adjacent to the $p=-5/2$ branch point at $|w|\rightarrow\infty$, and the instanton charge, represented as the jump of $\tilde C_{(8)}$ across the $p=-5/2$ branch point, is non-vanishing.

\subsection{Hyper-elliptic solutions with $L>3$}\label{sec:hyper}

For $L>3$ the differentials $\p \cA_\pm$ are elementary, but the functions $\cA_\pm$ lead to hyper-elliptic integrals. In this section we spell out the solutions and constraints for this case. The differentials are given by evaluating (\ref{4.d.1}) and (\ref{4.d.2}) in section~\ref{sec:differentials} for $K=1$, $N=(L+1)/2$ and $P=Q=0$. We again use $SL(2,\RR)$ transformations of the upper half plane to map the $p=-5/2$ transition point to $w=\infty$. This yields,
\begin{align}
	\partial_w\cA_+&=a_+ \frac{\prod_{a=1}^{(L+1)/2}(w-s_a)}{\prod_{\ell=1}^L\sqrt{w-b_\ell}}
	&
	\partial_w\cA_-&=a_+^\star\frac{\prod_{a=1}^{(L+1)/2}(w-s_a^\star)}{\prod_{\ell=1}^L\sqrt{w-b_\ell}}
\end{align}
Similar to the expressions for the $L=3$ case in (\ref{eq:dApm-L3-decomp}), the degree of the polynomial in the numerator can be reduced by separating out exact terms, with the remaining contour integrals determining $\cA_\pm$ now being hyper-elliptic.
The parameters are constrained by the matching conditions in (\ref{eq:lift}) which connect the boundary segments with $S^6$ collapsing.

\sm

The number of parameters available for given $L$ can be obtained from (\ref{eq:calN}) with $P=Q=0$ and $K=1$, which yields $L+2$.
Fixing the asymptotic behavior at the $p=-5/2$ transition point to match the D-instanton solution again leads to the constraint (\ref{eq:L3-ap-rel}), so that the number of free parameters within this class of solutions is
\begin{equation}\label{eq:hyper-parms}
\cN=L+1	
\end{equation}
For a generic solution the boundary $\partial\Sigma$ is partitioned into $L+1$ segments. On $(L+1)/2$ of them $S^6$ collapses and on the other half $S^2$ collapses. The independent 3-cycles are formed by $S^2$ fibered over curves connecting consecutive boundary segments with $S^2$ collapsing, while the independent 7-cycles are formed by $S^6$ fibered over curves connecting consecutive segments with $S^6$ collapsing. We obtain $(L-1)/2$ 3-cycles and $(L-1)/2$ 7-cycles. With the 3-cycles carrying $(p,q)$ 5-brane charge and the 7-cycles carrying $(p,q)$ string charge, the string and 5-brane charges of a solution with $L$ transition points are,
\begin{align}\label{eq:hyper-charges}
	(Q_{\rm F1}^{(i)},Q_{\rm D1}^{(i)})\,,&\quad i=1,\ldots,(L-1)/2\,,
	&
	(Q_{\rm NS5}^{(i)},Q_{\rm D5}^{(i)})\,,&\quad i=1,\ldots,(L-1)/2
\end{align}
With a single $p=-5/2$ transition point, there is only one D-instanton charge associated with these solutions regardless of $L$.

\subsection{Connection to the polarized IKKT model}
\label{sec:IKKT}

In this section we compare the solutions with D-instanton asymptotics and general $L$ to the saddle points of the Euclidean polarized IKKT model. We start with a brief review of the polarized IKKT integral defined in \cite{Bonelli:2002mb}, and refer to \cite{Hartnoll:2024csr,Komatsu:2024bop} for more detailed discussions.

\sm
	
The action of the polarized IKKT model is given in terms of $N\times N$ matrices $X_\mu$ with $\mu=1,\ldots,10$ and $\Psi_\alpha$ with $\alpha=1,\ldots,16$ by,
\begin{align}\label{IKKTa}
S &= \tr \left ( - { 1 \over 4} [X_\mu, X_\nu] [X^\mu , X^\nu] + \half \Psi _\a (\cC \G^\mu) _{\a \b} [ X_\mu , \Psi _\b] \right )
\no\\
&\hphantom{=}+ \tr \left ( { \Omega ^2\over 4^3} X^A X_A 
+  {3 \Omega ^2 \over 4^3} X_a X^a 
+ i \Omega  X_8 [X_9, X_{10}] 
+ i {\Omega \over 8} \Psi _\a (\cC \, \mN) _{\a \b} \Psi _\b +\eta\mathds{1}\right )
\end{align}
where $\Omega$ is the parameter controlling the polarization deformation and $\eta$ parametrizes a constant that may be added without breaking additional symmetries. As shown in \cite{Hartnoll:2024csr,Komatsu:2024bop}, the solutions to the saddle point equations for the Euclidean model are characterized by,
\begin{align}
	X_i&=\frac{3}{8}\Omega J_i\,, \quad i=8,9,10 & X_p&=0\,,\quad p=1,.,7
\end{align}
where $J_i$ satisfy the $SU(2)$ commutation relation,
\begin{align}
	[J_i,J_j]=i\epsilon_{ijk}J_k
\end{align}
The solutions to the saddle point equations are therefore in one-to-one correspondence with $N$-dimensional representations of $SU(2)$. A general representation may be decomposed into irreducible representations, which consist of some number, $n$, of groups of representations of the same dimension. We denote the dimension of the representations in the $i^{\rm th}$ group by $N_i$, and their multiplicity by $k_i$. A saddle point with $n$ groups of representations is therefore characterized by $2n$ integer parameters $(N_i,k_i)$, $i=1,\ldots n$.

\sm

Holographic descriptions for these saddle points were proposed in \cite{Komatsu:2024bop}. Here we show that the parameters of the solutions in section \ref{sec:sol} can be matched to those of the polarized IKKT saddles in a similar way. To this end, we start from a general (hyper-)elliptic solution with D-instanton asymptotics. Following the discussion in section \ref{sec:hyper}, these solutions have an equal number of 3-cycles carrying $(p,q)$ 5-brane charge and 7-cycles carrying $(p,q)$ string charge. We denote this number by $\tilde n$. They then have $L+1=2\tilde n+2$ boundary segments and are characterized by 
\begin{equation}
	\cN=L+1=2+2\tilde n
\end{equation}
free real parameters (see (\ref{eq:hyper-parms})). These parameters can be matched to the parameters of the saddle points of the IKKT model as follows. Two parameters correspond to the deformation parameter $\Omega$ and to $\eta$ in the polarized IKKT model, while the remaining number is consistent with the identification $\tilde n=n$, with $n$ denoting the number of groups of $SU(2)$ representations. With this identification the number of parameters matches.

\sm

The proposal of \cite{Komatsu:2024bop} identifies the D-string and NS5-brane charges with the dimension and multiplicity of the representations in the $n$ groups of $SU(2)$ representations, and gives further evidence for the validity of their prescription. The solutions constructed here, expressed in terms of the two holomorphic functions $\cA_\pm$, can be connected to the electrostatics formulation in terms of a single real harmonic function used in \cite{Komatsu:2024bop}, and we provide explicit formulas for translating our solutions to this formulation in appendix \ref{appc}.

\sm

We find a discrepancy with the analysis in \cite{Komatsu:2024bop} regarding the charges carried by the solutions. As noted in section \ref{sec:6.2} and shown in a concrete example in section \ref{sec:L3plots}, we find the F1 charges of our solutions to be generally non-vanishing. The analysis in \cite{Komatsu:2024bop}, on the other hand, concluded that the solutions constructed there do not carry F1 charge. With matching numbers of free parameters once the asymptotic behavior is fixed, these statements are not compatible. We believe that, rather than signaling an incompatibility between the solutions, the discrepancy arises in the evaluation of the F1 charge. As we show in appendix \ref{app:F1-charge-electrostatics}, the expression for the F1 charge given in \cite{Komatsu:2024bop} does not generically vanish for the solutions given there and yields non-zero F1 charge for the high-disc limit of the single-disc solution. For a discussion of the ramifications we refer to the Introduction.

\clearpage

\appendix

\section{Useful relations}
\label{sec:A}

We collect some useful relations valid for the general complex solutions as reviewed in section~\ref{sec:exactsolution}. Explicit expressions for the derivatives of $\cG$ are
\begin{align}\label{eq:dG}
	\partial_w\cG&=i\left[-(\cA_2+\tilde\cA_2)\partial_w\cA_1+(\cA_1+\tilde\cA_1)\partial_w\cA_2\right]
	\nonumber\\
	\partial_\tw\cG&=i\left[+(\cA_2+\tilde\cA_2)\partial_\tw\tilde\cA_1-(\cA_1+\tilde\cA_1)\partial_\tw\tilde\cA_2\right]
	\nonumber\\
	\kappa^2&=i\left(+\partial_w\cA_1\partial_\tw\tilde\cA_2-\partial_\tw \tilde\cA_1\partial_w\cA_2\right)
\end{align}
Using these expressions we can derive the following relations which follow directly from the definitions of the composite quantities,
\begin{align}\label{eq:dGcomb}
	\partial_w\cG\partial_\tw\tilde\cA_1+\partial_\tw \cG\partial_w\cA_1&=-(\cA_1+\tilde\cA_1)\kappa^2
	\nonumber\\
	\partial_w\cG\partial_\tw\tilde\cA_2+\partial_\tw \cG\partial_w\cA_2&=-(\cA_2+\tilde\cA_2)\kappa^2
\end{align}
and
\begin{align}
	\frac{\partial_w\cG\partial_\tw\cA_2-\partial_\tw \cG\partial_w\cA_2}{\cA_2+\tilde\cA_2}
	-
	\frac{\partial_w\cG\partial_\tw\cA_1-\partial_\tw \cG\partial_w\cA_1}{\cA_1+\tilde\cA_1}
	&=\frac{2i\partial_w\cG\partial_\tw\cG}{(\cA_1+\tilde\cA_1)(\cA_2+\tilde\cA_2)}
\end{align}

\bigskip

The axion-dilaton combinations  $\tau_\pm$ in (\ref{2.f.2}) can be re-expressed using the relations (\ref{eq:dGcomb}) as
\begin{align}
	\tau_\pm&=\frac{\kappa^2T (\cA_2+\tilde\cA_2)\mp(\partial_w\cG\partial_\tw\tilde\cA_2-\partial_\tw\cG\partial_w\cA_2)}{\kappa^2 T (\cA_1+\tilde\cA_1)\mp(\partial_w\cG\partial_\tw\tilde\cA_1-\partial_\tw\cG\partial_w\cA_1)}
\end{align}
The dilaton and axion themselves are given by
\begin{align}\label{eq:expphi}
	e^{-\phi}&=\frac{2\kappa^2T\partial_w\cG\partial_\tw\cG}{(\partial_\tw\tilde\cA_1\partial_w\cG-\partial_w\cA_1\partial_\tw\cG)^2-\kappa^4 T^2 (\cA_1+\tilde\cA_1)^2}
\nonumber\\
	\chi&=\frac{(\partial_\tw\tilde\cA_1\partial_w\cG-\partial_w\cA_1\partial_\tw\cG) (\partial_\tw\tilde\cA_2\partial_w\cG-\partial_w\cA_2\partial_\tw\cG)-\kappa^4T^2 (\cA_1+\tilde\cA_1) (\cA_2+\tilde\cA_2)}{(\partial_\tw\tilde\cA_1\partial_w\cG-\partial_w\cA_1\partial_\tw\cG)^2-\kappa^4 T^2 (\cA_1+\tilde\cA_1)^2}
\end{align}

\clearpage

\section{Flux potentials}
\label{sec:potentials}

In this appendix 
we determine the RR and NS-NS gauge potentials, starting with the two-forms and  the six-form potentials associated with five and 1-brane charges respectively. In addition we calculate  the RR eight-form potential which  is related to the D-instanton charge. We will also show that the parameter $\eta$ which could take  two possible values $\eta=\pm1$  in \cite{DHoker:2025nid} is fixed to be $\eta=-1$ by demanding integrability of the three form field strength.

\subsection{2-form potentials}

The general expression for the three-form field strengths in our ansatz is given in \cite[(3.3), below (B.4)]{DHoker:2025nid} as
\begin{align}\label{eq:2-form-ansatz-app}
	\tilde F_{(3)}&=F_{(3)}-\chi H_{(3)}=g_a e^a\wedge e^{67} & H_{(3)}&=h_a e^a \wedge e^{67}
\end{align}
where $e^{67}$ is the volume form on $\cM_{2\CC}$ and $e^a$ with $a=z,\tz$ the frame on $\Sigma_\CC$, given by
\begin{align}
	e^{67}&=f_2^2\vol_{\hat \cM_{2\CC}}
	&
	e^z&=\rho dw
	&
	e^\tz=\rho d\tw
\end{align}
The components were given in equation (3.50) in \cite{DHoker:2025nid} as
\begin{align}\label{eq:g-h}
	\rho \big ( h_z \pm i e^\phi g_z) & = 
	\pm 4i \nu \left({ \tau_\mp - \tilde \xi \over \tau_\pm - \tilde \xi} \right )^\frac{1}{2} { \p_w \tau_\pm \over \tau_+ - \tau_-} 
	&
	\tilde\xi&=\frac{\partial_\tw\tilde\cA_2}{\partial_\tw\tilde\cA_1}
	\nonumber \\
	\rho \big ( h_\tz \pm i e^\phi g_\tz) & =  
	\pm 4i \nu  \left({ \tau_\mp -  \xi \over \tau_\pm -  \xi}  \right )^\frac{1}{2} { \p_\tw \tau_\pm \over \tau_+ - \tau_-} 
	&
	\xi&=\frac{\partial_w\cA_2}{\partial_w\cA_1}
\end{align}
This leads to
\begin{align}
	H_{(3)}&=\left(b_w dw+b_\tw d\tw\right)\wedge\vol_{\hat \cM_{2\CC}}
	&
	F_{(3)}&=\left(c_w dw+c_\tw d\tw\right)\wedge\vol_{\hat \cM_{2\CC}}
\end{align}
where the components of $H_{(3)}$ are
\begin{align}
	b_w&=\frac{2i\nu \; e^{\phi\over 2} f_2^2}{\tau_+-\tau_-} \left ( { \tau_- - \tilde \xi \over \tau_+ - \tilde \xi} \right )^\frac{1}{2}
	\left[
	\partial_w\tau_+
	-
	{ \tau_+ - \tilde \xi \over \tau_- - \tilde \xi}\,\partial_w\tau_-
	\right]
	\nonumber\\
	b_\tw&=\frac{2i\nu\;  e^{\phi\over 2} f_2^2}{\tau_+-\tau_-} \left ( { \tau_- - \xi \over \tau_+ - \xi} \right )^\frac{1}{2}
	\left[
	\partial_\tw\tau_+
	-
	{ \tau_+ - \xi \over \tau_- - \xi}\,\partial_\tw\tau_-
	\right]
\end{align}
and the components of $F_{(3)}$ are
\begin{align}\label{eq:cform}
	c_w&=\frac{2i\nu \; e^{\phi\over 2} f_2^2}{\tau_+-\tau_-} \left ( { \tau_- - \tilde \xi \over \tau_+ - \tilde \xi} \right )^\frac{1}{2}
	\left[
	\tau_-\partial_w\tau_+
	-
	{ \tau_+ - \tilde \xi \over \tau_- - \tilde \xi}\,\tau_+\partial_w\tau_-
	\right]
	\nonumber\\
	c_\tw&=
	\frac{2i\nu\;  e^{\phi\over 2} f_2^2}{\tau_+-\tau_-} \left ( { \tau_- - \xi \over \tau_+ - \xi} \right )^\frac{1}{2}
	\left[
	\tau_-\partial_\tw\tau_+
	-
	{ \tau_+ - \xi \over \tau_- - \xi}\,\tau_+\partial_\tw\tau_-
	\right]
\end{align}

The first step in evaluating the components is to eliminate the square roots. This can be accomplished noting that $e^\phi f_2^4(\tau_--\tilde\xi)/(\tau_+-\tilde\xi)$ and $e^\phi f_2^4(\tau_--\xi)/(\tau_+-\xi)$, are complete squares.  Here  $f_2$ is the Einstein frame metric function (\ref{eq:metric}).
\begin{align}
	e^\phi f_2^4\frac{\tau_--\tilde\xi}{\tau_+-\tilde\xi}&=
	4c_2^4k_1^2 T^{-4}\left(\frac{\partial_w\cA_1\partial_\tw\cG (T-1)+\partial_\tw\tilde\cA_1\partial_w\cG (T+1)}{c_6\kappa^2(T-1)}\right)^2
	\nonumber\\
	e^\phi f_2^4\frac{\tau_--\xi}{\tau_+-\xi}&=
	4c_2^4k_1^2 T^{-4}\left(\frac{\partial_\tw\tilde\cA_1\partial_w\cG (T+1)+\partial_w\cA_1\partial_\tw\cG (T-1)}{c_6\kappa^2(T+1)}\right)^2
\end{align}
Taking the square root introduces sign choices which we parametrize by $\nu_{1/2}$ as
\begin{align}
	\nu\sqrt{e^\phi f_2^4\frac{\tau_--\tilde\xi}{\tau_+-\tilde\xi}}&=\nu_1 \frac{2c_2^2k_1}{c_6}T^{-2}\frac{\partial_\tw\tilde\cA_1\partial_w\cG(T+1)+\partial_w\cA_1\partial_\tw\cG(T-1)}{(T-1)\kappa^2}
	\nonumber\\
	\nu\sqrt{e^\phi f_2^4\frac{\tau_--\xi}{\tau_+-\xi}}&=\nu_2 \frac{2c_2^2k_1}{c_6}T^{-2}\frac{\partial_\tw\tilde\cA_1\partial_w\cG(T+1)+\partial_w\cA_1\partial_\tw\cG(T-1)}{(T+1)\kappa^2}
\end{align}
The only remaining square roots are implicit in $T$; they can be eliminated in the complete expressions for the fields strengths components producing the following expressions
\begin{align}
	b_w&=\frac{4 i c_2^2k_1\nu_1}{c_6T^{1-\eta}}\left[\frac{\partial_w\cA_1\partial_\tw\cG}{\kappa^2\partial_w\cG}\partial_w^2\cG-\frac{\partial_\tw\cG}{\kappa^2}\partial_w^2\cA_1+\partial_w\cA_1
		-\left(\frac{T^2+1}{T^2-1}\partial_w\cA_1\partial_\tw\cG+\partial_\tw\tilde\cA_1 \partial_w\cG
		\right)\frac{\partial_w T^2}{2\kappa^2 T^2}\right]
	\nonumber\\
	b_\tw&=\frac{4 i c_2^2k_1\nu_2}{c_6T^{1-\eta}}\left[\frac{\partial_\tw\tilde\cA_1\partial_w\cG}{\kappa^2\partial_\tw\cG}\partial_\tw^2\cG-\frac{\partial_w\cG}{\kappa^2}\partial_\tw^2\tilde\cA_1+\partial_\tw\tilde\cA_1
	-\left(\frac{T^2+1}{T^2-1}\partial_\tw\tilde\cA_1\partial_w\cG+\partial_w\cA_1 \partial_\tw\cG
	\right)\frac{\partial_\tw T^2}{2\kappa^2 T^2}\right]
\end{align}
These expressions can be partially integrated and written more compactly as
\begin{align}
	b_w&=+\nu_1 T^{1+\eta}\partial_w\left[
	\frac{6 i c_2^2k_1\nu_1}{c_6}\left(\frac{\partial_\tw\tilde\cA_1\partial_w\cG-\partial_w\cA_1\partial_\tw\cG}{3\kappa^2T^2}+\cA_1-\tilde\cA_1\right)
	\right]
	\nonumber\\
	b_\tw&=-\nu_2 T^{1+\eta}\partial_\tw\left[
	\frac{6 i c_2^2k_1\nu_1}{c_6}\left(\frac{\partial_\tw\tilde\cA_1\partial_w\cG-\partial_w\cA_1\partial_\tw\cG}{3\kappa^2T^2}+\cA_1-\tilde\cA_1\right)
	\right]
\end{align}
With these expressions the integrability condition $\partial_\tw b_w-\partial_w b_\tw=0$ leads to the following conditions
\begin{align}
	\nu_2&=-\nu_1 & \eta&=-1
\end{align}
This in particular fixes the branch choice parameter $\eta\in\{\pm1\}$ that was left undetermined in  our previous paper \cite{DHoker:2025nid}.

\sm

The calculations  to determine the potential $C_{(2)}$  for the field strength $F_{(3)}$  (\ref{eq:cform}) are analogous, and we can integrate to obtain the potentials
\begin{align}
b_w=\partial_w f_{B_{(2)}}, \quad  b_\tw=\partial_\tw f_{B_{(2)}}, \quad  c_w=\partial_w f_{C_{(2)}}, \quad c_\tw=\partial_\tw f_{C_{(2)}}
\end{align} 
where  the potential two form  is given by
\begin{align}
	B_{(2)}&=f_{B_{(2)}} \vol_{\hat \cM_{2\CC}}
	&
	C_{(2)}&=f_{C_{(2)}} \vol_{\hat \cM_{2\CC}}
\end{align}
The final result for the potentials can be expressed compactly as 
\begin{align}\label{eq:2-form-potentials}
	f_{B_{(2)}}&=\frac{6 i c_2^2k_1\nu_1}{c_6}
	\left(\frac{\cU_1}{3T^2}+\cA_1-\tilde\cA_1\right)
	&
	f_{C_{(2)}}&=\frac{6 i c_2^2k_1\nu_1}{c_6}
	\left(\frac{\cU_2}{3T^2}+\cA_2-\tilde\cA_2\right)	
\end{align}
with
\begin{align}\label{eq:U-def}
	\cU_i&=\frac{\partial_\tw  \tilde\cA_i\, \partial_w\cG-\partial_w  \cA_i\, \partial_\tw\cG}{\kappa^2}
    \hskip 1in i=1,2
\end{align}
Note that the $SL(2)$ transformation rules for the potentials follow directly from the fact that the holomorphic functions $\cA_i$ transform as a doublet.

\subsection{6-form potentials}

We next turn to the 6-form potential. For the complexification of Type IIB we take the view that the dualized fields in Type IIB$_\RR$ are complexified. This means the dual field strengths and potentials in complex Type IIB have the same rank as in standard Type IIB$_\RR$ but are complex.\footnote{This is different from dualizing the fields after complexification within Type IIB$_\CC$.} We parametrize the 6-form potentials as follows,
\begin{align}\label{eq:6-form-def}
	dC_{(6)}&=\star\left(\chi \, e^{\phi}\,  dB_{(2)}- e^{\phi} \, dC_{(2)}\right)
	&
	C_{(6)}&=f_{C_{(6)}} \, \hat e^{\, 0 \cdots 5}
	\nonumber\\
	dB_{(6)}&= \star\left( (e^{-\phi} + e^\phi \chi^2) dB_{(2)} -e^\phi \chi dC_{(2)}\right)
	& B_{(6)}&=f_{B_{(6)}} \, \hat e^{\, 0 \cdots 5}
\end{align}
The expressions in components become
\begin{align}
	\partial_wf_{C_{(6)}}&=+if_6^6f_2^{-2} e^{\phi} \, (\chi b_w-c_w)
	& \qquad
	\partial_wf_{B_{(6)}}&=+if_6^6f_2^{-2} e^{\phi}(\tau_+\tau_-b_w-\chi c_w)
	\nonumber\\
	\partial_\tw f_{C_{(6)}}&=-if_6^6f_2^{-2} \, e^{\phi} \, (\chi b_\tw-c_\tw)
	& \qquad
	\partial_\tw f_{B_{(6)}}&=-if_6^6f_2^{-2} \, e^{\phi}(\tau_+\tau_-b_\tw-\chi c_\tw)
\end{align}
where $f_6$, $f_2$ are in Einstein frame and the factors of $i$ are from taking the dual form on $\Sigma$ before complexification.
The Bianchi identity is  satisfied and the dual potentials are given by
\begin{align}\label{eq:6-form-potentials}
	f_{C_{(6)}}=40c_6^3k_1^3\nu  \bigg(&\frac{3i}{5}\cG \, \cU_1
	+4(\cW_1+\tilde\cW_1)
	-(\cA_1-\tilde\cA_1)(2\tilde\cA_1\cA_2-2\cA_1\tilde\cA_2+3i\cG)\bigg)
	\nonumber\\
	f_{B_{(6)}}=40c_6^3k_1^3\nu  \bigg(&\frac{3i}{5}\cG \, \cU_2
	+4(\cW_2+\tilde\cW_2)
	-(\cA_2-\tilde\cA_2)(2\tilde\cA_1\cA_2-2\cA_1\tilde\cA_2+3i\cG)\bigg)
\end{align}
where the auxiliary functions $\cW_i$ for $i=1,2$ are defined  by the following relations
\begin{align}
	\partial_w\cW_i&=\cA_i\, \partial_w\cB & \partial_\tw\tilde\cW_i&=\tilde\cA_i \, \partial_\tw\tilde\cB
\end{align}
which can be integrated due to  holomorphicity of $\cA_i$ and $\cB$.

\subsection{8-form potential} 

For the 8-form potentials we start from
\begin{align}
	dC_{(8)}&= e^{2\phi} \, \star d\chi+dB_{(2)}\wedge C_{(6)}
	&
	C_{(8)}&=f_{C_{(8)}} \, \hat e^{\, 0\cdots 7}
\end{align}
The $\star$ is again understood as taking the dual before complexification.
In components this becomes (compare \cite[appendix C]{Gutperle:2018vdd})
\begin{align}
	\partial_w f_{C_{(8)}}&=i f_6^6f_2^2 \, e^{2\phi} \, \partial_w\chi+f_{C_{(6)}}\partial_w f_{B_{(2)}}
&
	\partial_\tw f_{C_{(8)}}&=-i f_6^6f_2^2 \, e^{2\phi} \, \partial_\tw\chi+f_{C_{(6)}}\partial_\tw f_{B_{(2)}}
\end{align}
where the metric functions are in the Einstein frame. Spelling this out more explicitly,
\begin{align}
i f_6^6f_2^2 \, e^{2\phi} \, \partial_w\chi&=\frac{288\, i \, k_1^4\, c_6^2\, c_2^2\, \cG^2}{(\tau_+-\tau_-)^2}\partial_w(\tau_++\tau_-)
\end{align}
With the expressions for $f_{B_{(2)}}$ and $f_{C_{(6)}}$ derived before this gives an explicit expression for the field strength which is algebraic in $\cG$, $T^2$ and the holomorphic functions. To integrate it we use a grading by the powers of $T^2$ in the denominator and $\cG$ in the numerator. Starting with the highest powers the expression can be integrated systematically.
This leads to 
\begin{align}
	f_{C_{(8)}}=f_{B_{(2)}}f_{C_{(6)}}+48c_2^2c_6^2k_1^4\bigg ( &
	3\cG  \, \cU_1 (\cA_1-\tilde\cA_1)
	-\frac{6\cG^2\partial_w\cA_1\partial_\tw\tilde\cA_1}{\kappa^2}
	-15i (\cY_1-\tilde\cY_1)
	\nonumber\\&
	-5i(\cA_1^2+\tilde\cA_1^2-\cA_1\tilde\cA_1) (\cA_1\tilde\cA_2-\tilde\cA_1\cA_2-2 i\cG)
	\bigg )
\end{align}
where the auxiliary functions $\cY_{1,2}$ are defined  by the following relations
\begin{align}
	\partial_w\cY_1&=\cA_1^2\partial_w\cB
	&
	\partial_\tw\tilde\cY_1&=\tilde\cA_1^2\partial_\tw\tilde\cB
\end{align}

In the notation of \cite{Bergshoeff:2001pv}, $C_{(8)}$  corresponds to the potentials in the ${\bf C}$-basis. The corresponding $\mathbf{A}$-basis 8-form potentials, whose field strengths involve the 6-form potentials only through their field strength, are given by 
\begin{align}
	\tilde C_{(8)}&=C_{(8)}-B_{(2)}\wedge C_{(6)}
\end{align}
We denote them here by a tilde rather than a new letter.
These potentials are then given by
\begin{align}\label{eq:C8B8tilde-potential}
	\tilde C_{(8)}=
	48\, c_2^2\, c_6^2\, k_1^4\bigg (&
	3\cG \, \cU_1 (\cA_1-\tilde\cA_1)
	-\frac{6\cG^2\partial_w\cA_1\partial_\tw\tilde\cA_1}{\kappa^2}
	-15i (\cY_1-\tilde\cY_1)
	\nonumber\\&
	-5i(\cA_1^2+\tilde\cA_1^2-\cA_1\tilde\cA_1) (\cA_1\tilde\cA_2-\tilde\cA_1\cA_2-2 i\cG)
	\bigg ) \, \hat e^{\, 0 \cdots 7}
\end{align}
For the 6-form potentials the expressions in the two bases agree, since $C_{(4)}=0$.

\clearpage

\section{Closed form of $\cB$ for elliptic solutions}
\label{app:cB-polylogs}

In this appendix we  give a closed form expression for $\cB$ for the elliptic $L=3$ solutions presented in section \ref{sec:sol}. These are expressed in terms of elliptic polylogarithms. We start by collecting some definitions, for which we follow \cite{Broedel:2017kkb}. We start with an elliptic curve defined by,
\begin{align}
	y^2&=P_3(w) & P_3(w)&=(w-a_1)(w-a_2)(w-a_3)
\end{align}
Associated elliptic polylogarithms are defined through iterated integration by
\begin{align}
	E_3\big(\substack{n_1,..,n_k\\c_1,..,c_k};w,\vec{a}\big)
	&=
	\int_0^w \varphi_{n_1}(c_1,t,\vec{a})E_3\big(\substack{n_2,..,n_k\\c_2,..,c_k};w,\vec{a}\big)
	&E_3(;w,\vec{a})&=1
\end{align}
with integration kernels $\varphi_n(c,w,\vec{a})$.
To define the integration kernels we introduce
\begin{align}
	a_{ij}&=a_i-a_j &
	c_3&=\frac{\sqrt{a_{31}}}{2} 
	&
	\tilde\Phi_3(w)&=\frac{-3w+s_1(\vec{a})}{3c_3 y}
\end{align}
where $s_1(\vec{a})=a_1+a_2+a_3$.
For the definitions of the periods $\omega_{1/2}$ and quasi-periods $\eta_{1/2}$ we refer to \cite{Broedel:2017kkb}.
The integration kernels we will use are
\begin{align}
	\varphi_1(0,w)&=\frac{c_3}{y} \\ 
	\varphi_1(\infty,w)&=\frac{c_3}{y}Z_3(w)
	&
	Z_3(w)&=\int_{a_3}^w dt \Phi_3(t)
	&
	\Phi_3(w)&=\tilde\Phi_3(w)-8c_3\frac{\eta_1}{\omega_1 y}
\end{align}
We will use these definitions to express the holomorphic one-form $\partial_w\cB$ and integrate for $\cB$.
For the case at hand, we have $a_1=0$, $a_2=1$ and $a_3=b$ and $P_3(w)=w(w-1)(w-b)$.
Using $y=\sqrt{P_3(w)}$ the differentials in (\ref{eq:dApm-L3-decomp}) can be expressed straightforwardly as
\begin{align}
	\partial_w\cA_+&=\partial_w\left(\frac{2}{3}a_+ y\right)
	+\alpha\frac{1}{y}+\beta \frac{w}{y}
	\nonumber\\
	\partial_w\cA_-&=\partial_w\left(\frac{2}{3}a_+^\star y\right)
	+\alpha^\star\frac{1}{y}+\beta^\star \frac{w}{y}
\end{align}
where the constants $\alpha$, $\beta$ are related to $A$, $B$ by simply matching the terms.
Using the above definitions this becomes
\begin{align}
	\partial_w\cA_+&=\partial_w\left(\frac{2}{3}a_+ y\right)
	+\tilde \alpha\varphi_1(0,w)+\tilde\beta \Phi_3(w)
	\nonumber\\
	\partial_w\cA_-&=\partial_w\left(\frac{2}{3}a_+^\star y\right)
	+\tilde \alpha^\star\varphi_1(0,w)+\tilde\beta^\star \Phi_3(w)
\end{align}
where $\tilde\alpha$, $\tilde\beta$ can be expressed straightforwardly in terms of $\alpha$, $\beta$ and the periods and quasi-periods.
The functions themselves are then
\begin{align}
	\cA_+&=\cA_+^0 + \frac{2}{3}a_+y+\tilde \alpha E_3\big(\substack{1\\0};w,\vec{a}\big)+\tilde\beta Z_3(w)
	\nonumber\\
	\cA_-&=\cA_-^0 + \frac{2}{3}a_+^\star y+ \tilde \alpha^\star E_3\big(\substack{1\\0};w,\vec{a}\big)+\tilde\beta^\star Z_3(w)
\end{align}
\sm
We now come to the integration for $\cB$. To this end we start with the differential
\begin{align}
	\partial_w\cB&=-\cA_+\partial_w\cA_-+\cA_-\partial_w\cA_+
\end{align}
We have
\begin{align}
	\cA_+\partial_w\cA_-&=\partial_w(\cA_+^0\cA_-) + \frac{2}{3}a_+y\partial_w\cA_-
	+\tilde\alpha\tilde\alpha^\star E_3\big(\substack{1\\0};w,\vec{a}\big)\varphi_1(0,w)+\tilde\beta\tilde\beta^\star Z_3(w)\Phi_3(w)
	\nonumber\\
	&\hphantom{=}
	+\tilde\alpha\tilde\beta^\star E_3\big(\substack{1\\0};w,\vec{a}\big) \Phi_3(w)
	+\tilde\beta\tilde\alpha^\star Z_3(w)\varphi_1(0,w)
\end{align}
Upon combining with the $\pm$-swapped expression, the terms proportional to $|\tilde\alpha|^2$ and $|\tilde\beta|^2$ drop out, and we are left with
\begin{align}
	\partial_w\cB&=\partial_w\left(-\cA_+^0\cA_-+\cA_-^0\cA_+\right) + \frac{2}{3}y\left(a_+^\star\partial_w\cA_+-a_+\partial_w\cA_-\right)
	\nonumber\\
	&\hphantom{=}+\big(\tilde\alpha\tilde\beta^\star-\alpha^\star\beta\big)
	\left(
	Z_3(w)\varphi_1(0,w)-E_3\big(\substack{1\\0};w,\vec{a}\big) \Phi_3(w)
	\right)
\end{align}
The first term is a total derivative, and the second term is too, as can be seen from
\begin{align}
	y\partial_w\cA_+&=\partial_w\left(\frac{1}{3}a_+ y^2+\alpha w+\frac{1}{2}\beta w^2\right)
\end{align}
Using $\varphi_1(0,w)=\partial_wE_3\big(\substack{1\\0};w,\vec{a}\big)$, we can write
\begin{align}
	Z_3(w)\varphi_1(0,w)-E_3\big(\substack{1\\0};w,\vec{a}\big) \Phi_3(w)&=
	2Z_3(w)\partial_wE_3\big(\substack{1\\0};w,\vec{a}\big)-	\partial_w\left[E_3\big(\substack{1\\0};w,\vec{a}\big)Z_3(w)\right]
\end{align}
where
\begin{align}
	Z_3(w)\partial_wE_3\big(\substack{1\\0};w,\vec{a}\big)=Z_3(w)\varphi_1(0,w)=\frac{c_3}{y}Z_3(w)=\varphi_1(\infty,w)
\end{align}
Putting all pieces together, we arrive at
\begin{align}
	\partial_w\cB&=\partial_w\left(-\cA_+^0\cA_-+\cA_-^0\cA_+
	+\frac{2}{3}(a_+^\star\alpha-a_+\alpha^\star) w+\frac{1}{3}(a_+^\star\beta-a_+\beta^\star) w^2\right)
	\nonumber\\
	&\hphantom{=}
	-\partial_w\left[\big(\tilde\alpha\tilde\beta^\star-\alpha^\star\beta\big)E_3\big(\substack{1\\0};w,\vec{a}\big)Z_3(w)\right]
	+2\big(\tilde\alpha\tilde\beta^\star-\alpha^\star\beta\big)\varphi_1(\infty,w)
\end{align}
From this form we finally get $\cB$ itself as
\begin{align}
	\cB&=\cB_0
	-\cA_+^0\cA_-+\cA_-^0\cA_+
	+\frac{2}{3}(a_+^\star\alpha-a_+\alpha^\star) w+\frac{1}{3}(a_+^\star\beta-a_+\beta^\star) w^2
	\nonumber\\&\hphantom{=}
	+\big(\tilde\alpha\tilde\beta^\star-\alpha^\star\beta\big)\left(2E_3\big(\substack{1\\ \infty};w,\vec{a}\big)-E_3\big(\substack{1\\0};w,\vec{a}\big)Z_3(w)\right)
\end{align}
When forming $\cG$ one may expect single-valued elliptic polylogarithms, similar to the expression for $\cG$ in terms of standard single-valued polylogarithms for $AdS_6\times S^2\times\Sigma$ in \cite[appendix C]{Uhlemann:2020bek}.

\clearpage

\section{Comparison with Komatsu et al. in  \cite{Komatsu:2024bop}}
\label{appc}

In this section we first connect the formulation of the $dS_{1,5}\times S^2\times \Sigma$ and $S^6\times S^2\times \Sigma$ solutions in terms of two holomorphic functions $\cA_\pm$ to the formulation of $S^6\times S^2\times\Sigma$ solutions in \cite{Komatsu:2024bop}, where a single real function $V$ satisfying the Laplace equation associated with a four-dimensional  axially symmetric setup was used to parametrize the solutions. We then work within the solutions discussed in \cite{Komatsu:2024bop} and discuss the computation of the fundamental string charge.

\subsection{Gauge-fixed formulation}

We start from the formulation of the $dS_{1,5}\times S^2\times\Sigma$ solutions used in section \ref{sec:3}, in terms of two locally holomorphic functions $\cA_{1,2}$. This pair of functions can locally be reduced to a single holomorphic function $f$ by using $\cA_1$ as coordinate as follows, 
\begin{align}
	\cA_1&=\sqrt{2}w & \cA_2&=-\sqrt{2}if
\end{align}
The expressions in (\ref{2.g.2}) and (\ref{2.g.2b}) reduce to\footnote{We note that our sign convention for $\hat \cG$ differs from the convention for $\cG^\prime$ in \cite[(A.27)]{Komatsu:2024bop}, with our definitions leading to $\hat\kappa^2=-\partial_w\partial_{\bar w}\cG$ as compared to ${\kappa'}^2=+\partial_w\partial_{\bar w}\cG'$.}
\begin{align}
	\hat\kappa^2&=4\, \Im(\partial_w f) & \partial_w\cB&=\partial_w(2iwf) -4i f
&
	\hat\cG&=8\, \Im(w)\Re(f)-8\; \Im(\int f)
\end{align}
We can now set, as in \cite[Appendix A]{Komatsu:2024bop},
\begin{align}
	w&=z+i\rho & 
	\int f&=U+i W
	&
	W&=-\rho V
\end{align}
with harmonic functions $U$, $W$. Using holomorphicity of $f$,
\begin{align}
	\hat\cG&=-8W+8\rho \partial_\rho W = 8\rho\partial_\rho V
\end{align}
$W$ being harmonic, $\nabla^2W=0$, is equivalent to $V$ being harmonic in 4d and axially symmetric,
\begin{align}
	\partial_z^2V+2\rho^{-1}\partial_\rho V+\partial_\rho^2V=0
\end{align}
The real coordinates $(\rho,z)$ were used to parametrize the upper right quadrant in \cite{Komatsu:2024bop}, with $S^6$ shrinking along the positive $\rho$ axis and finite horizontal cuts starting on the positive $z$ axis along which $S^2$ shrinks. In the electrostatics description these can be viewed as grounded conducting discs.

In order to locally map a generic solution parametrized by $\cA_{1,2}(w)$ to the formulation used in \cite{Komatsu:2024bop}, one has to trace through the above identifications. The first step is to define a new coordinate $u$ by,
\begin{equation}
	u=\frac{1}{\sqrt{2}}\cA_1(w)\qquad\longleftrightarrow\qquad
	w=\cA_1^{-1}(\sqrt{2}u)
\end{equation}
For the global solutions discussed in the main part, the function $\cA_1$ thus provides the map from the solutions on the upper half plane to the geometries that appeared in \cite{Komatsu:2024bop}. Then the functions $f$ and $W$ are expressed in terms of this coordinate $u$ as
\begin{align}
	f&=\frac{i}{\sqrt{2}}\cA_2 \Big (\cA_1^{-1}(\sqrt{2}u) \Big )
	&
	W&=\Im \int du\frac{i}{\sqrt{2}}\cA_2 \Big (\cA_1^{-1}(\sqrt{2}u) \Big )
\end{align}
The formulation in terms of $V$ is then recovered by setting
\begin{align}
u&=z+i\rho & V=-W/\rho
\end{align}
The formulation in terms of two locally holomorphic functions can be viewed as a uniformization of the cut quadrant to the upper half plane. It has the advantages that the geometry of the underlying Riemann surface can be kept simple and that $(\cA_1,\cA_2)$ have natural transformation behavior under $SL(2,\RR)$, namely as doublet.

\subsection{Fundamental string charge}\label{app:F1-charge-electrostatics}

The analysis in \cite{Komatsu:2024bop} concluded that the fundamental string charge vanishes for the $S^6\times S^2\times\Sigma$ solutions constructed there. By contrast, we find that the solutions constructed in section \ref{sec:sol} generally carry non-vanishing F1 charge. Demanding the F1 charges to vanish would constrain the parameters in the solutions of section \ref{sec:sol} and lead to a mismatch with the number of free parameters found in \cite{Komatsu:2024bop}. It would also prevent the parameter count from matching that of the IKKT configurations discussed in section \ref{sec:IKKT}.\footnote{For $L=3$ a useful perspective can be gained from the residual $SL(2,\RR)$ transformations that preserve the D-instanton asymptotics, discussed below (\ref{eq:p5half-dilaton-asympt}). Such a transformation with $a=d=1$ can be used to transform the F1 charge to zero while preserving the D1 and NS5 charges as well as the leading asymptotics of $dC_{(2)}$, which was linked in \cite{Komatsu:2024bop} to the parameter $\Omega$ in (\ref{IKKTa}). This transformation shifts the constant part in the $\tau_-$ asymptotics (cf.~(\ref{eq:p5half-tau})), which was identified in \cite{Komatsu:2024bop} with the $\eta$ parameter in (\ref{IKKTa}), thus linking $\eta$ to the remaining parameters. For $L>3$, an $SL(2,\RR)$ transformation can only remove all F1 charges if the $(Q_{\rm F1},Q_{\rm D1})$ charge vectors in (\ref{eq:hyper-charges}) are all proportional to each other.}

\sm

To address this tension, we revisit the fundamental string charge in the solutions of \cite{Komatsu:2024bop}. We  evaluate the general expression for the F1 charge given in \cite{Komatsu:2024bop} using the integral representations for the potential $V$ for generic disc configurations as given there, and then explicitly derive the F1 charge for a single high disc, for which we find a non-zero F1 charge. 
Thus, our evaluation does not reproduce the vanishing F1 charge reported in~\cite{Komatsu:2024bop}.

\subsubsection{General expression}

Our starting point is the expression for the F1 charge given in equation \cite[(3.39)]{Komatsu:2024bop}, 
\begin{equation}
	N_{\rm F1}=A\int d\rho\,\rho^3 \Big ( 3\rho(\partial_z\partial_\rho V)^2-\partial_\rho(\partial_z V)^2 \Big )\Big\vert_{z=z_s-\epsilon}^{z=z_s+\epsilon}
\end{equation}
We have absorbed all numerical prefactors  into the constant $A$. Here,  $z_s$ is the vertical location of one of the conducting discs, and $V$ is the electrostatics potential. Contrary to the conclusion below \cite[(3.39)]{Komatsu:2024bop},  we find that this expression does not generally vanish.

\sm

To evaluate this expression more explicitly, we split the potential into the singular part associated with the disc at $z=z_s$ and the remainder, which is regular at $z=z_s$,
\begin{equation}
		V(\rho,z)=V_{\rm reg}(\rho,z) + V_s(\rho,z)
\end{equation}
The expression for a single disc was given in \cite[(3.42)]{Komatsu:2024bop}, from which we identify,
\begin{equation}
\label{eq:V-s}
	V_s(\rho,z)=\frac{1}{4\pi^2\rho}\int_0^{\rho_s} dr\, f_s(r)\frac{\rho-r}{(\rho-r)^2+(z-z_s)^2}
\end{equation}
In this expression $f_s'(r)=-2\pi r \sigma_s(r)$ with $\sigma_s(r)$ the charge density of the disc at $z=z_s$.

\sm

The expression for the F1 charge only involves $\partial_zV$, which can be evaluated as follows,
\begin{align}
	\partial_z V_s\big\vert_{z=z_s\pm \epsilon}&=\mp \frac{\epsilon}{2\pi^2\rho}\int_0^{\rho_s}dr f_s(r)\frac{\rho-r}{\left((\rho-r)^2+\epsilon^2\right)^2}
\end{align}
Using integration by parts,
\begin{align}
	\partial_z V_s\big\vert_{z=z_s\pm \epsilon}&=\mp \frac{\epsilon}{4\pi^2\rho}\left[\frac{f_s(r)}{(\rho-r)^2+\epsilon^2} \right]_0^{\rho_s}
	\pm \frac{\epsilon}{4\pi^2\rho}\int_0^{\rho_s}dr \frac{f_s'(r)}{(\rho-r)^2+\epsilon^2} 
\end{align}
The boundary terms are $\mathcal O(\epsilon)$ and, using $r=\rho+\epsilon y$ for the remaining integral, yields,
\begin{align}
	\partial_z V_s\big\vert_{z=z_s\pm \epsilon}&=\pm\frac{1}{4\pi^2\rho}\int_{-\rho/\epsilon}^{(\rho_s-\rho)/\epsilon} dy\,\frac{f_s'(\rho+\epsilon y)}{y^2+1} 
	+\mathcal O(\epsilon)
	= \pm \frac{1}{4\pi\rho}f_s'(\rho)+\mathcal O(\epsilon)
\end{align}
For the second equality we assumed $\rho<\rho_s$.

\sm

The remaining task is to evaluate the combinations in the expression for the F1 charge, $(\partial_z V)^2$ and $(\partial_\rho\partial_zV)^2$. Using the split of $V$ into $V_{\rm reg}$ and $V_s$ we find
\begin{align}
	\left(\partial_z V\right)^2 \Big\vert_{z=z_s-\epsilon}^{z=z_s+\epsilon}
	&=\frac{1}{\pi\rho}f_s'(\rho)\partial_z V_{\rm reg}
	\nonumber\\
	\left(\partial_\rho\partial_z V\right)^2 \Big\vert_{z=z_s-\epsilon}^{z=z_s+\epsilon}
	&=\frac{1}{\pi}(\partial_z\partial_\rho V_{\rm reg})\partial_\rho\left(\frac{f_s'(\rho)}{\rho}\right)
\end{align}
The resulting expression for the string charge becomes,
\begin{equation}
	N_{\rm F1}=\frac{A}{\pi}\int d\rho\,\rho^3 \left \{ 3\rho \Big (\partial_z\partial_\rho V_{\rm reg} \Big )\partial_\rho\left(\frac{f_s'(\rho)}{\rho}\right )
	-\partial_\rho\left(\frac{f_s'(\rho)}{\rho}\partial_z V_{\rm reg}\right)\right \} 
\end{equation}
In terms of the charge density $\sigma_s$ defined below (\ref{eq:V-s}) this becomes
\begin{equation}
\label{eq:app-C-F1-charge}
	N_{\rm F1}=-2A\int d\rho\,\rho^3 \Big \{ 3\rho \big (\partial_z\partial_\rho V_{\rm reg} \big )\sigma_s'(\rho)
	-\partial_\rho\left(\sigma_s(\rho)\partial_z V_{\rm reg}\right) \Big \}
\end{equation}
The value of this integral depends on the charge density $\sigma_s(\rho)$ of the disc at $z=z_s$ and on the remaining part of the potential.

\subsubsection{Explicit evaluation for a high disc}

To test whether the general expression for the F1 charge in (\ref{eq:app-C-F1-charge}) is non-zero, we evaluate it for the case of a single disc at large $z_s$. The equation determining the charge density $\sigma_s(r)$ for a high disc is \cite[(C.5)]{Komatsu:2024bop}, which reads,
\begin{equation}
	\rho\sigma_s(\rho)=\frac{1}{\pi^2}\text{p.v.}\int_{-\rho_s}^{\rho_s}du\sqrt{\frac{\rho_s^2-u^2}{\rho_s^2-\rho^2}}\frac{\tilde f'(u)}{\rho-u}
\end{equation}
where $\tilde f$ is a function defined by \cite[(C.4)]{Komatsu:2024bop} as
\begin{equation}
	\tilde f(\rho)=2\pi\rho(V_0-V_{\rm bg}-V_\text{other discs})
\end{equation}
In this expression $V_0$ is a constant to be determined, $V_{\rm bg}=\alpha(\rho^2z-z^3)$ is a background potential with a coefficient $\alpha$ whose value is identified with the strength of the polarization deformation in the IKKT model. In particular, $\alpha$ is non-zero. $V_\text{other discs}$ is to be ignored in the limit of a high disc. We thus find
\begin{equation}
	\rho\sigma_s(\rho)=\frac{2}{\pi\sqrt{\rho_s^2-\rho^2}}\left((V_0+\alpha z_s^3)I_0-3\alpha z_s I_2\right)
\end{equation}
where
\begin{equation}
	I_n=\text{p.v.}\int_{-\rho_s}^{\rho_s}du\sqrt{\rho_s^2-u^2}\frac{u^n}{\rho-u}
\end{equation}
By decomposing the fraction, $I_2$ can be expressed in terms of $I_0$, and we find
\begin{align}
	I_2&=\rho^2 I_0-\frac{1}{2}\rho \rho_s^2\pi & I_0&=\pi \rho
\end{align}
The result is
\begin{equation}
	\sigma_s(\rho)=\frac{2(V_0+\alpha z_s^3)+3\alpha z_s(\rho_s^2-2\rho^2)}{\sqrt{\rho_s^2-\rho^2}}
\end{equation}
Fixing $V_0$ such that $\sigma(\rho)$ is finite as $\rho\rightarrow\rho_s$ leads to
\begin{equation}
	\sigma_s(\rho)=6\alpha z_s\sqrt{\rho_s^2-\rho^2}
\end{equation}
With this charge density the F1 charge in (\ref{eq:app-C-F1-charge}) can be evaluated straightforwardly, and we find
\begin{equation}
	N_{\rm F1}=\frac{27}{8}A\pi\alpha^2\rho_s^4z_s(3\rho_s^2+2z_s^2)
\end{equation}
With $A$, $\alpha$, $z_s$ and $\rho_s$ all non-zero, this is a non-vanishing F1 charge.


\clearpage

\bibliographystyle{utphys}
\bibliography{s6s2sigma}
\end{document}